\documentclass{aa}  

\usepackage{natbib}
\usepackage{graphicx}
\usepackage{txfonts}
\usepackage[draft]{hyperref}
\usepackage{color}
\usepackage{upgreek} \newcommand\micron{\ensuremath{\rm \upmu m}\@\xspace} 
\usepackage{color}
\usepackage[version=3]{mhchem}
\begin{document}

   \title{An extensive grid of DARWIN models for M-type AGB stars}

   \subtitle{I. Mass-loss rates and other properties of dust-driven winds}

 \author{S. Bladh\inst{1,2}
          \and
          S. Liljegren\inst{1,3}
          \and
          S. H\"ofner\inst{1}
          \and 
          B. Aringer\inst{2}
         \and
                  P. Marigo\inst{2}
                    }
           \institute{Theoretical Astrophysics, Department of Physics and Astronomy, Uppsala University, Box 516, SE-751 20 Uppsala, Sweden\label{inst1}\\
    \email{sara.bladh@physics.uu.se}
        \and  
   Dipartimento di Fisica e Astronomia Galileo Galilei, Università di Padova, Vicolo dell’Osservatorio 3, 35122 Padova, Italy\label{inst2}
           \and
Department of Astronomy, Stockholm University, Oscar Klein Centre, AlbaNova University Centre, 106 91 Stockholm, Sweden\label{inst3}
           }

   \date{Received 26 February 2019/Accepted 11 April 2019}

 
  \abstract
   {The stellar winds of asymptotic giant branch (AGB) stars are commonly attributed to radiation pressure on dust grains, formed in the wake of shock waves that arise in the stellar atmospheres. The mass loss due to these outflows is substantial, and modelling the dynamical properties of the winds is essential both for studies of individual stars and for understanding the evolution of stellar populations with low to intermediate mass.
}
   {The purpose of this work is to present an extensive grid of dynamical atmosphere and wind models for M-type AGB stars, covering a wide range of relevant stellar parameters. 
}
   {We used the DARWIN code, which includes frequency-dependent radiation-hydrodynamics and a time-dependent description of dust condensation and evaporation, to simulate the dynamical atmosphere. The wind-driving mechanism is photon scattering on submicron-sized Mg$_2$SiO$_4$ grains. The grid consists of $\sim4000$ models, with luminosities from $L_\star=890\,{\mathrm{L}}_\odot$ to  $L_\star=40000\,{\mathrm{L}}_\odot$ and effective temperatures from 2200\,K to 3400\,K. For the first time different current stellar masses are explored with M-type DARWIN models, ranging from 0.75\,M$_\odot$ to 3\,M$_\odot$. The modelling  results are radial atmospheric structures, dynamical properties such as mass-loss rates and wind velocities, and dust properties (e.g. grain sizes, dust-to-gas ratios, and degree of condensed Si). 
}
   {We find that the mass-loss rates of the models correlate strongly with luminosity. They also correlate with the ratio $L_*/M_*$: increasing $L_*/M_*$ by an order of magnitude  increases the mass-loss rates by about three orders of magnitude, which may naturally create a superwind regime in evolution models. There is, however, no discernible trend of mass-loss rate with effective temperature, in contrast to what is found for C-type AGB stars. We also find that the mass-loss rates level off at luminosities higher than $\sim14000\,{\mathrm{L}}_\odot$, and consequently at pulsation periods longer than $\sim800$ days. The final grain radii range from 0.25\,\micron to 0.6\,\micron. The amount of condensed Si is typically between 10\% and 40\%, with gas-to-dust mass ratios between 500 and 4000.
}
   {}

    \keywords{stars: AGB and post-AGB 
   – stars: late-type 
   – stars: mass-loss 
   – stars: winds, outflows 
   – stars: atmospheres
   - stars: evolution
               }
   \maketitle

\section{Introduction}

The atmospheres of cool AGB stars are inherently dynamical due to large-amplitude radial pulsations, leading to a periodically contracting and expanding photosphere. The radial motions of the photospheric gas layers give rise to propagating sound waves that turn into strong shocks in the steep density gradient of the atmosphere. In the wake of these atmospheric shock waves compressed gas is elevated to regions with cooler temperatures, thereby creating favourable conditions for dust formation. The newly formed dust grains interact with the strong radiation field from the star, and momentum is transferred by absorption or scattering of numerous stellar photons. The radiatively accelerated dust particles collide with the surrounding gas and, if sufficient momentum is transferred from the radiation field, a general outflow of both dust and gas can be triggered. 

The presence of silicates in the circumstellar envelopes of M-type AGB stars can be inferred from observations of the characteristic silicate features at 9.7\,$\mu$m and 18\,$\mu$m \citep{wolfney69,low1970,molster2002}. Silicates have therefore been a long-standing candidate as wind-driving dust species in M-type AGB stars. If silicates are Fe-free they can condense comparably close to the star (at $\sim2R_*$) and grow to sizes of about 0.1-1 \micron, comparable to the wavelength of the stellar flux maximum. At these grain sizes, photon scattering contributes substantially to the radiative acceleration and can generate sufficient momentum to drive a stellar wind.

Photon scattering on Fe-free silicates as a wind-driving mechanism was suggested by \cite{hoefner2008}, in response to the conundrum presented by \cite{woitke2006} that radiative heating prevents Fe-bearing silicates and metallic Fe from condensing sufficiently close to the star to trigger an outflow. The wind models presented by \citet[][computed with the DARWIN code]{hoefner2008} were proof of concept for a small set of stellar parameters typical of M-type AGB stars. They showed that photon scattering on Fe-free silicates can produce winds, and that the resulting wind properties are reasonable. This wind scenario was further explored in \cite{bladh2012} and \cite{bladh2013}, and \cite{bladh2015} presented the first large grid of dynamical models for M-type AGB stars, systematically covering a wide range of effective temperature and luminosities but still restricted to models of solar mass. Synthetic observables from the models in the grid were successfully compared to observed wind properties and near-IR photometry, indicating that the DARWIN models capture the essential parts of  the momentum transfer and of the variable spectral energy distribution in these stars. Recent observations also show evidence of dust grains of sizes $0.1-0.5$\,$\mu$m in the close vicinity M-type AGB stars \citep[see e.g.][]{norris12,Ohnaka16,Ohnaka17}. Discussion about other dust species observed in the circumstellar environment in M-type AGB stars can be found in \cite{bladh2012} and \cite{bladh2013}, whereas \cite{hoefner2016} specifically investigated the potential role of Al$_2$O$_3$ in wind acceleration.

The main focus of the DARWIN models is to study wind formation and the properties of stellar winds by modelling the mass-loss process from first principles. This paper is the first in a series presenting a new extensive grid of DARWIN models for M-type AGB stars, exploring different stellar masses for the first time, but also covering a wider range of effective temperatures and stellar luminosities. The focus here is on presenting this new grid and the resulting dynamical properties, whereas forthcoming papers will focus on mass-loss descriptions, photometry, and the morphology of the outflows. The dynamical output from this extensive model grid can be used in stellar evolution codes modelling the AGB phase, and  to interpret data from detailed observational studies of individual AGB stars.

The paper is organised in the following way. In Sect.~\ref{modmeth} we summarise the main features of the DARWIN code used to describe the atmospheres and winds of M-type AGB stars. In Sect.~\ref{gridpar} we present the input parameters of the model grid, with a short motivation of how they were chosen. The resulting wind characteristics (mass-loss rates and wind velocity)  and grain properties are presented in Sect.~\ref{modresult}; we also investigate how these properties depend on different input parameters. In Sect.~\ref{compobssyn} we compare the observed and synthetic wind properties, in Sect.~\ref{stellarevo} we present a simple mass-loss routine for M-type AGB stars to be used in stellar evolution codes, and in Sect.~\ref{outlook} we provide an outlook on future applications. Finally, in Sect.~\ref{conclusions} we give a short summary of our conclusions.


\begin{figure*}
\centering
\includegraphics[width=0.49\textwidth]{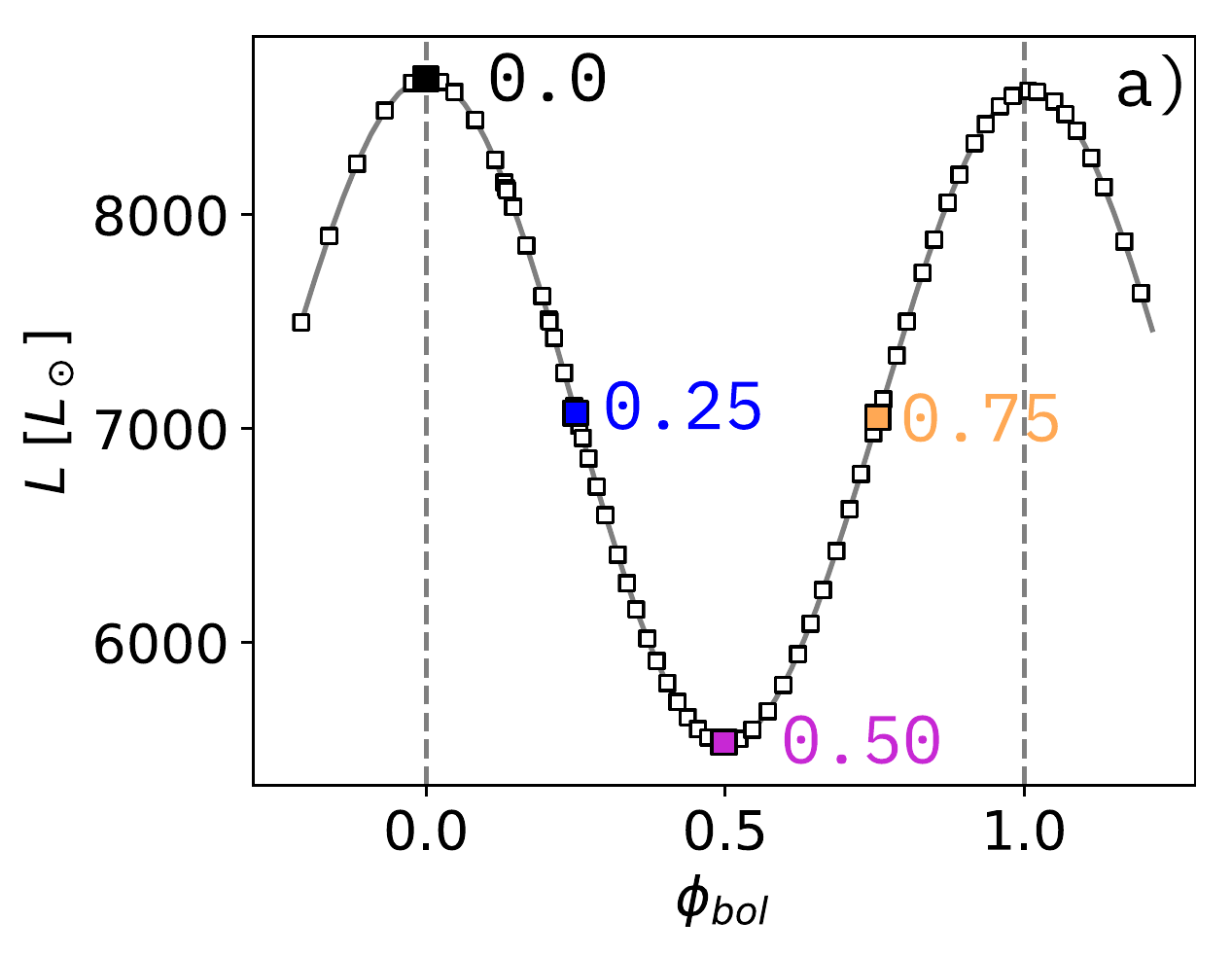} 
\includegraphics[width=0.49\textwidth]{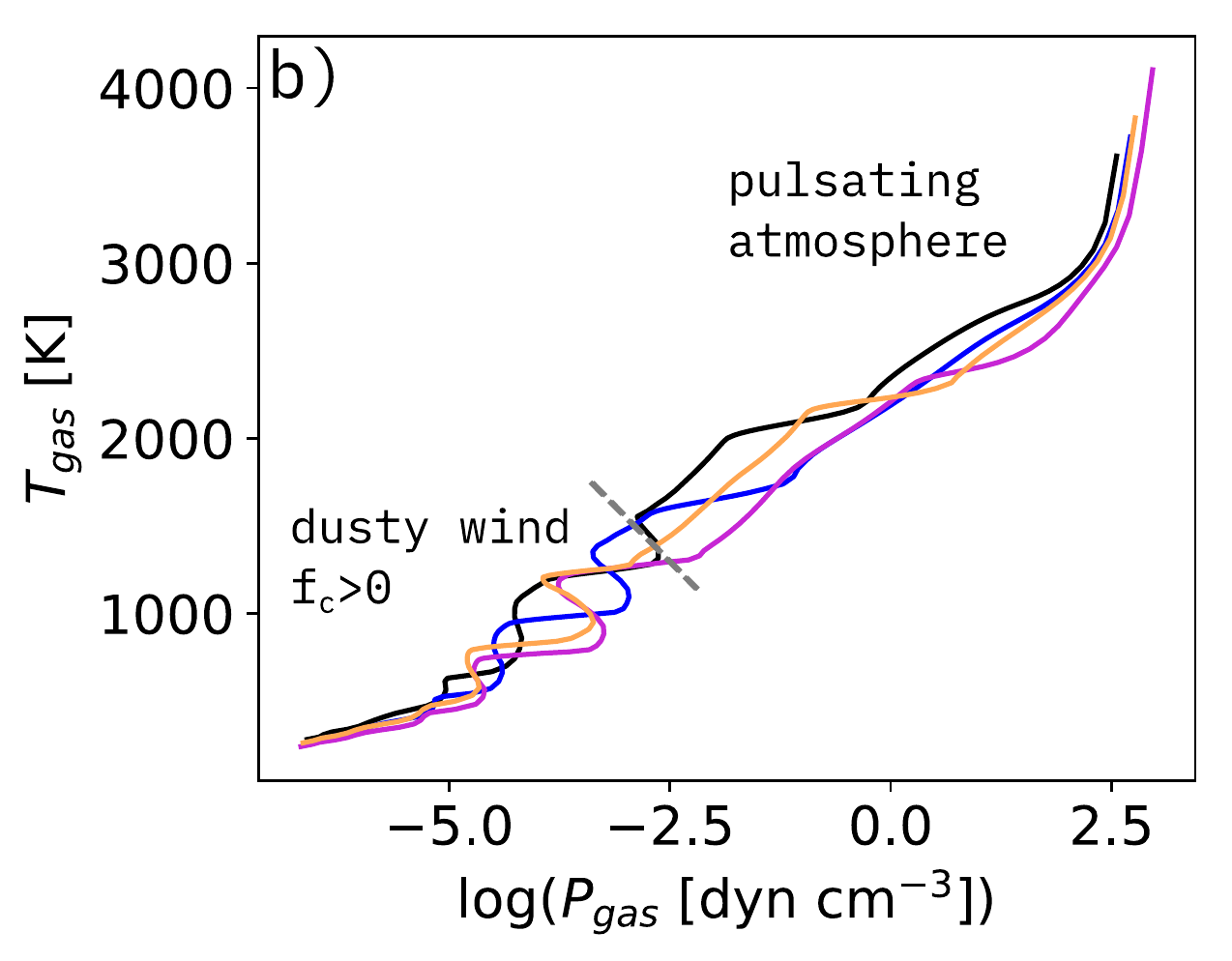} 
\includegraphics[width=0.49\textwidth]{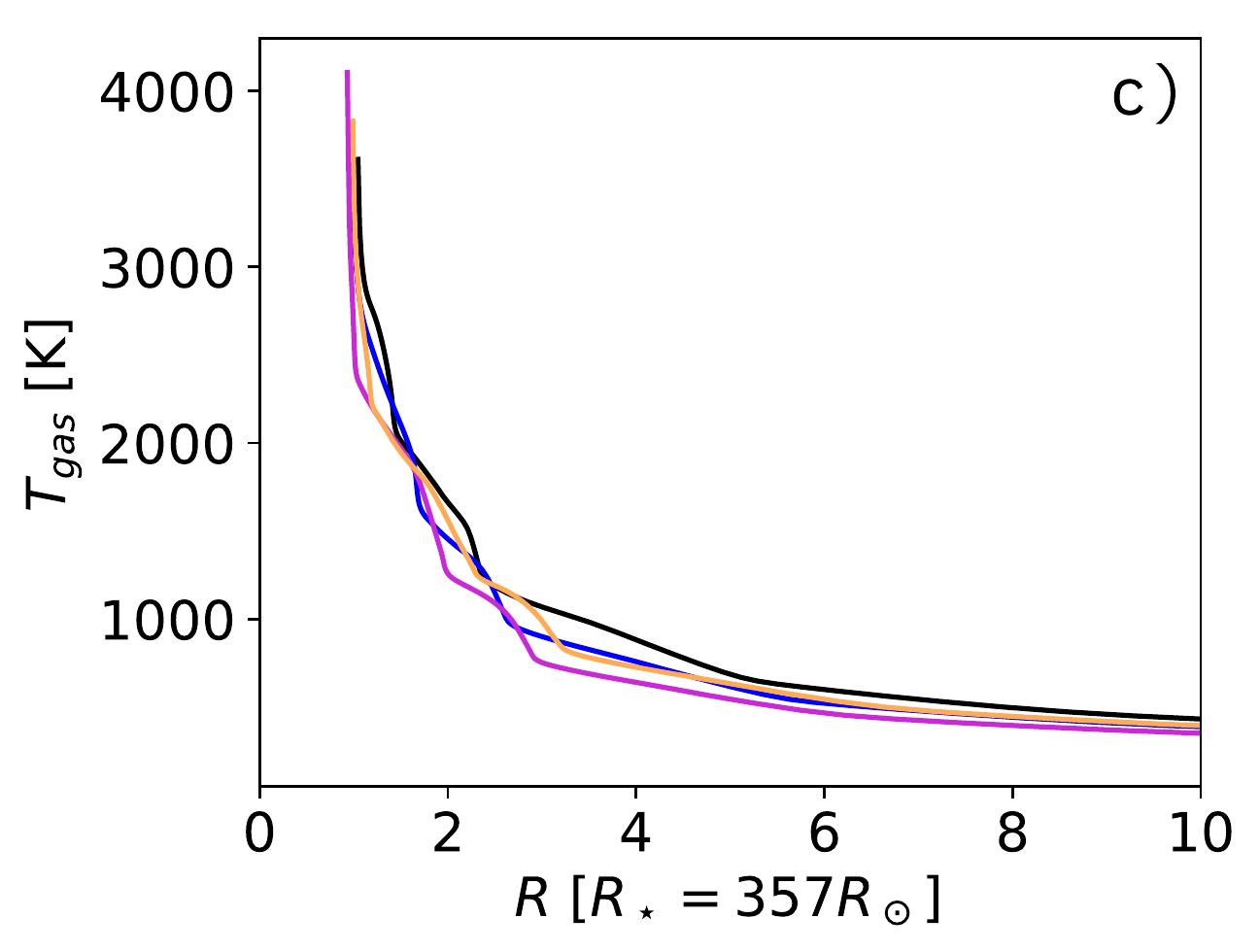} 
\includegraphics[width=0.49\textwidth]{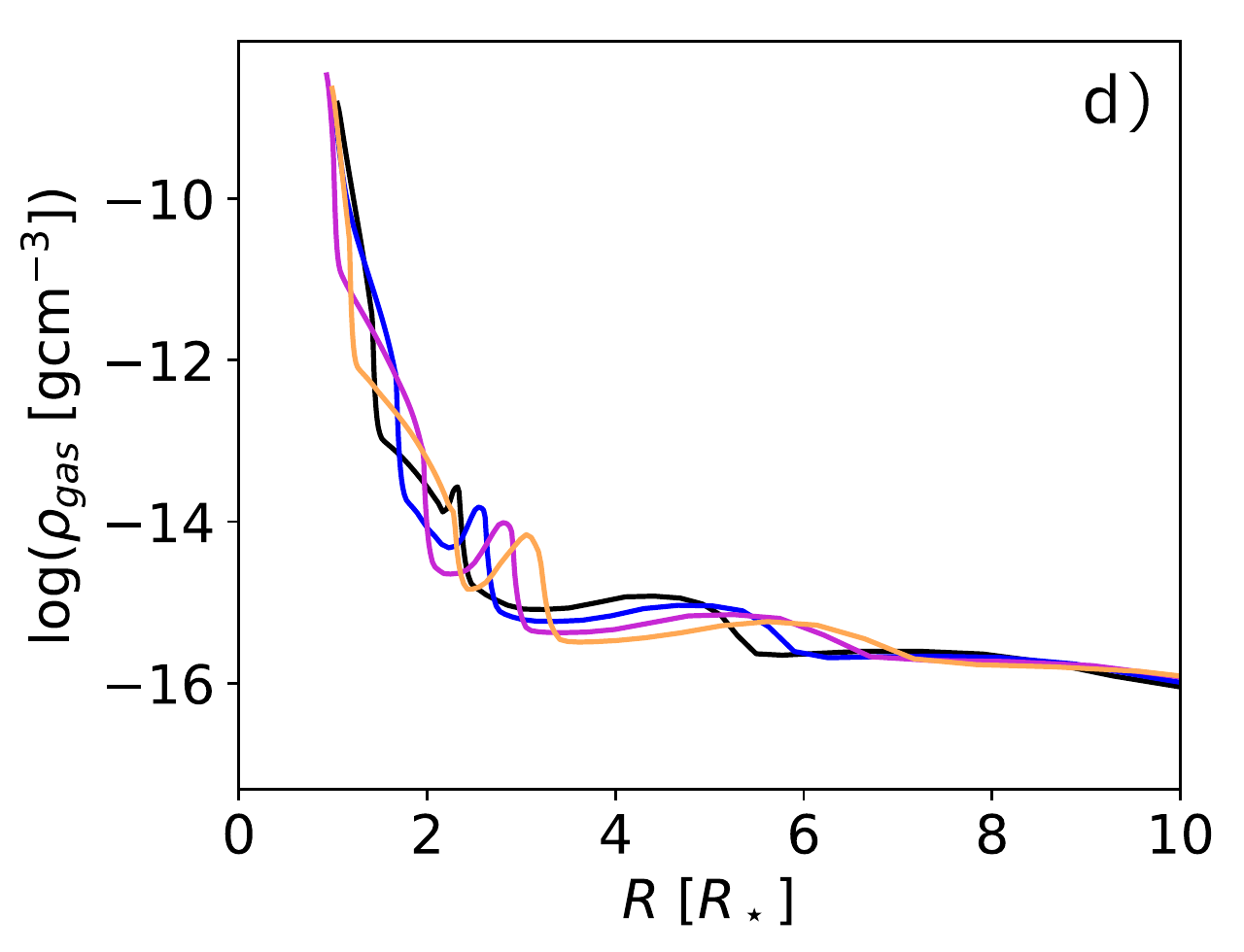} 
\includegraphics[width=0.49\textwidth]{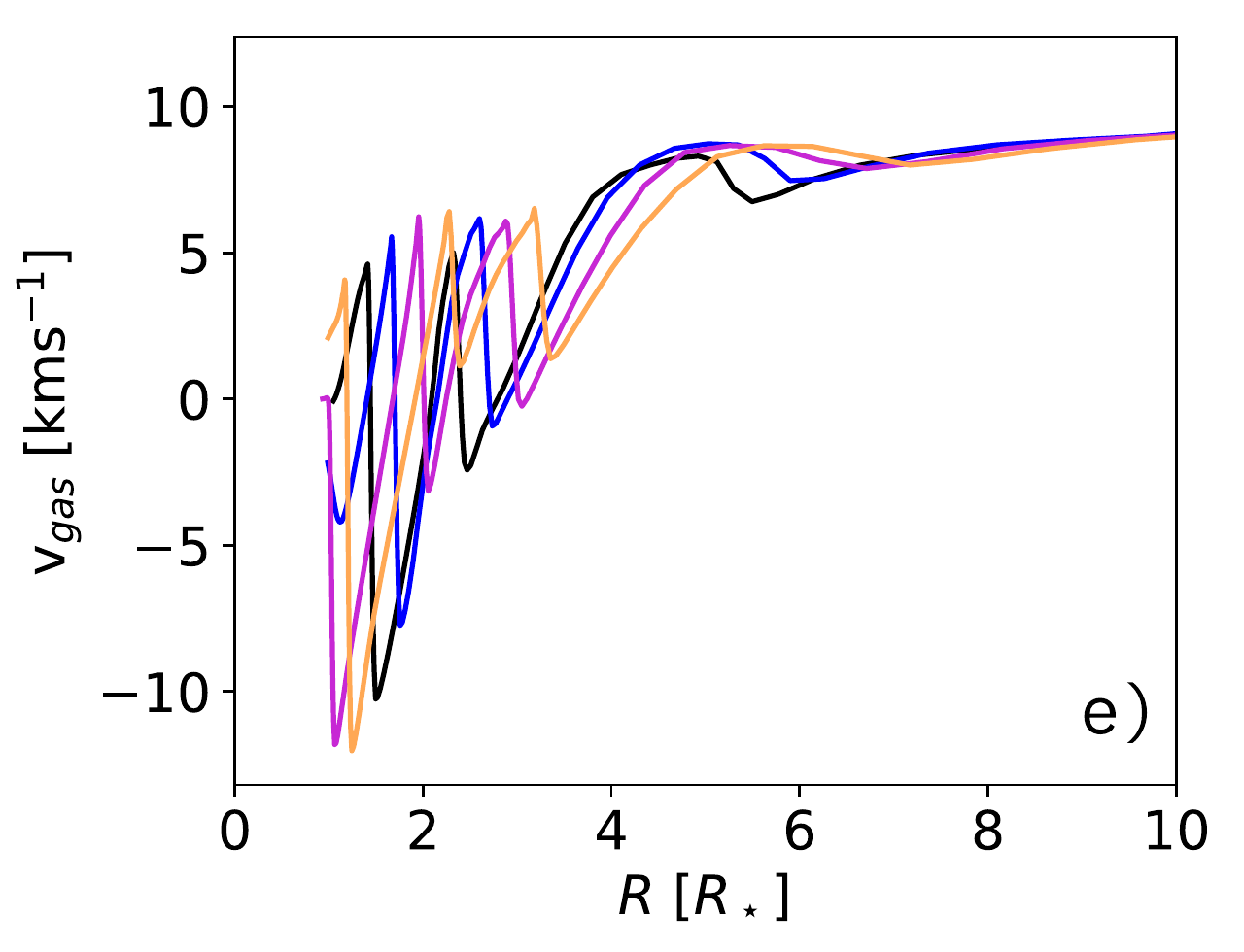} 
\includegraphics[width=0.49\textwidth]{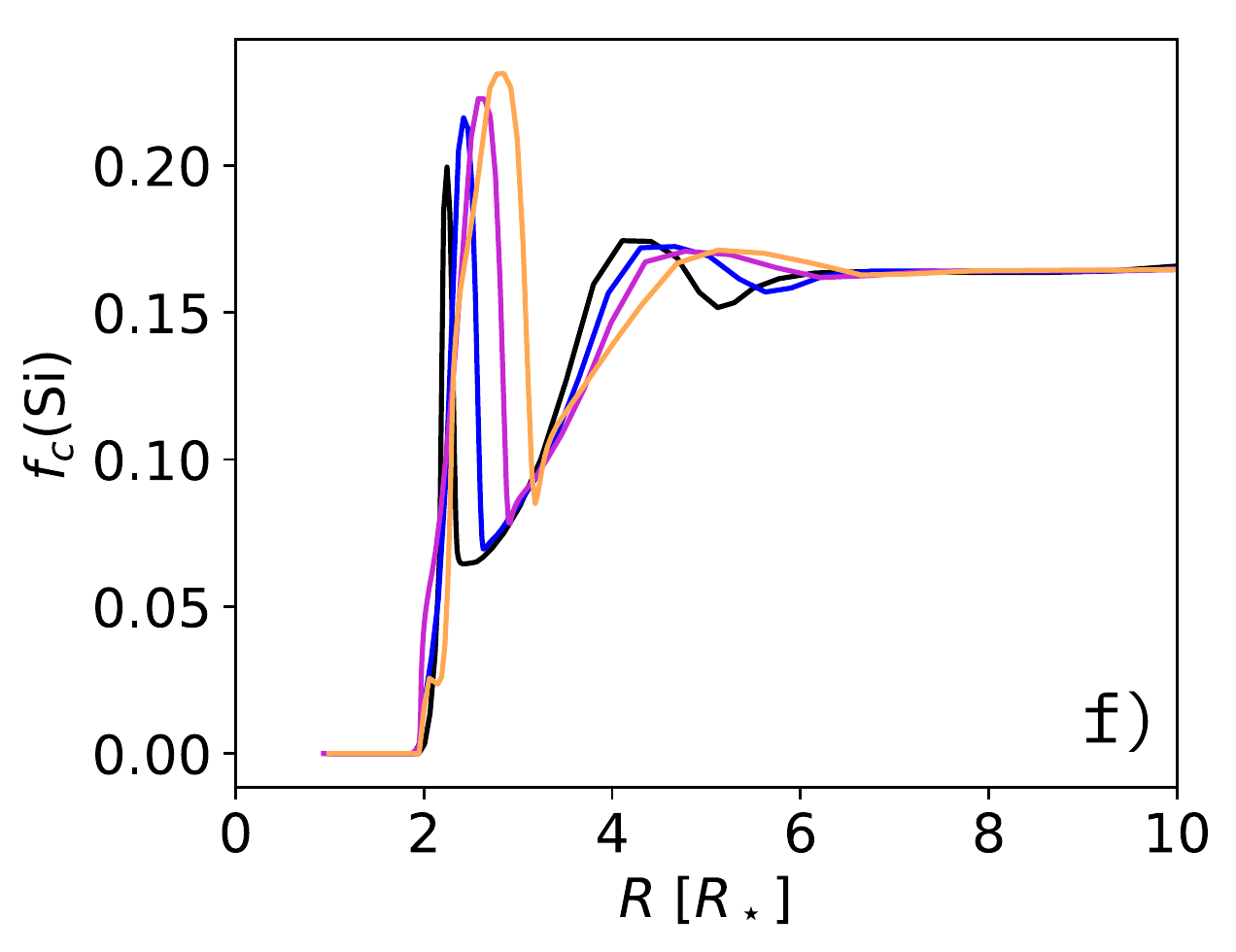}
\caption{Characteristic properties of a DARWIN model with $M_*=1\,\mathrm{M}_{\odot}$, $\log L_*=3.85\,\mathrm{L}_{\odot}$, $T_*=2700\,$K, $u_{\mathrm{p}}=2$\,km/s, and $\log (n_{\mathrm{d}}/n_{\mathrm{H}})=-15.0$. Plotted in panel a is the bolometric light curve resulting from the variable inner boundary. The boxes indicate instances of time for which snapshots of the atmospheric structure were stored, partly labelled with the corresponding bolometric phase. The dotted vertical lines indicate the points in time of $\Phi_{\mathrm{bol}}=0.0$ and 1.0. The other panels show the atmospheric structures of selected phases during one pulsation cycle (colour-coded in the same way as the phase labels in panel a). Panel b shows gas temperature vs. gas pressure used to characterise classic hydrostatic stellar atmospheres. The middle and lower panels illustrate the radial structures of gas temperatures (panel c), gas densities (panel d), gas velocities (panel e), and degree of Si condensed into Mg$_2$SiO$_4$ grains (panel f). The plots are cut at 10 stellar radii to provide better resolution of the dynamically important region.}
\label{fig:struct}
\end{figure*}

\section{Modelling methods}
\label{modmeth}
\subsection{DARWIN  atmosphere and wind models}

\begin{figure*}
\centering
\includegraphics[width=0.45\textwidth]{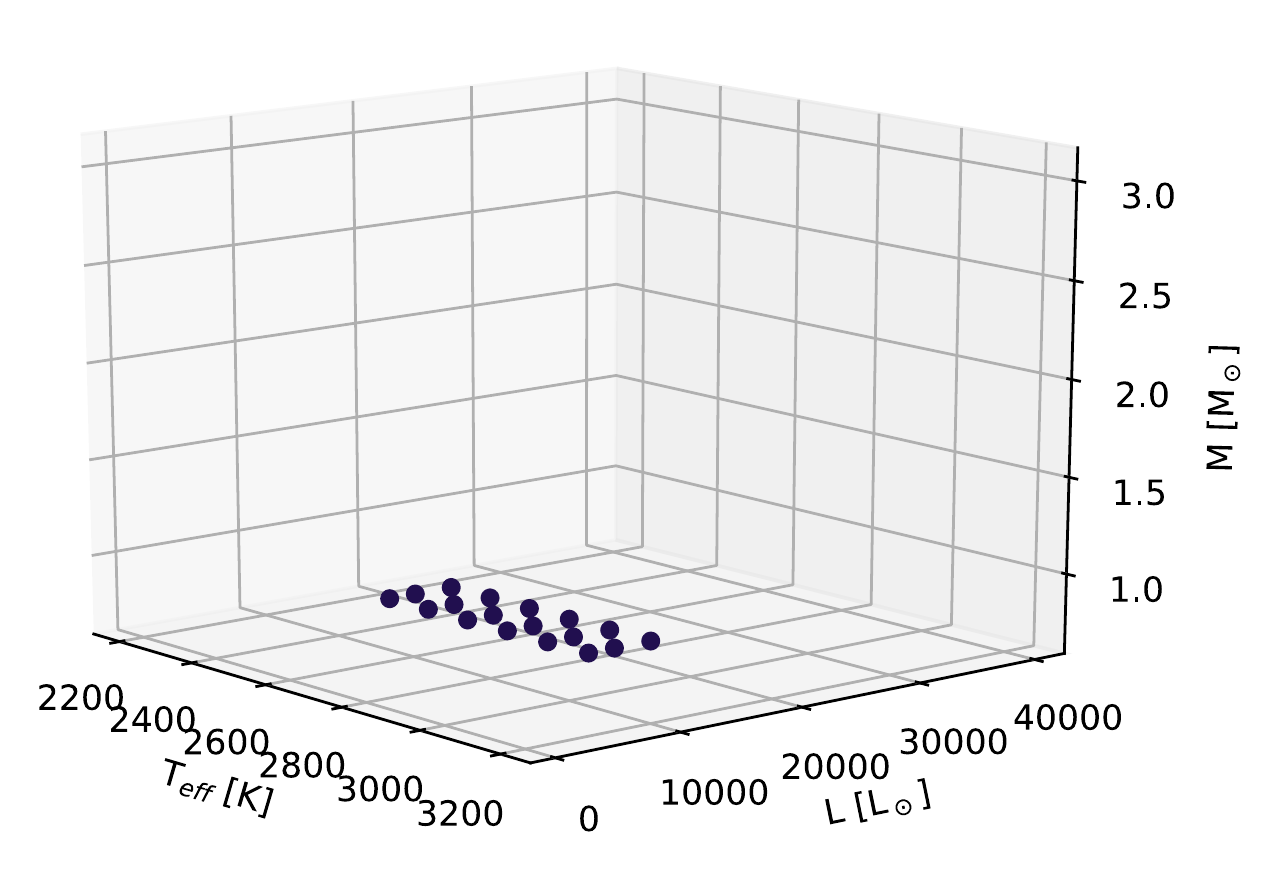} 
\includegraphics[width=0.45\textwidth]{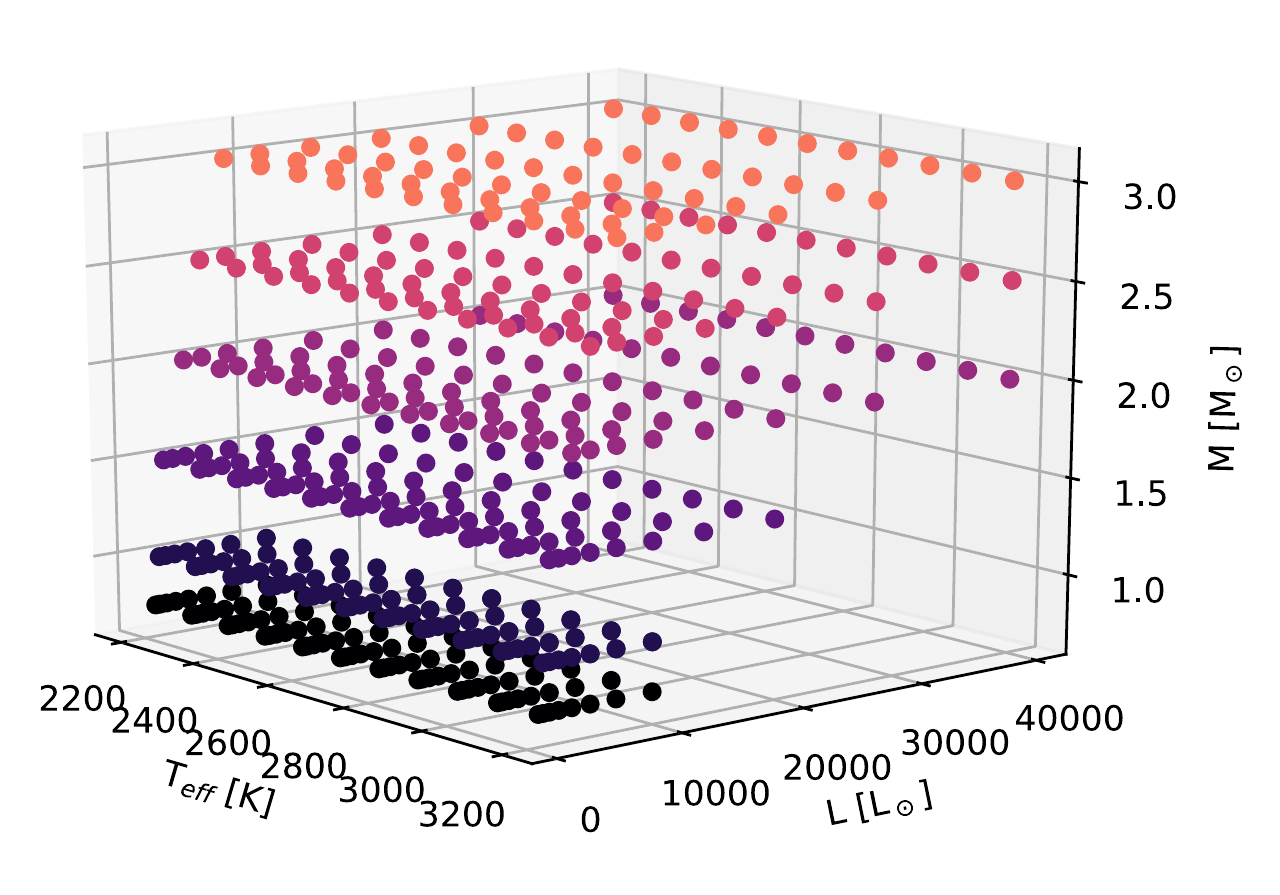} 
\caption{Overview of the stellar parameters covered in the grid of DARWIN models for M-type AGB stars published in \citet[][left panel]{bladh2015} compared to stellar parameters covered in the grid presented in this paper (right panel). The grid points are colour-coded according to mass. 
\label{fig:grid_comp}}
\end{figure*}

We modelled the atmospheres and winds of M-type AGB stars using   Dynamic
Atmosphere and Radiation-driven Wind models based on Implicit Numerics  \citep[DARWIN;][]{hoefner2016}. This 1D radiation-hydrodynamics code produces time-dependent radial structures of the atmosphere, and wind properties, assuming spherical symmetry. The DARWIN models cover a spherical shell with the inner boundary situated just below the photosphere of the star and the outer boundary defined by the dynamics of the atmosphere, allowing for an outflow. If a stellar wind develops, the outer boundary is set at 25\,R$_*$  where the flow velocities have typically  reached their terminal value. Otherwise, the outer boundary follows the periodic motion of the outermost layers of the atmosphere. The starting structure is a hydrostatic dust-free model atmosphere, with the input parameters mass $M_*$, luminosity $L_*$, effective temperature $T_*$, and chemical composition. Stellar pulsation is simulated by a sinusoidal variation in radius and luminosity at the inner boundary. This variation is gradually ramped up, turning the initially hydrostatic atmosphere into a dynamical atmosphere by simultaneously solving the hydrodynamic equations (conservation of mass, momentum, and energy), frequency-dependent radiation transfer, and time-dependent grain growth at each time step. The system of partial differential equations is solved with a Newton-Raphson scheme on an adaptive spatial grid \citep{dorfi1987}, taking temperature and density gradients into account. The gas opacities are calculated with the opacity generation code COMA \citep{aringer2016}, assuming solar abundances according to \cite{grevesse1989} that are complemented by C, N, and O abundances from \cite{grevesse1994} (see Sect.~\ref{stellarevo} for further discussion). Since the opacities of molecules and dust forming in the extended atmosphere dominate the radiation field, the inclusion of frequency-dependent radiative transfer is crucial for achieving realistic atmospheric structures \citep{hoefner2003}, and the models presented here use 319 frequency points. For a more detailed description of the DARWIN code see \cite{hoefner2016}, and references therein.

There are two types of results available from the DARWIN simulations:
(i) mass-loss rates, wind velocities, dust yields, and radial snapshots of the atmosphere structures as a direct output, and (ii) synthetic spectra, light-curves, interferometric visibilities calculated {a posteriori} with the COMA code \citep{aringer2016} using the radial snapshots.
Figure~\ref{fig:struct} shows characteristic properties of a DARWIN model with stellar parameters $M_*=1\,\mathrm{M}_{\odot}$, $\log (L_*/L_{\odot})=3.85$, and $T_*=2700\,$K. 

\subsection{Pulsation description}
The  variability of AGB stars is vital to wind driving, as pulsation induces shock waves that elevate the gas to distances where dust can form. In the DARWIN models the effects of pulsation, i.e. the expansion and contraction of the stellar surface layers and the luminosity variation, are modelled in a parameterised fashion. The radial variation $R_{in}(t)$ at an impermeable inner boundary is given by
\begin{equation}
\label{eqn1}
R_\mathrm{in}(t) = R_0 + \frac{\Delta u_p P}{2 \pi} \sin{\left ( \frac{2 \pi }{P} t \right)},
\end{equation}
where $\Delta u_p$ is the velocity amplitude and $P$ is the pulsation period. This sinusoidal form was introduced by \citet[][]{bowen88}, and the corresponding velocity variation is given by
\begin{equation}
\label{eqn11}
u_\mathrm{in}(t) = \Delta u_p \cos\left ( \frac{2 \pi }{P} t \right).
\end{equation}
The luminosity at the inner boundary is proportional to the square of the radial variation, $L_\mathrm{in} \propto R^2_\mathrm{in}$, if a constant radiative flux is assumed at $R_\mathrm{in}$, as was done in early models. However, the resulting bolometric variation was found to be too small when compared to observations \citep[see][]{gautschy04}. A free parameter ($f_{\mathrm{L}}$) was therefore introduced, giving the following luminosity variation at the inner boundary:
\begin{equation}
\label{eqn2}
\Delta L_\mathrm{in}(t) = L_\mathrm{in} - L_0= f_L \left (\frac{R^2_{in}(t) - R^2_0}{R^2_0} \right ) \times L_0.
\end{equation}
The variability of  radius and luminosity at the inner boundary is visualised in the embedded panels in Fig.~\ref{ts_ex}. The validity of this inner boundary condition has been investigated. \cite{freytag08} tested models with different descriptions for $R_\textrm{in}$, based on exploratory 3D `star-in-a-box' models, and found that the amplitude is the important parameter for wind-driving, as the information of the shape of $R_\mathrm{in}(t)$ is lost when shock waves develop. In contrast, DARWIN models are sensitive to the shape of $L_\textrm{in}(t)$ \citep[explored in][]{liljegren17} as changes to the luminosity also change the temperature structure, which in turn influences where and when dust can form. It was concluded, however,  that using the standard boundary condition described above results in realistic mass-loss rates, wind velocities, and photometric colours in the case of models for M-type AGB stars. It is also important to keep in mind that the standard inner boundary condition used here is meant to describe pulsation properties of Mira variables pulsating in the fundamental mode (see also Sect.~\ref{sec:pulswind}).

\subsection{Dust description}
Grain growth in AGB stars usually takes place far from chemical equilibrium, on timescales that are comparable to the pulsation period and the ballistic motions in the atmosphere. In order to realistically model this process we include a time-dependent description of grain growth and evaporation, based on the method of \cite{gail1999}. In contrast to that paper, however, the wind-driving dust species in our DARWIN models for M-type AGB stars is Mg$_2$SiO$_4$, and the growth and evaporation of these grains are modelled by adding magnesium (Mg), silica (SiO), and water (H$_2$O) from the gas phase to the grain surface according to the net reaction
\begin{equation}
\ce{2Mg + SiO + 3H2O <-> Mg2SiO4 + 3H2}
\end{equation}
with the availability of SiO limiting the total growth rate. The abundances of the molecular species are determined by chemical equilibrium, and the material condensed into dust particles is subtracted from the gas phase at each time step. The Mg$_2$SiO$_4$ grains are assumed to be spherical and the optical properties are calculated from data by \cite{jaeger2003}, using Mie theory. 

Since the nucleation processes  (i.e. the formation of the first seed particles out of the gas phase) is still not fully understood in oxygen-rich environments \citep[see e.g.][]{gail16,gobrecht16} we assume pre-existing seed particles that start to grow when thermodynamical conditions are favourable. These seed particles consist of about 1000 monomers onto which molecules can condense according to the net reaction given above. The abundance of the seed particles $n_\mathrm{d}/n_\mathrm{H}$, defined as the number density of the seed particles $n_\mathrm{d}$ divided by the number density of hydrogen $ n_\mathrm{H}$, is a free parameter in the models \citep[for a discussion, see][]{hoefner2016}.

\begin{figure}
\centering
\includegraphics[width=\hsize]{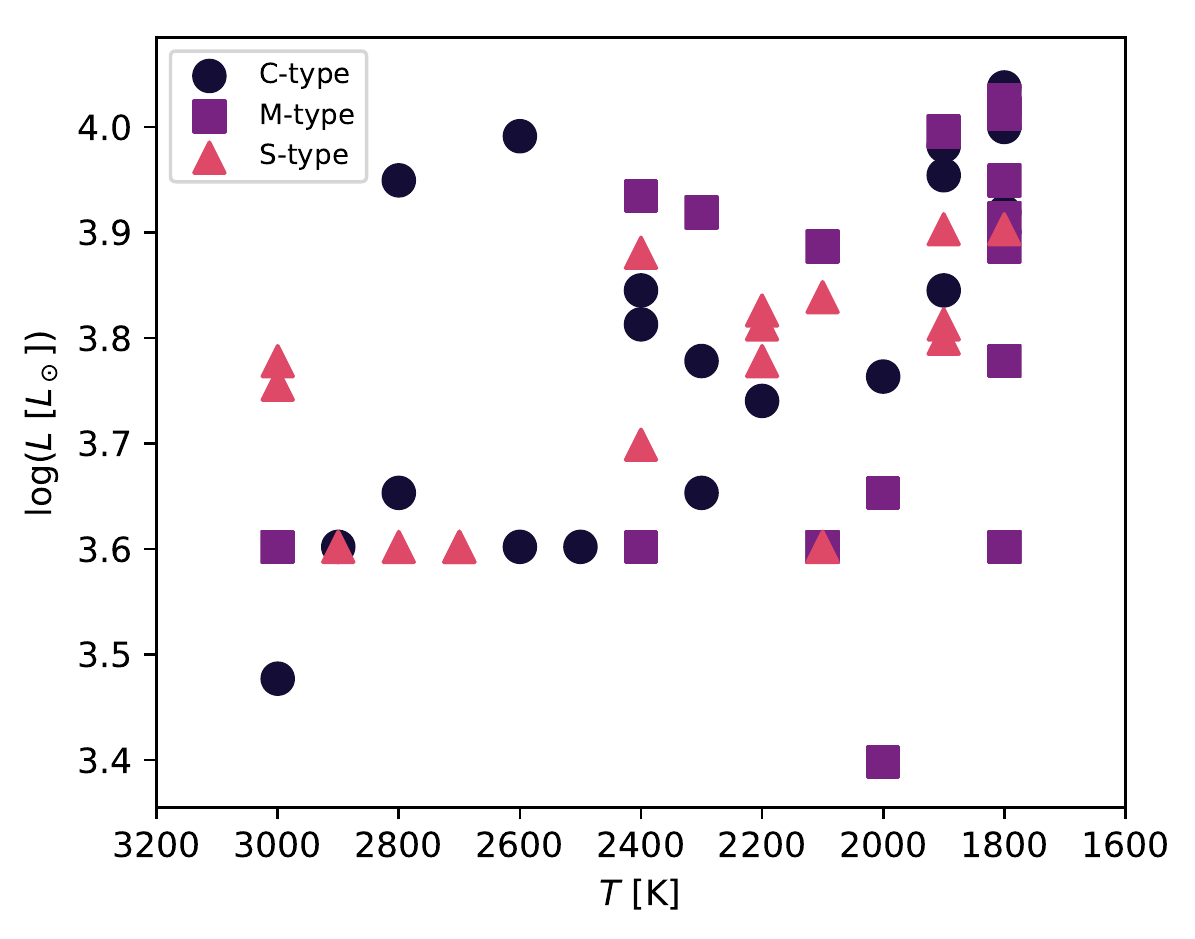}
   \caption{Observed stellar luminosities and effective temperatures for a set of galactic AGB stars of different types \citep{ramstedt14}.}
      \label{fig:obs_lt}
\end{figure}

\begin{figure*}
\centering
\includegraphics[width=0.49\textwidth]{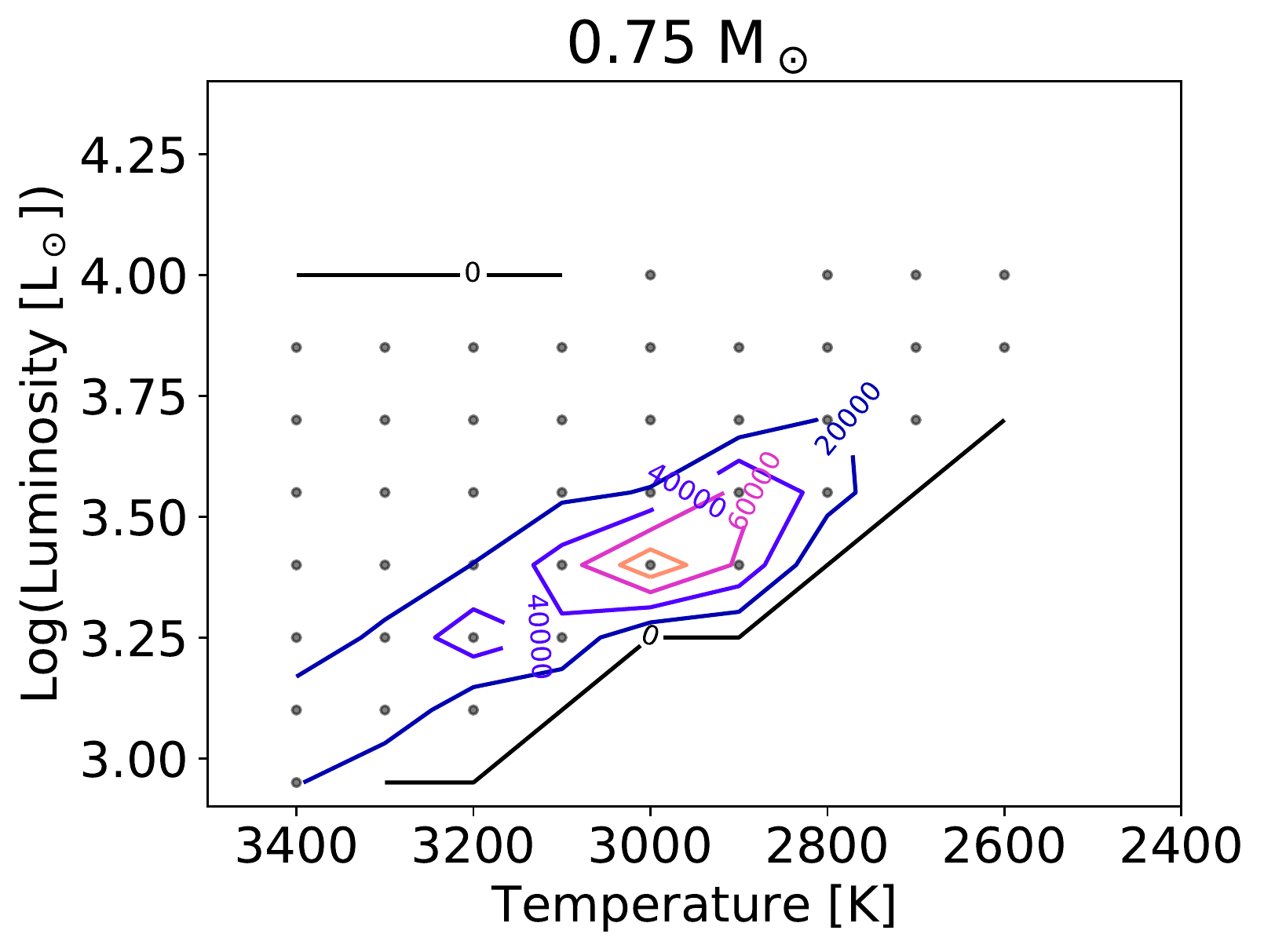} 
\includegraphics[width=0.49\textwidth]{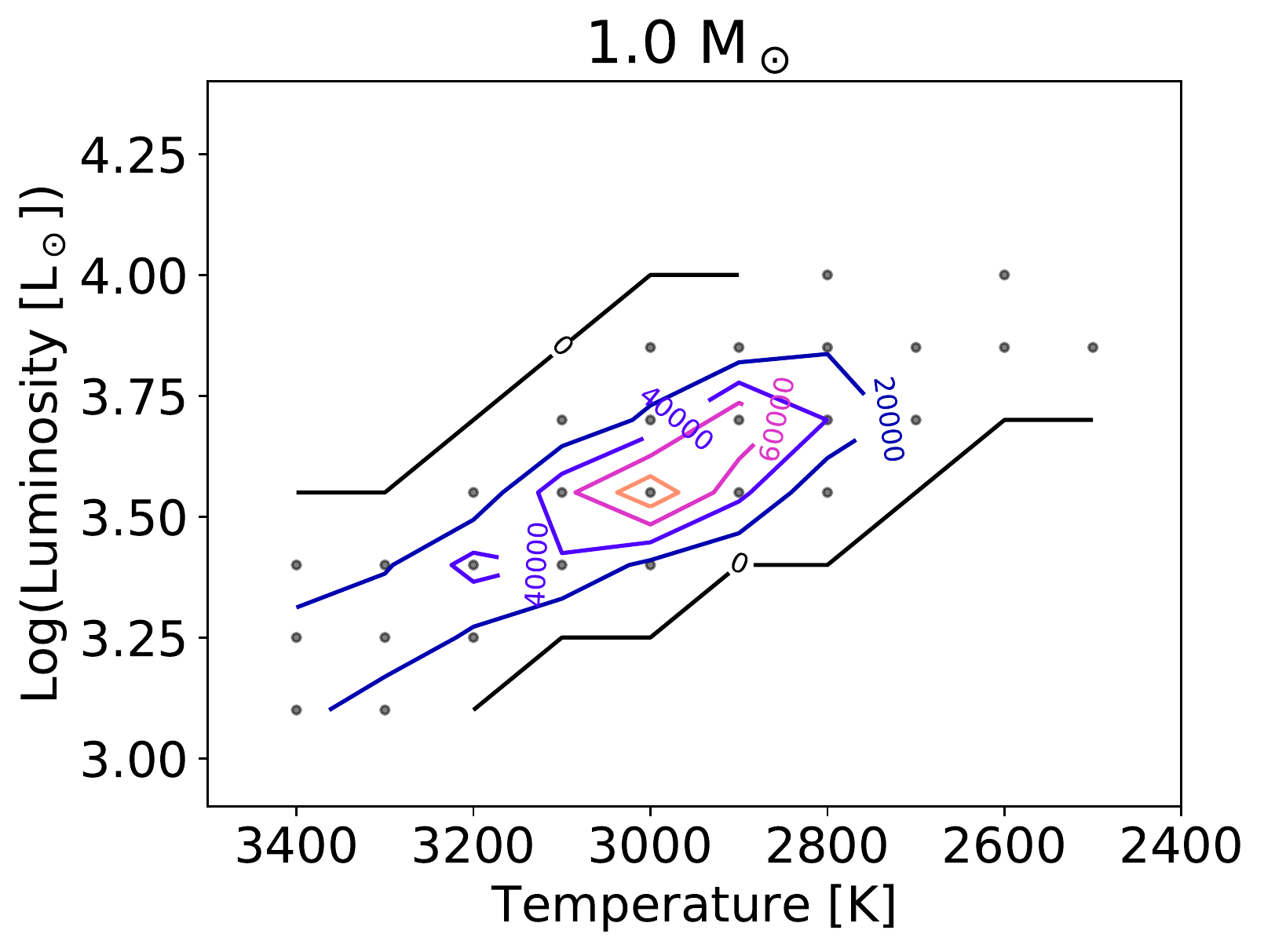} 
\includegraphics[width=0.49\textwidth]{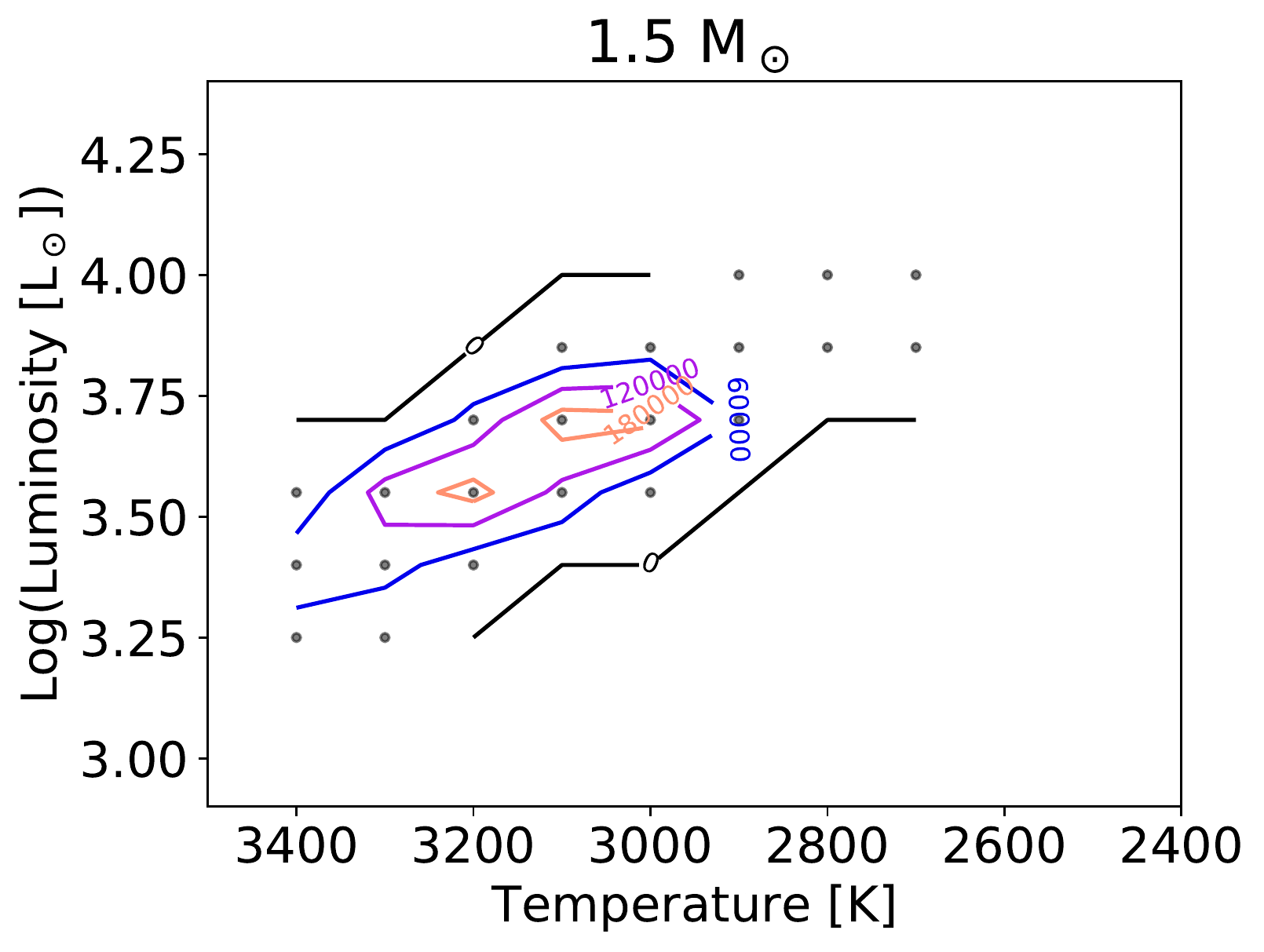}
\includegraphics[width=0.49\textwidth]{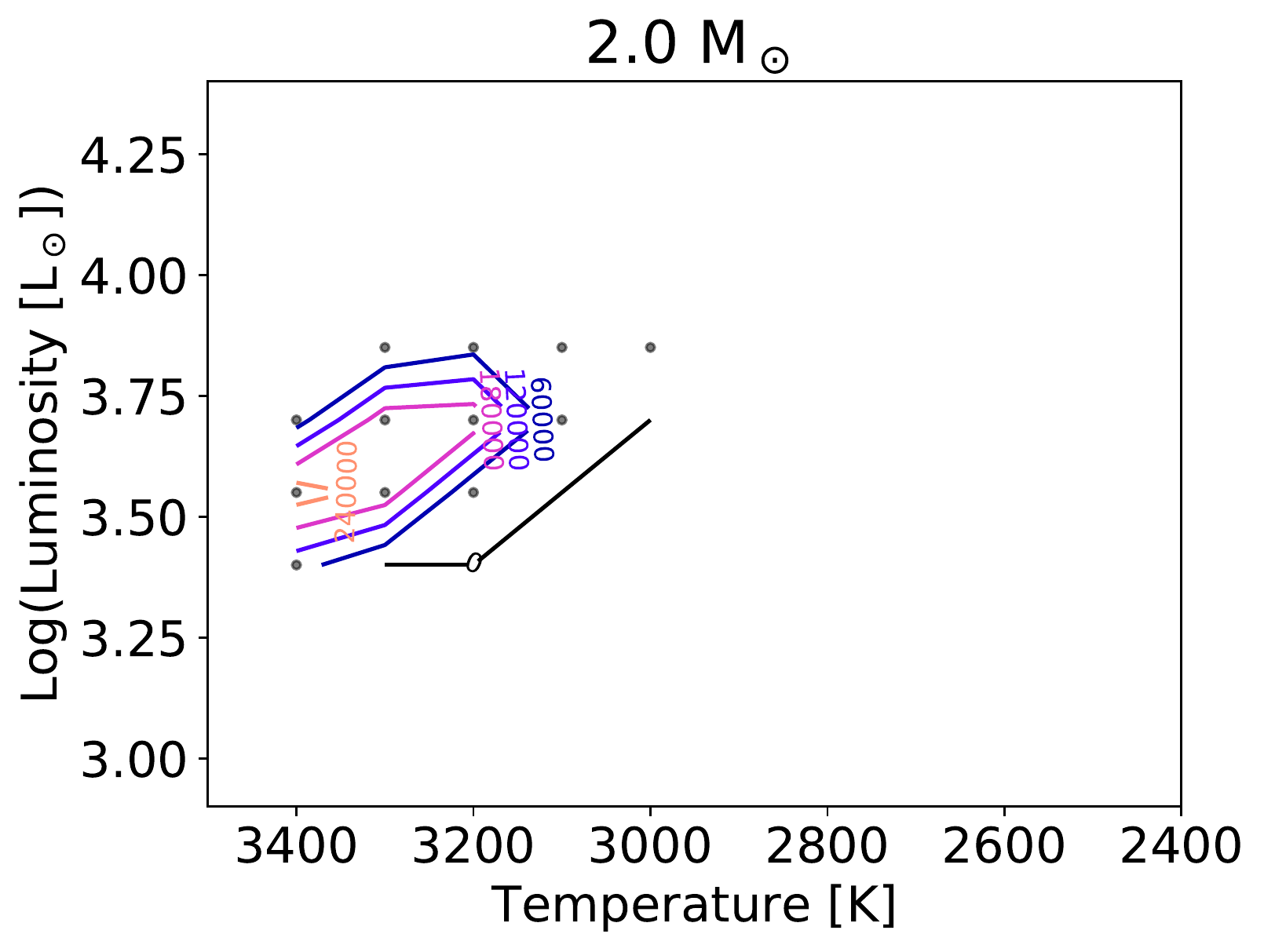}
\includegraphics[width=0.49\textwidth]{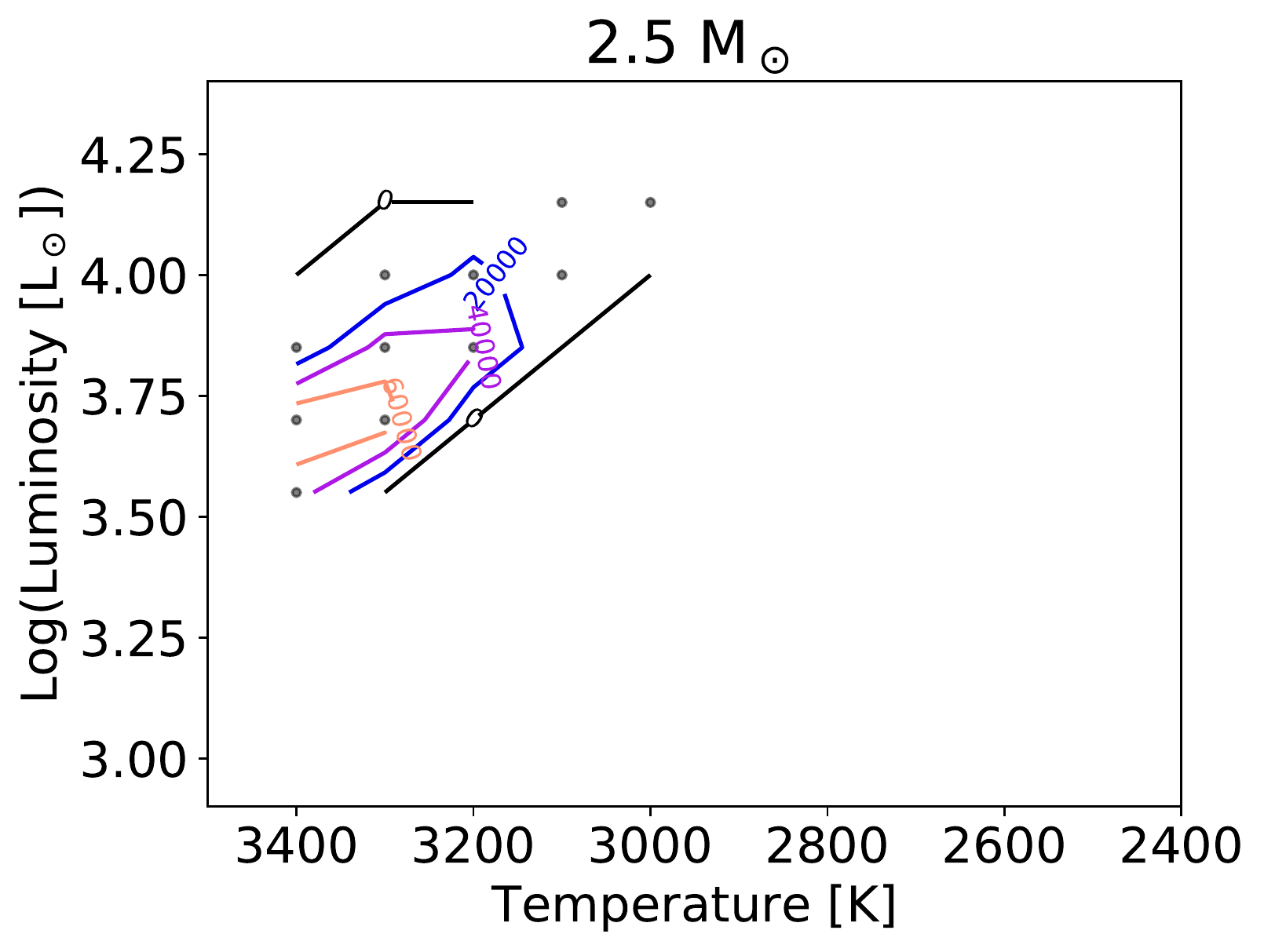}
\includegraphics[width=0.49\textwidth]{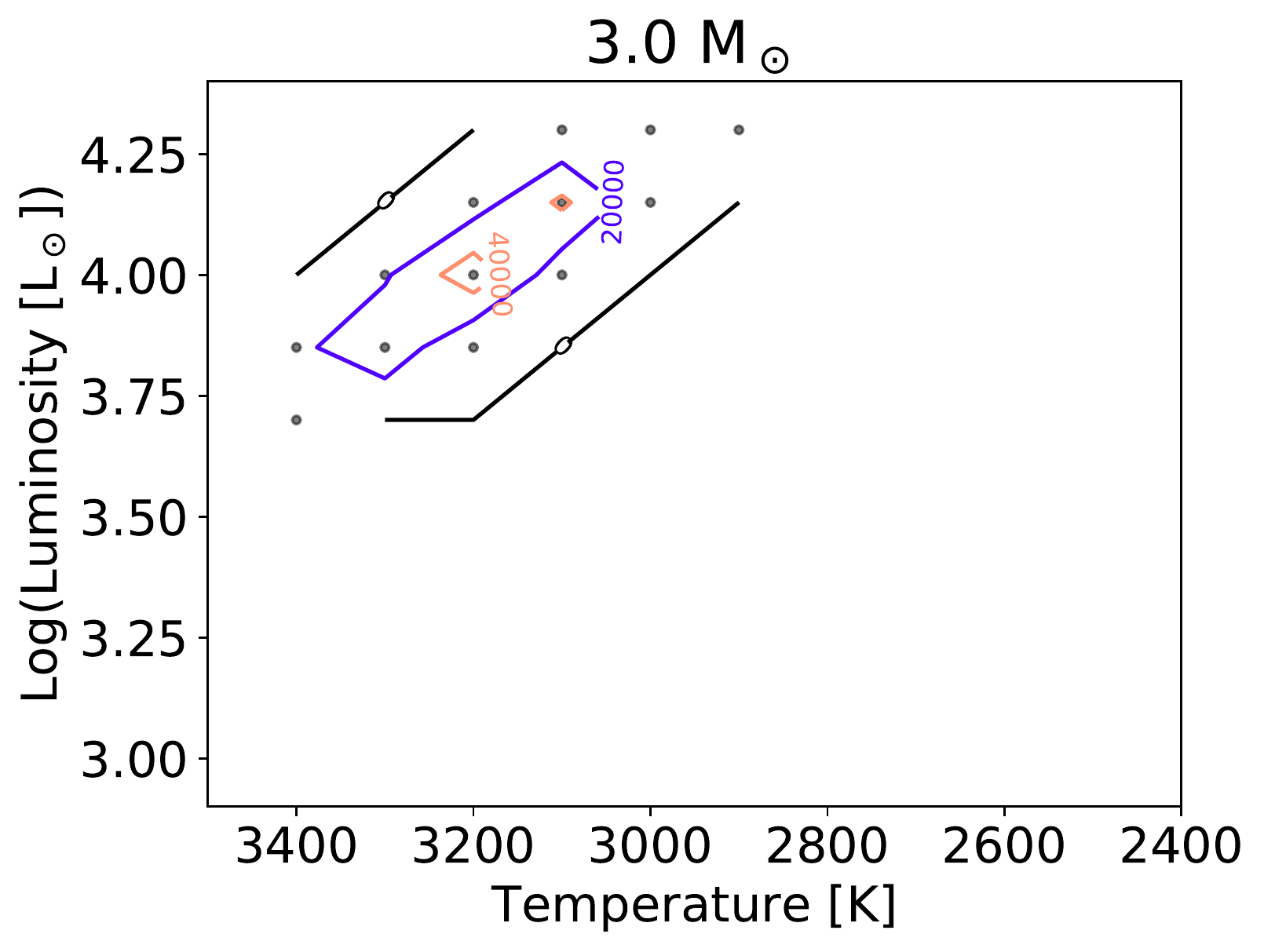}
\caption{Overview of the average time (in years) spent per initial mass at each model grid point (corresponding to a combination of stellar parameters), estimated from evolutionary tracks at solar metallicity (sorted by current mass $M_*\leq 3.25\,$M$_{\odot}$ and $\mathrm{C/O}<1$) calculated with the stellar evolution code COLIBRI \citep{marigo13,marigo17}. The solid lines are iso contours of the time spent in years at each grid point.
\label{fig:evo_comp}}
\end{figure*}

\section{Grid parameters}
\label{gridpar}
In contrast to the previous study presented in \cite{bladh2015}, the new extensive grid of models for M-type AGB stars includes models with different current stellar mass, ranging from 0.75\,M$_{\odot}$  to 3\,M$_{\odot}$. As can be seen in Fig.~\ref{fig:grid_comp}, which shows the stellar parameters covered in the grid by \cite{bladh2015} (left panel) and the new grid presented in this paper (right panel), the parameter space has increased significantly in the new study.  

The lowest luminosity that leads to outflows varies with mass since the ability to produce stellar winds largely depends on the ratio between the radiative and gravitational acceleration. The chosen parameter space in luminosity therefore varies with mass. The lower limit at a given mass is always set as the luminosity where the DARWIN models no longer produce a stellar wind. The upper limit is set as close to $\log(L_*/L_{\odot})=4.60$ as converging starting models can be constructed. This is easier for higher masses since an increased gravitational potential makes the atmospheres more compact. The upper luminosity limit of $\log(L_*/L_{\odot})=4.60$ was chosen so that the grid also includes wind models suitable for OH/IR stars. It should be noted, however, that the surface abundances and pulsation properties of M-type AGB stars with high initial masses (typically >\,5\,M$_{\odot}$ at solar metallicity) can be quite different to those properties  of oxygen-rich AGB stars with lower initial masses. Consequently, the results from this model grid should be applied with caution to AGB stars with high initial masses.

The decision regarding which effective temperatures to include is based on information from stellar evolution models, but also on observations of individual M-type AGB stars. Stellar evolution models calculated with the code COLIBRI \citep{marigo13,marigo17} predict effective temperatures down to 2400\,K for M-type AGB stars with solar metallicity, whereas observational evidence indicates that there are galactic AGB stars with even lower effective temperatures. An example of this is given in Fig.~\ref{fig:obs_lt}, showing luminosities and effective temperatures for a set of galactic AGB stars of different type \citep{ramstedt14}. The effective temperatures of this sample are determined by SED fitting, using DUSTY and assuming the star has a Planckian radiation field, with an estimated uncertainty of $\pm 300\,$K. Given that this method tends to systematically underestimate the effective temperature due to the re-processing of stellar flux in the circumstellar envelope, we set the lower limit of the grid to 2200\,K. The upper limit is set between 3000 and 3400\,K, depending on the wind properties. Higher temperatures are included if the DARWIN models still produce a stellar wind at 3000\,K. 

The pulsation period $P$ of the sinusoidal variation at the inner boundary is derived from the period-luminosity relation presented in \cite{whitelock09}. The sensitivity of the derived mass-loss rates to changes in the period-luminosity relation is tested by calculating a subset of models ($M_{\odot}=1,2$, and $3\,\mathrm{M}_{\odot}$, $n_{\mathrm{d}}/n_{\mathrm{H}}=1\cdot10^{-15}$ and $\Delta u_{\mathrm{p}}=2$\,km/s) with a pulsation period that is changed with $\pm 10\%$. On average such a change in the pulsation periods changes the  mass-loss rates by approximately $\pm 20\%$. This is small compared to the uncertainty of observed mass-loss rates, which are estimated to be as high as a factor of three \citep{Ramstedt2008}.

The piston velocity $\Delta u_{\mathrm{p}}$, which describes the velocity amplitude of the variations, ranges from 2 to 4\,km/s. This results in shock amplitudes of about 15–20\,km/s in the inner atmosphere, which agrees well with observed radial velocity amplitudes derived from second overtone CO lines, probing the deep photosphere \citep{lebzelter02,nowotny10}. The scaling factor for the luminosity amplitude $f_\mathrm{L}$ is set to 2, in accordance with the previous grid of DARWIN models for M-type AGB stars.

The parameter describing the seed particle abundance, $n_{\mathrm{d}}/n_{\mathrm{H}}$, is constrained by comparing the dynamical properties (mass-loss rate and wind velocity) from models calculated with different seed particle abundances to empirical data derived from observations of CO-lines \citep{olofsson02,gondel03}. The observational values were best reproduced with a seed particle abundance of $1\times 10^{-15}$ and $3\times 10^{-15}$. The physical parameters (i.e. stellar, pulsation, and dust parameters) covered in this grid of DARWIN models for M-type AGB stars are listed in Table~\ref{tab:grid}. The pulsation period $P$ is not treated as an independent parameter since it is set by a P-L relation, but it is listed in Table~\ref{tab:grid} for easy reference.

It is important to remember that not all combinations of stellar parameters are equally probable. Figure~\ref{fig:evo_comp} shows the average time spent per initial mass at each grid point (corresponding to a combination of stellar parameters), estimated from evolutionary tracks calculated with the stellar evolution code COLIBRI \citep{marigo13,marigo17}, assuming solar metallicity, a current mass $M_*\leq 3.25$, and $\mathrm{C/O}<1$. This is not an estimate of the average lifetime a star spends at each grid point since the evolutionary tracks are not convolved with an initial mass function, but it gives some indication of the combinations of stellar parameters that occur frequently. According to this estimate the lower right corner of the panels in Fig.~\ref{fig:evo_comp}, representing a low luminosity in combination with a low effective temperature, is not part of the parameter space reached by the COLIBRI tracks. Even so, we use a Cartesian grid, including combinations of parameters that are less likely, in order to facilitate interpolation in the produced mass-loss rates so they can easily be used in stellar evolution codes. A large model grid is also important given that the use of mass-loss rates derived from DARWIN models in stellar evolution models might change the predicted parameter space. 

\begin{table*}
\caption{Fundamental parameters (stellar, pulsation, and dust parameters) covered in the grid of DARWIN models for M-type AGB stars.}     
\label{tab:grid}      
\centering          
\begin{tabular}{c c c | c c c | c}   
\hline\hline     
$M_*$         & $\log (L_*/L_{\odot})$    & $T_*$ &  $P$ & $u_{\mathrm{p}}$ & $f_{\mathrm{L}}$ & $\log (n_{\mathrm{d}}/n_{\mathrm{H}})$\\ 
~[M$_{\odot}$] &         & [K]   & [d] & km/s & &\\
\hline
0.75    & 2.95 & 2200, 2300, \ldots, 3000 & 101 & 2, 3, 4 & $2$ & $-14.5$, $-15.0$ \\
        & 3.10 & 2200, 2300, \ldots, 3000 & 132 & 2, 3, 4 & $2$ & $-14.5$, $-15.0$ \\
        & 3.25 & 2200, 2300, \ldots, 3000 & 171 & 2, 3, 4 & $2$ & $-14.5$, $-15.0$ \\
        & 3.40 & 2200, 2300, \ldots, 3000 & 222 & 2, 3, 4 & $2$ & $-14.5$, $-15.0$ \\
        & 3.55 & 2200, 2300, \ldots, 3200 & 288 & 2, 3, 4 & $2$ & $-14.5$, $-15.0$ \\
        & 3.70 & 2200, 2300, \ldots, 3200 & 373 & 2, 3, 4 & $2$ & $-14.5$, $-15.0$ \\
        & 3.85 & 2200, 2300, \ldots, 3300 & 485 & 2, 3, 4 & $2$ & $-14.5$, $-15.0$ \\
\hline
1.0     & 3.10 & 2200, 2300, \ldots, 3000 & 132 & 2, 3, 4 & $2$ & $-14.5$, $-15.0$ \\
        & 3.25 & 2200, 2300, \ldots, 3000 & 171 & 2, 3, 4 & $2$ & $-14.5$, $-15.0$ \\
        & 3.40 & 2200, 2300, \ldots, 3000 & 222 & 2, 3, 4 & $2$ & $-14.5$, $-15.0$ \\
        & 3.55 & 2200, 2300, \ldots, 3000 & 288 & 2, 3, 4 & $2$ & $-14.5$, $-15.0$ \\
        & 3.70 & 2200, 2300, \ldots, 3000 & 373 & 2, 3, 4 & $2$ & $-14.5$, $-15.0$ \\
        & 3.85 & 2200, 2300, \ldots, 3200 & 485 & 2, 3, 4 & $2$ & $-14.5$, $-15.0$ \\
        & 4.00 & 2200, 2300, \ldots, 3200 & 629 & 2, 3, 4 & $2$ & $-14.5$, $-15.0$ \\
        & 4.15 & 2200, 2300, \ldots, 3200 & 817 & 2, 3, 4 & $2$ & $-14.5$, $-15.0$ \\
\hline
1.5     & 3.25 & 2200, 2300, \ldots, 3000 & 171 & 2, 3, 4 & $2$ & $-14.5$, $-15.0$ \\
        & 3.40 & 2200, 2300, \ldots, 3000 & 222 & 2, 3, 4 & $2$ & $-14.5$, $-15.0$ \\
        & 3.55 & 2200, 2300, \ldots, 3000 & 288 & 2, 3, 4 & $2$ & $-14.5$, $-15.0$ \\
        & 3.70 & 2200, 2300, \ldots, 3000 & 373 & 2, 3, 4 & $2$ & $-14.5$, $-15.0$ \\
        & 3.85 & 2200, 2300, \ldots, 3200 & 485 & 2, 3, 4 & $2$ & $-14.5$, $-15.0$ \\
        & 4.00 & 2200, 2300, \ldots, 3200 & 629 & 2, 3, 4 & $2$ & $-14.5$, $-15.0$ \\
        & 4.15 & 2200, 2300, \ldots, 3200 & 817 & 2, 3, 4 & $2$ & $-14.5$, $-15.0$ \\       
        & 4.30 & 2200, 2300, \ldots, 3200 & 1060 &2, 3, 4 & $2$ & $-14.5$, $-15.0$ \\  
\hline
2.0     & 3.55 & 2200, 2300, \ldots, 3000 & 288 & 2, 3, 4 & $2$ & $-14.5$, $-15.0$ \\
        & 3.70 & 2200, 2300, \ldots, 3000 & 373 & 2, 3, 4 & $2$ & $-14.5$, $-15.0$ \\
        & 3.85 & 2200, 2300, \ldots, 3200 & 485 & 2, 3, 4 & $2$ & $-14.5$, $-15.0$ \\
        & 4.00 & 2200, 2300, \ldots, 3100 & 629 & 2, 3, 4 & $2$ & $-14.5$, $-15.0$ \\
        & 4.15 & 2200, 2300, \ldots, 3200 & 817 & 2, 3, 4 & $2$ & $-14.5$, $-15.0$ \\        
        & 4.30 & 2200, 2300, \ldots, 3200 & 1060& 2, 3, 4 & $2$ & $-14.5$, $-15.0$ \\
        & 4.45 & 2200, 2300, \ldots, 3200 & 1376& 2, 3, 4 & $2$ & $-14.5$, $-15.0$ \\        
        & 4.60 & 2200, 2300, \ldots, 3300 & 1786& 2, 3, 4 & $2$ & $-14.5$, $-15.0$ \\
\hline
2.5     & 3.70 & 2200, 2300, \ldots, 3000 & 373 & 2, 3, 4 & $2$ & $-14.5$, $-15.0$ \\
        & 3.85 & 2200, 2300, \ldots, 3000 & 485 & 2, 3, 4 & $2$ & $-14.5$, $-15.0$ \\
        & 4.00 & 2200, 2300, \ldots, 3200 & 629 & 2, 3, 4 & $2$ & $-14.5$, $-15.0$ \\
        & 4.15 & 2200, 2300, \ldots, 3200 & 817 & 2, 3, 4 & $2$ & $-14.5$, $-15.0$ \\        
        & 4.30 & 2200, 2300, \ldots, 3200 & 1060& 2, 3, 4 & $2$ & $-14.5$, $-15.0$ \\
        & 4.45 & 2200, 2300, \ldots, 3200 & 1376& 2, 3, 4 & $2$ & $-14.5$, $-15.0$ \\        
        & 4.60 & 2200, 2300, \ldots, 3300 & 1786& 2, 3, 4 & $2$ & $-14.5$, $-15.0$ \\
\hline
3.0     & 3.85 & 2200, 2300, \ldots, 3000 & 485 & 2, 3, 4 & $2$ & $-14.5$, $-15.0$ \\
        & 4.00 & 2200, 2300, \ldots, 3100 & 629 & 2, 3, 4 & $2$ & $-14.5$, $-15.0$ \\
        & 4.15 & 2200, 2300, \ldots, 3200 & 817 & 2, 3, 4 & $2$ & $-14.5$, $-15.0$ \\        
        & 4.30 & 2200, 2300, \ldots, 3200 & 1060& 2, 3, 4 & $2$ & $-14.5$, $-15.0$ \\
        & 4.45 & 2200, 2300, \ldots, 3200 & 1376& 2, 3, 4 & $2$ & $-14.5$, $-15.0$ \\        
        & 4.60 & 2200, 2300, \ldots, 3200 & 1786& 2, 3, 4 & $2$ & $-14.5$, $-15.0$ \\
\hline
\end{tabular}
\tablefoot{The columns show stellar mass $M_*$, stellar luminosity $L_*$, effective temperature $T_*$, pulsation period $P$, piston velocity $u_{\mathrm{p}}$, the scaling factor for the luminosity amplitude $f_\mathrm{L}$, and the seed particle abundance $n_{\mathrm{d}}/n_{\mathrm{H}}$. The pulsation period is not an independent parameter, but is set according to the period-luminosity relation by \cite{whitelock09}.}
\end{table*}


\begin{figure}
\centering
\includegraphics[width=\hsize]{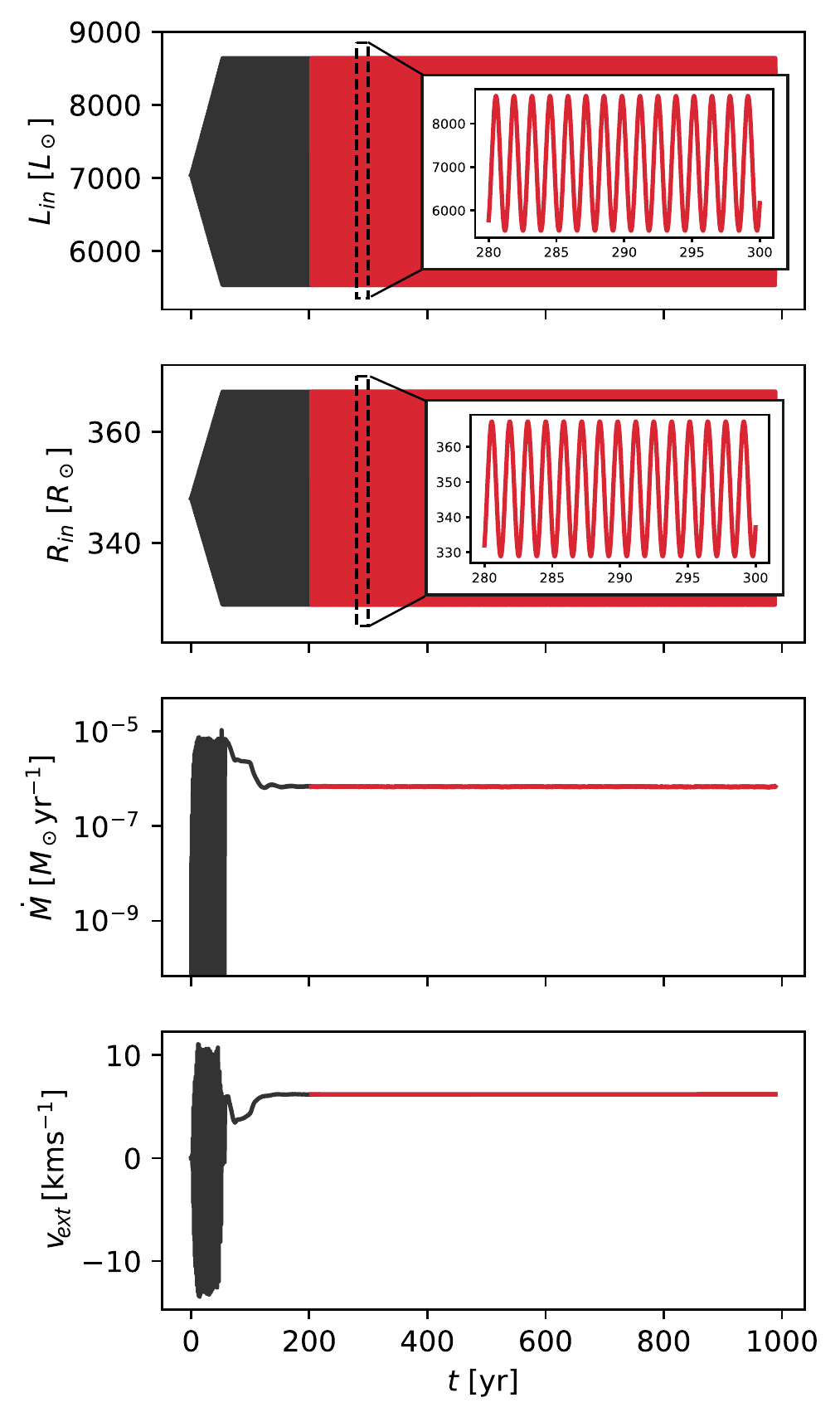}
      \caption{Temporal evolution of a DARWIN model with input parameters $M_*=1\,\mathrm{M}_{\odot}$, $\log(L_*/L_{\odot})=3.85$, $T_*=2700\,$K, $u_{\mathrm{p}}=2$\,km/s and $\log (n_{\mathrm{d}}/n_{\mathrm{H}})=-15.0$. The panels show, from top to bottom, luminosity and radius at the inner boundary (both input quantities) and the resulting mass-loss rate and wind velocity at the outer boundary. The wind properties for this model are calculated by averaging over the time interval marked with red. The two embedded panels show the variations at the inner boundary in enlarged format for a short time frame.}
\label{ts_ex}
\end{figure}

\section{Results of  mass-loss rates and wind velocities}
\label{modresult}
Each DARWIN model consists of snapshots of the atmospheric structure that  evolve over time. The wind properties are calculated by taking the mass-loss rate and velocity from the outermost layers and averaging the values, typically over  hundreds of pulsation periods. The early pulsation periods are excluded to avoid transient effects of ramping up the amplitude. Figure~\ref{ts_ex} shows an example of such a time series for a model with input parameters $M_*=1\,\mathrm{M}_{\odot}$, $\log(L_*/L_{\odot})=3.85$, $T_*=2700\,$K, $u_{\mathrm{p}}=2$\,km/s and $\log (n_{\mathrm{d}}/n_{\mathrm{H}})=-15.0$. The  sections in red show the time interval over which the mass-loss rate and wind velocity are averaged. The wind and dust properties resulting from the DARWIN models presented in this paper will be made available online, together with the input parameters of each model. A format guide to the online data is shown in Table~\ref{tab:online} in  Appendix~\ref{appendixA}.

\begin{figure*}
\centering
\includegraphics[width=0.49\textwidth]{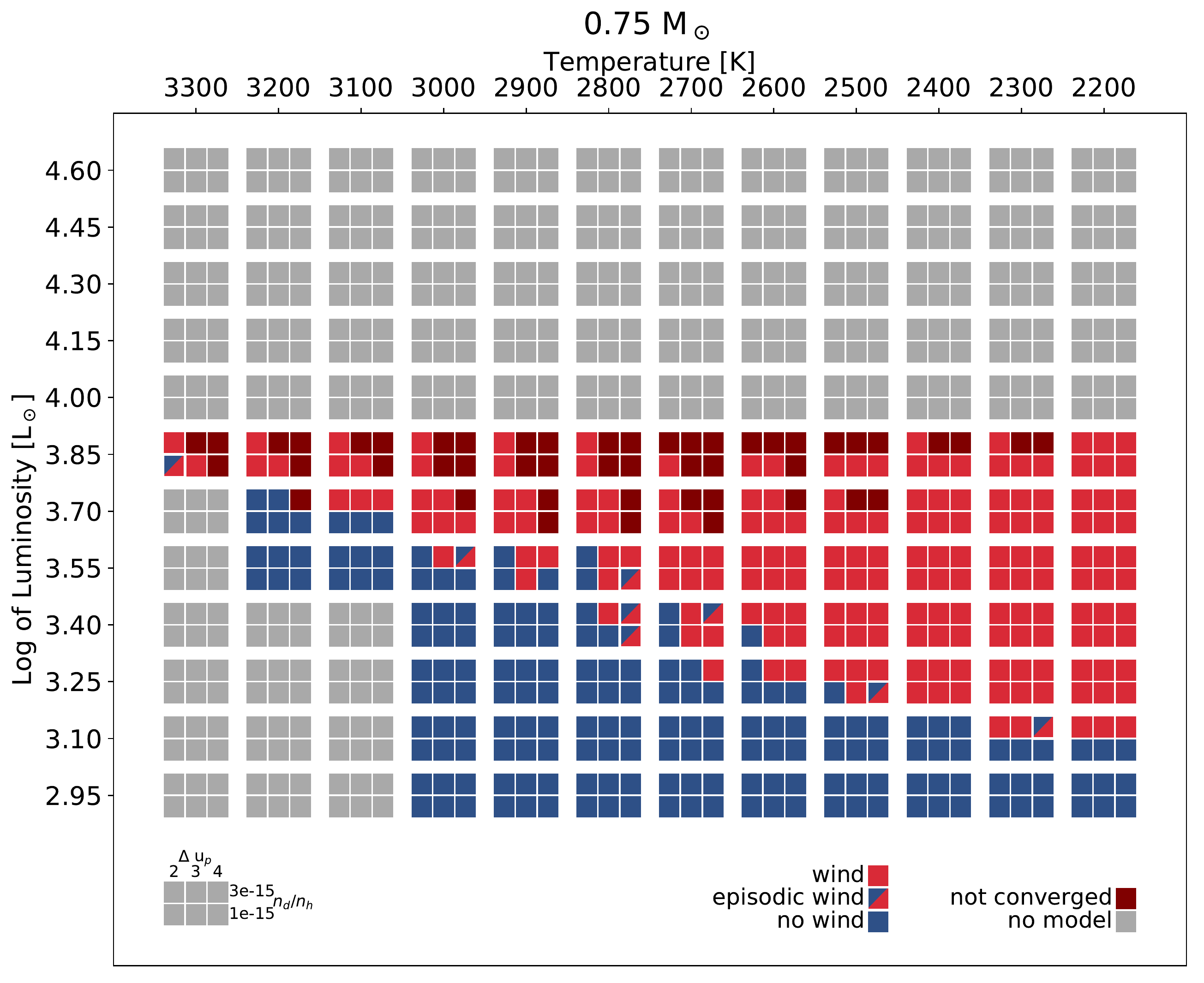} 
\includegraphics[width=0.49\textwidth]{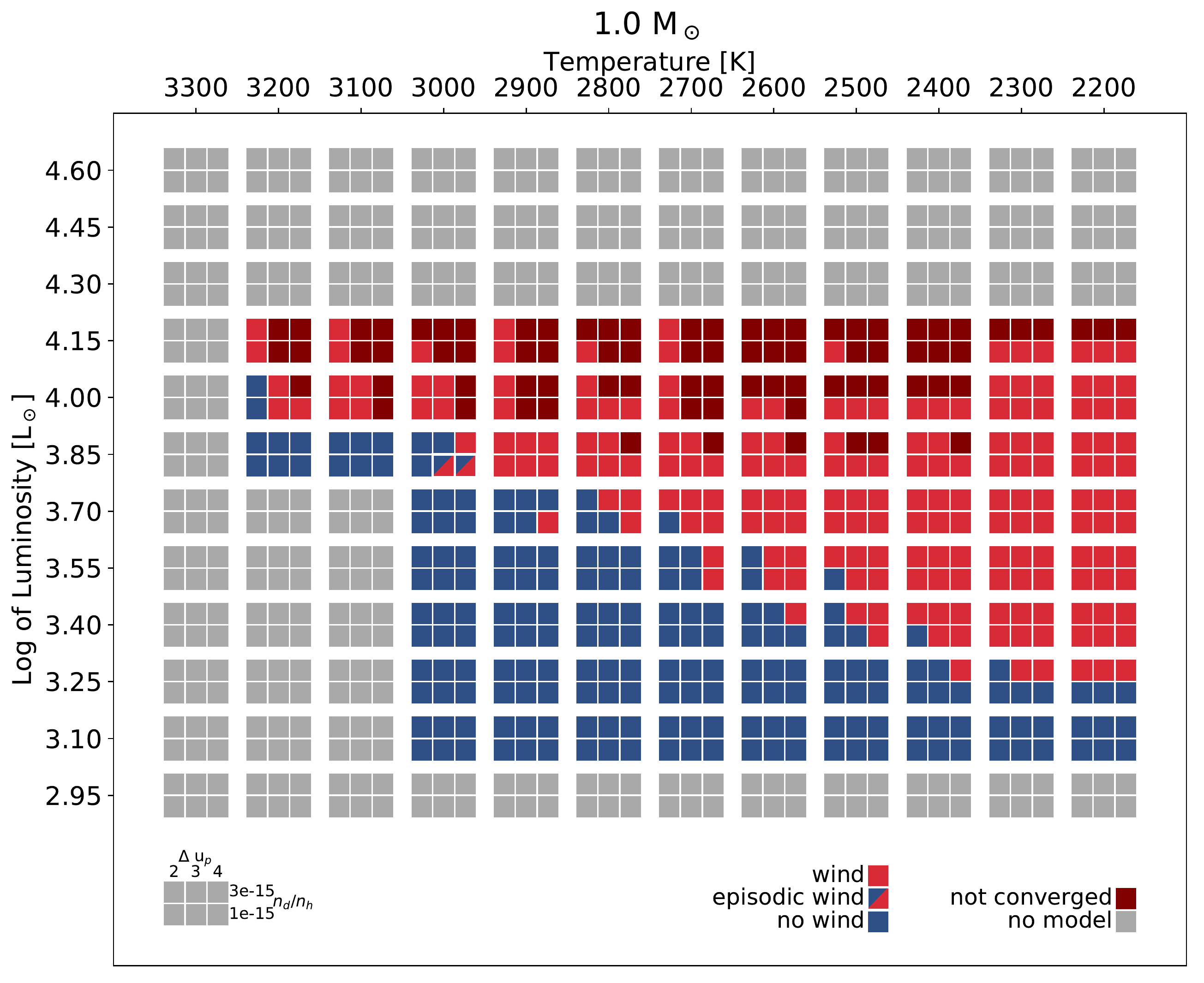} 
\includegraphics[width=0.49\textwidth]{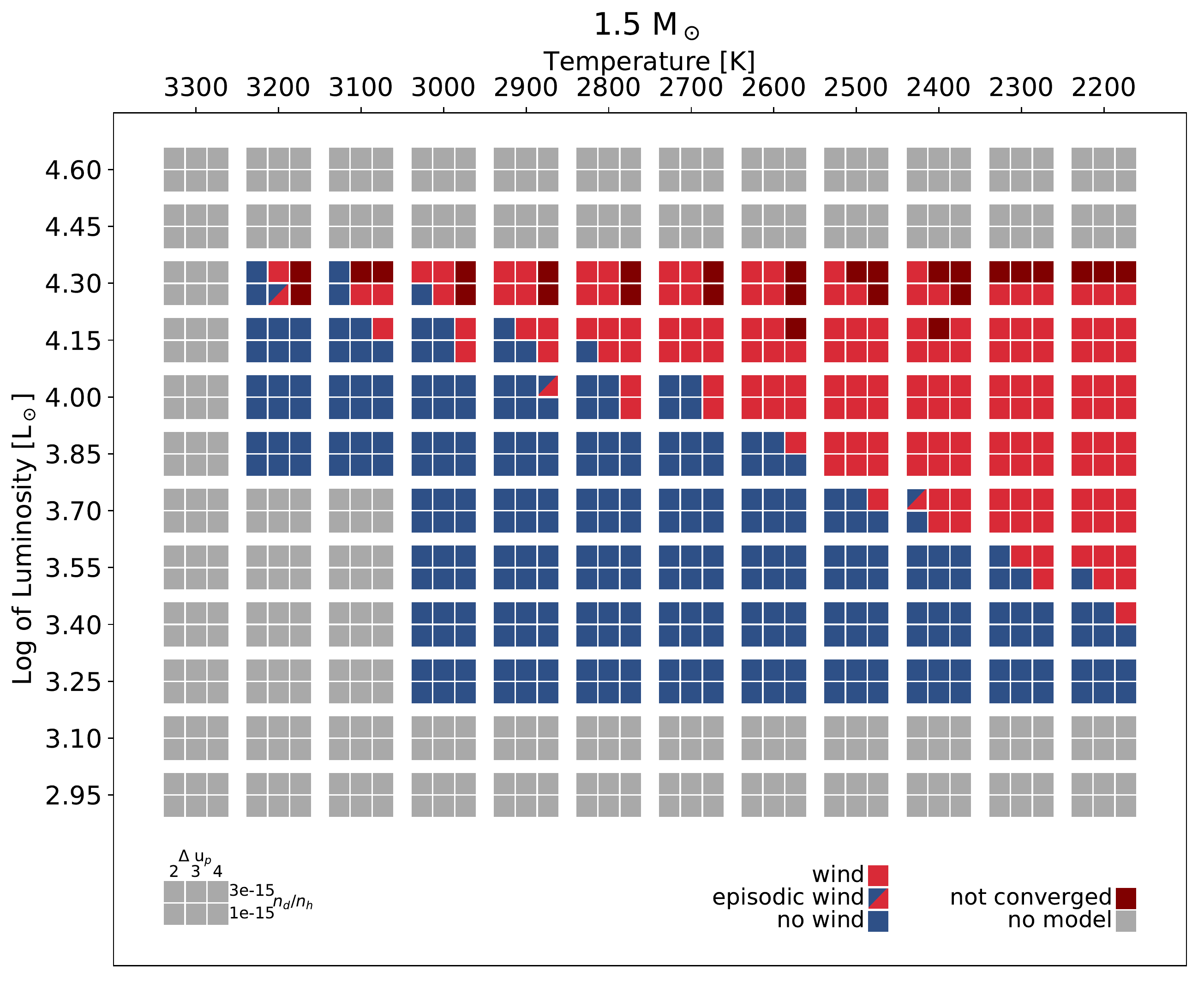} 
\includegraphics[width=0.49\textwidth]{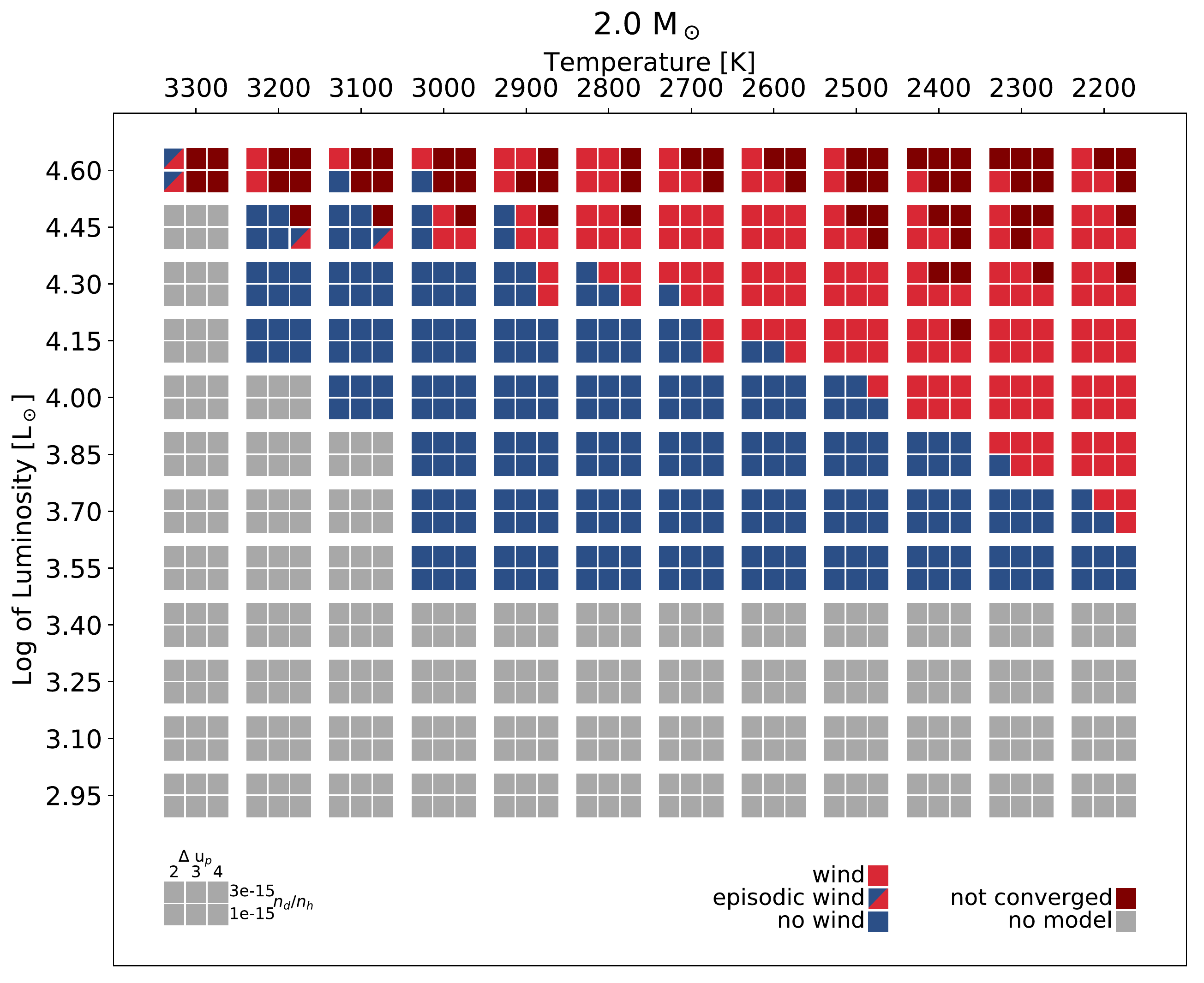}
\includegraphics[width=0.49\textwidth]{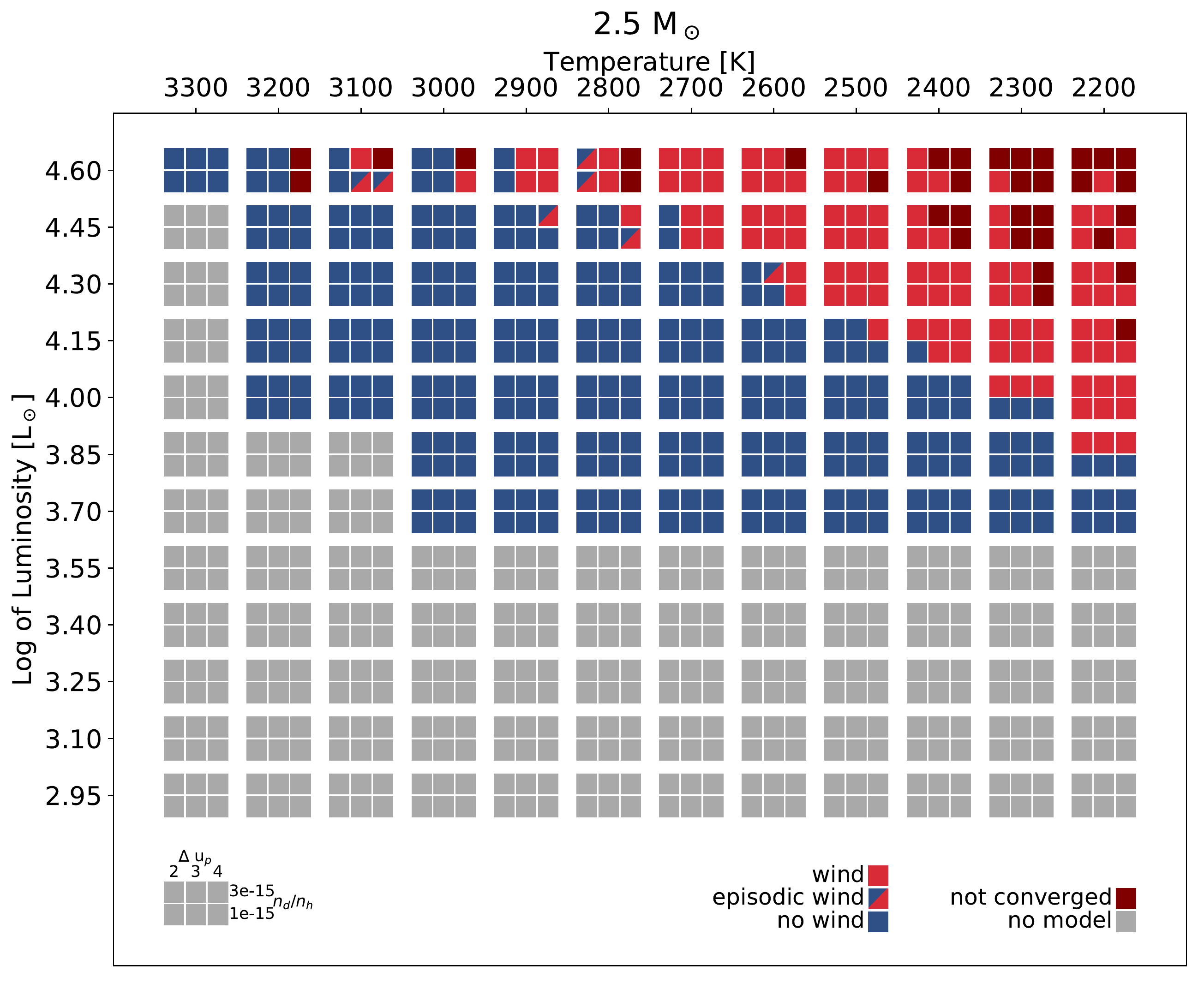} 
\includegraphics[width=0.49\textwidth]{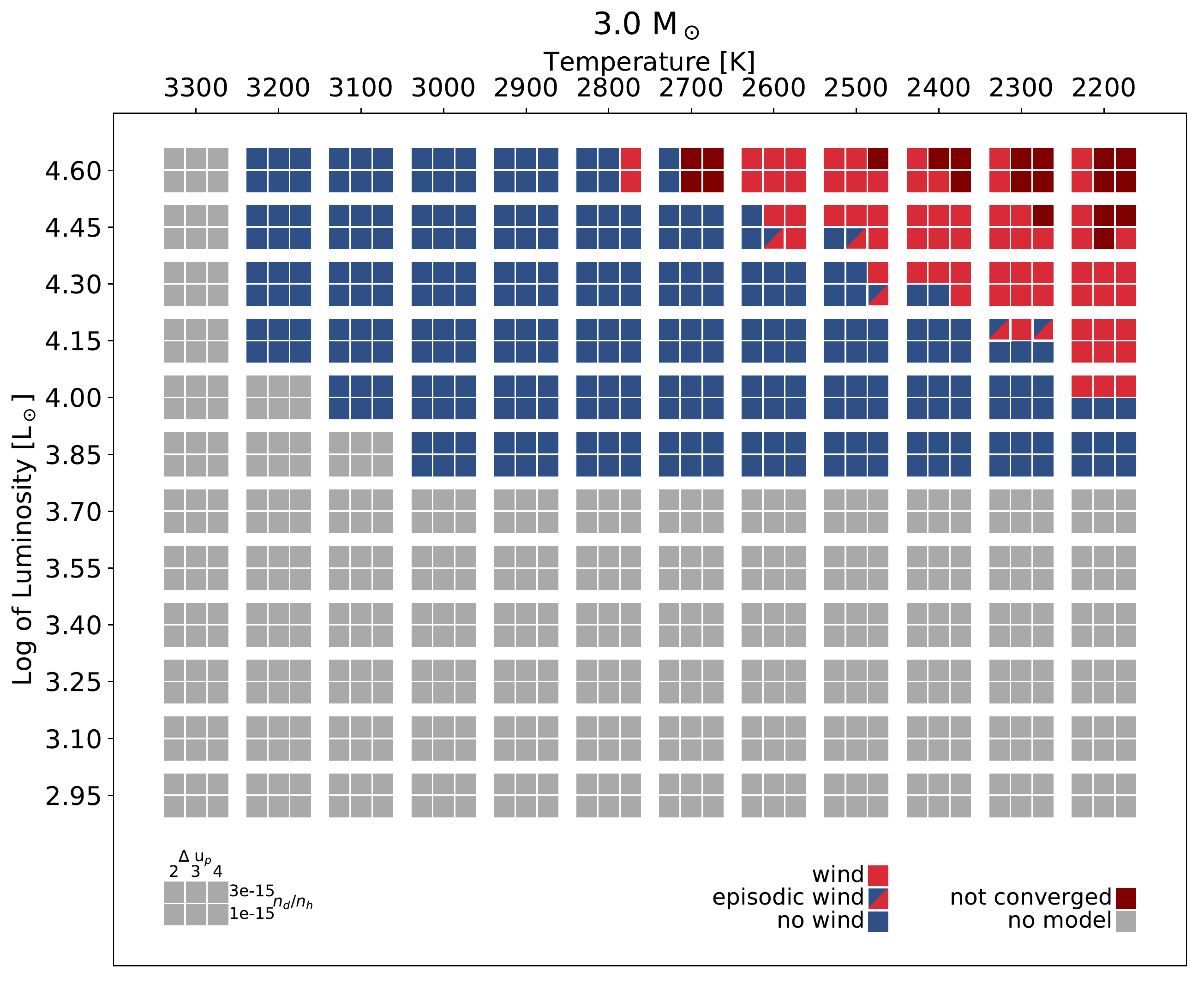}
\caption{Schematic overview showing the dynamic behaviour of the models in the grid as a function of the model parameters. The different panels show maps for different stellar masses. The temperature increases to the left in a format resembling a  HR diagram. The red boxes represent models with a stellar wind, the blue boxes represent models with no wind, the boxes with red/blue triangles represent models with episodic mass loss, the grey boxes indicate combinations of parameters not tested, and the dark red boxes represent models that develop a wind but fail to converge for numerical reasons. For each combination of luminosity and effective temperature the seed particle abundance and piston velocity are varied as indicated by the inset box in the bottom left corner of each panel.}
\label{fig:windmaps}
\end{figure*}

\subsection{Schematic overview of dynamical properties}
\label{sec:windmaps}
Figure~\ref{fig:windmaps} shows schematic overviews of the dynamical properties of the new extensive grid, sorted by stellar mass. These maps are organised like a Hertzsprung-Russell (HR) diagram, with decreasing temperature in the positive x-direction and increasing luminosity in the positive y-direction. A subset of boxes is assigned to every combination of effective temperature and stellar luminosity and each subset is organised such that the piston velocity is increasing upwards and the seed particle abundance is increasing towards the right. In the end, every box represents a model in the grid: the subset it belongs indicates the stellar parameters of the model and the placement within the subset indicates the piston velocity and seed particle abundance used. The dynamical properties of each model are indicated by the colour-coding of the box. Red boxes represent models that develop a stellar wind, blue boxes represent models with no mass loss, red-and-blue boxes  indicate models with episodic mass loss, grey boxes indicate combinations of parameters not covered in the grid, and dark red boxes represent models that clearly develop a wind but fail to converge at some point for numerical reasons. This occurs when dust formation and wind acceleration are too abrupt, leading to decreasing numerical time steps, and eventually non-converging structures. It is important to remember that this introduces a bias against high mass-loss rates (between $10^{-5}$ and $10^{-4}$\,M$_{\odot}$/yr) in the grid. 

These grid maps illustrate the combinations of stellar parameters (mass, luminosity, and effective temperature) that produce outflows. They also indicate where the boundary between models with and without a wind is situated. From this schematic overview alone it is clear that the DARWIN models for M-type AGB stars produce outflows for a wide range of stellar parameters, although it is generally more difficult to drive winds for high stellar masses, high effective temperatures and low luminosities. We note that the lower end of the luminosity range and the higher end of the effective temperature range that can produce outflows both change with stellar mass. In particular, the occurrence of stellar winds at high effective temperatures and low luminosities is facilitated by a smaller current mass.


\begin{figure*}
\centering
\includegraphics[width=0.49\textwidth]{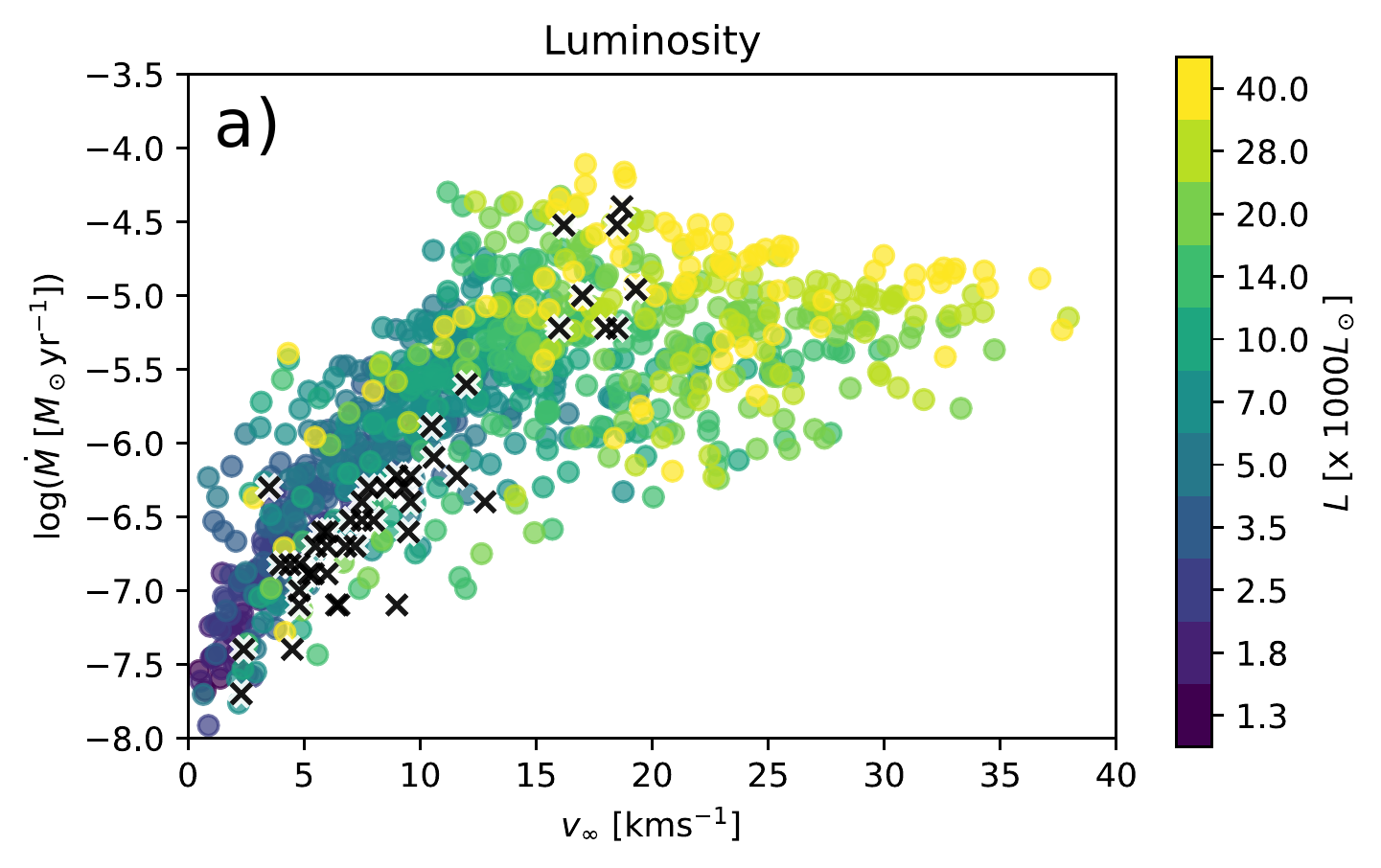}
\includegraphics[width=0.49\textwidth]{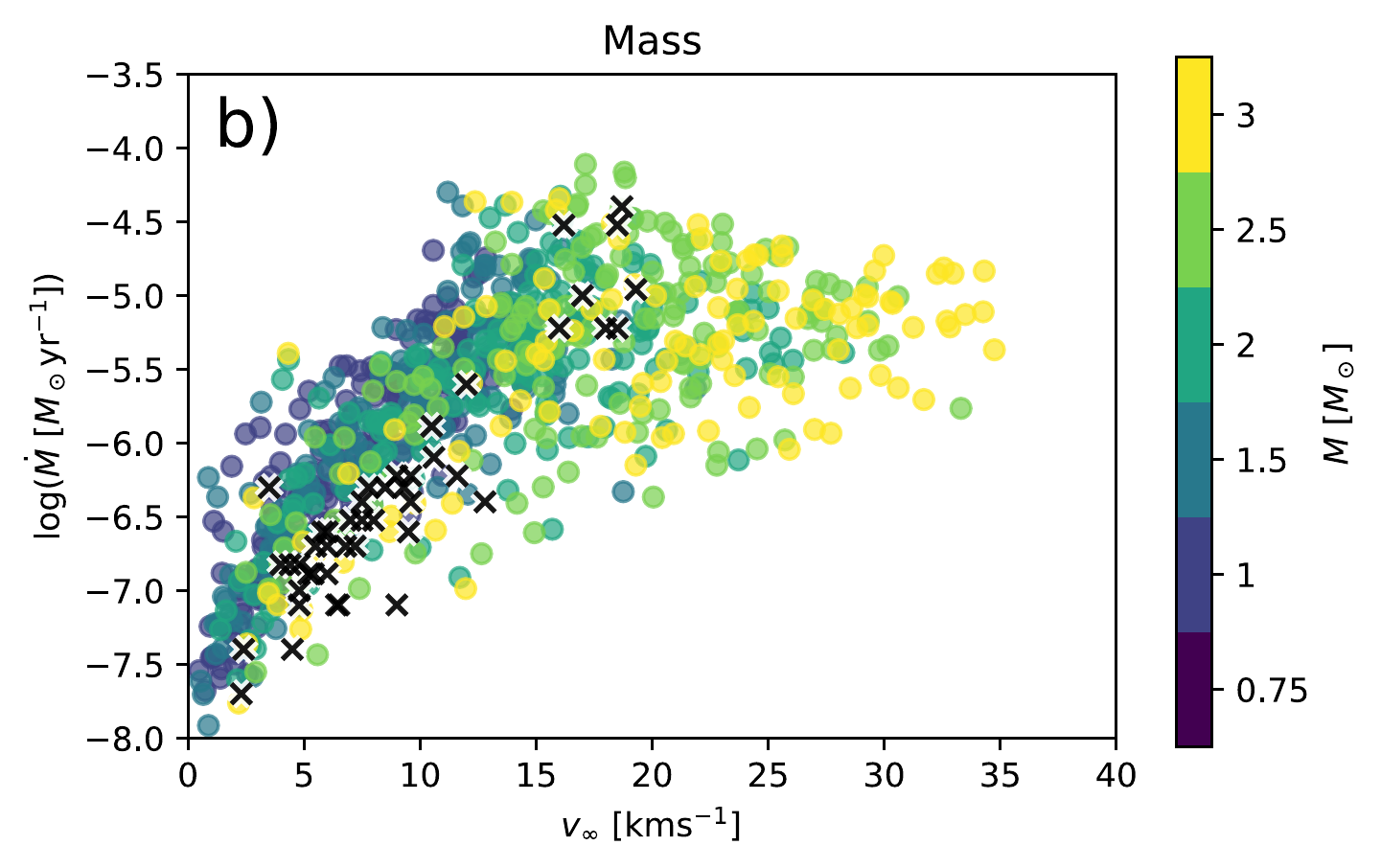}
\includegraphics[width=0.49\textwidth]{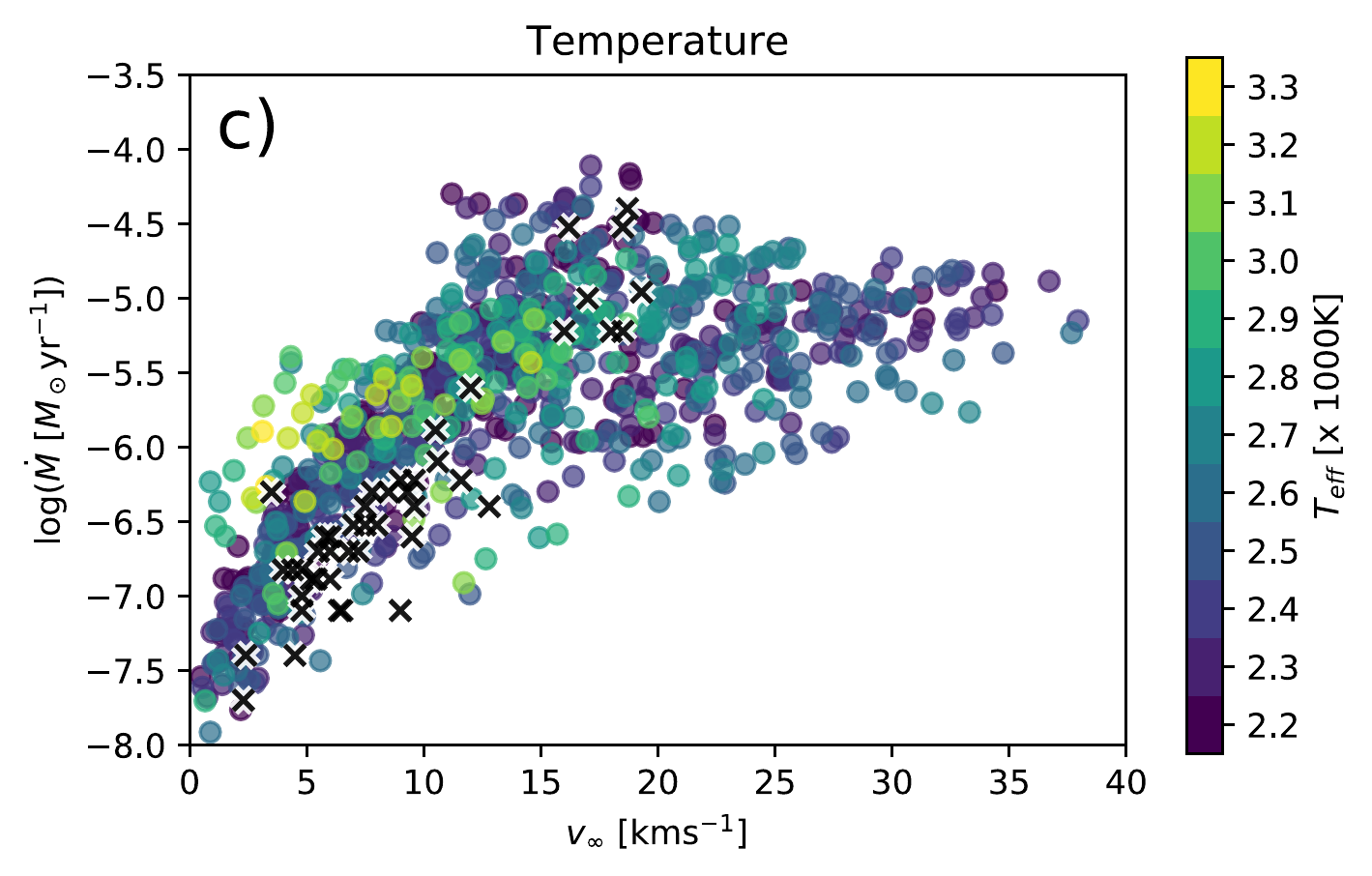}
\includegraphics[width=0.49\textwidth]{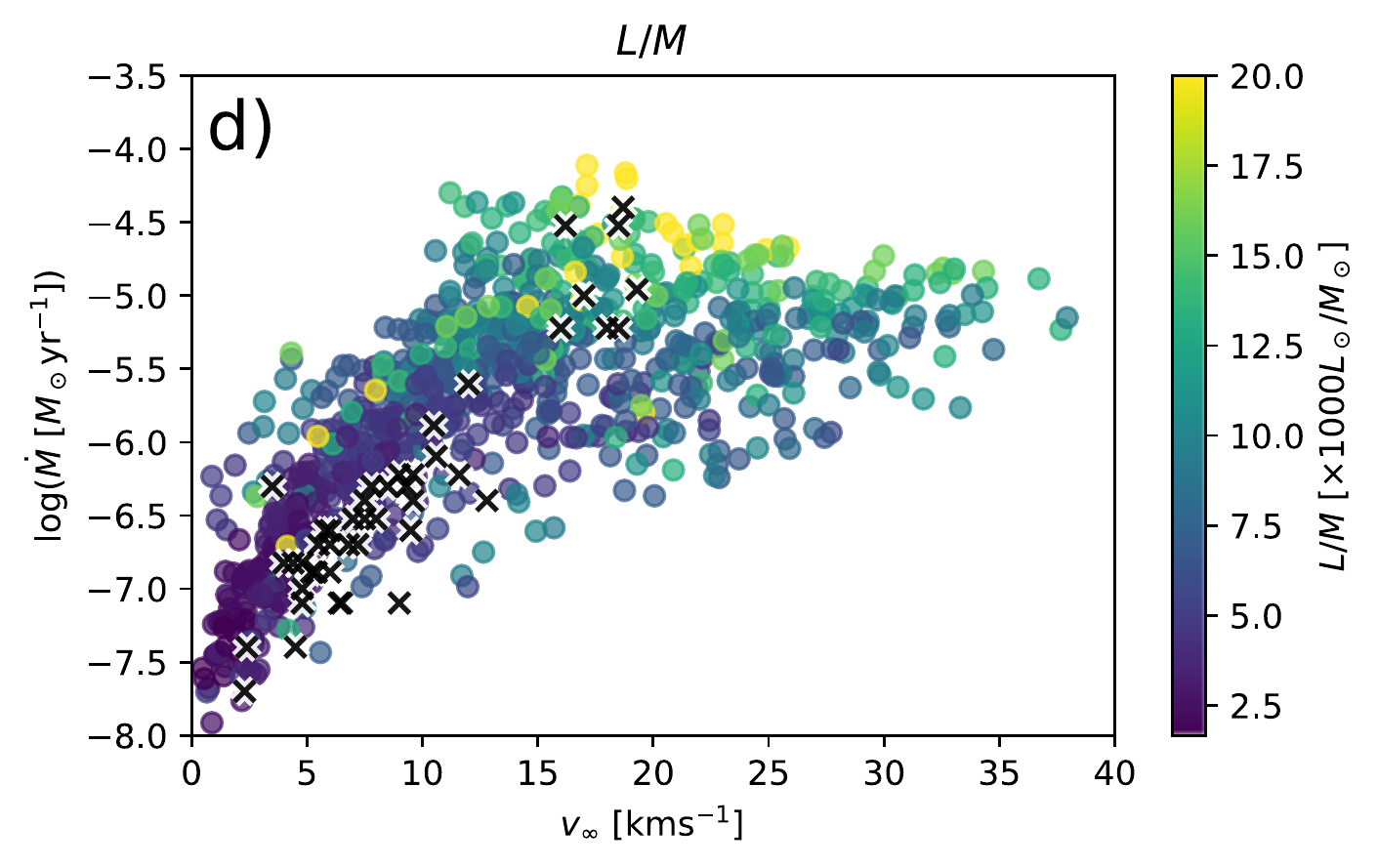} 
\includegraphics[width=0.49\textwidth]{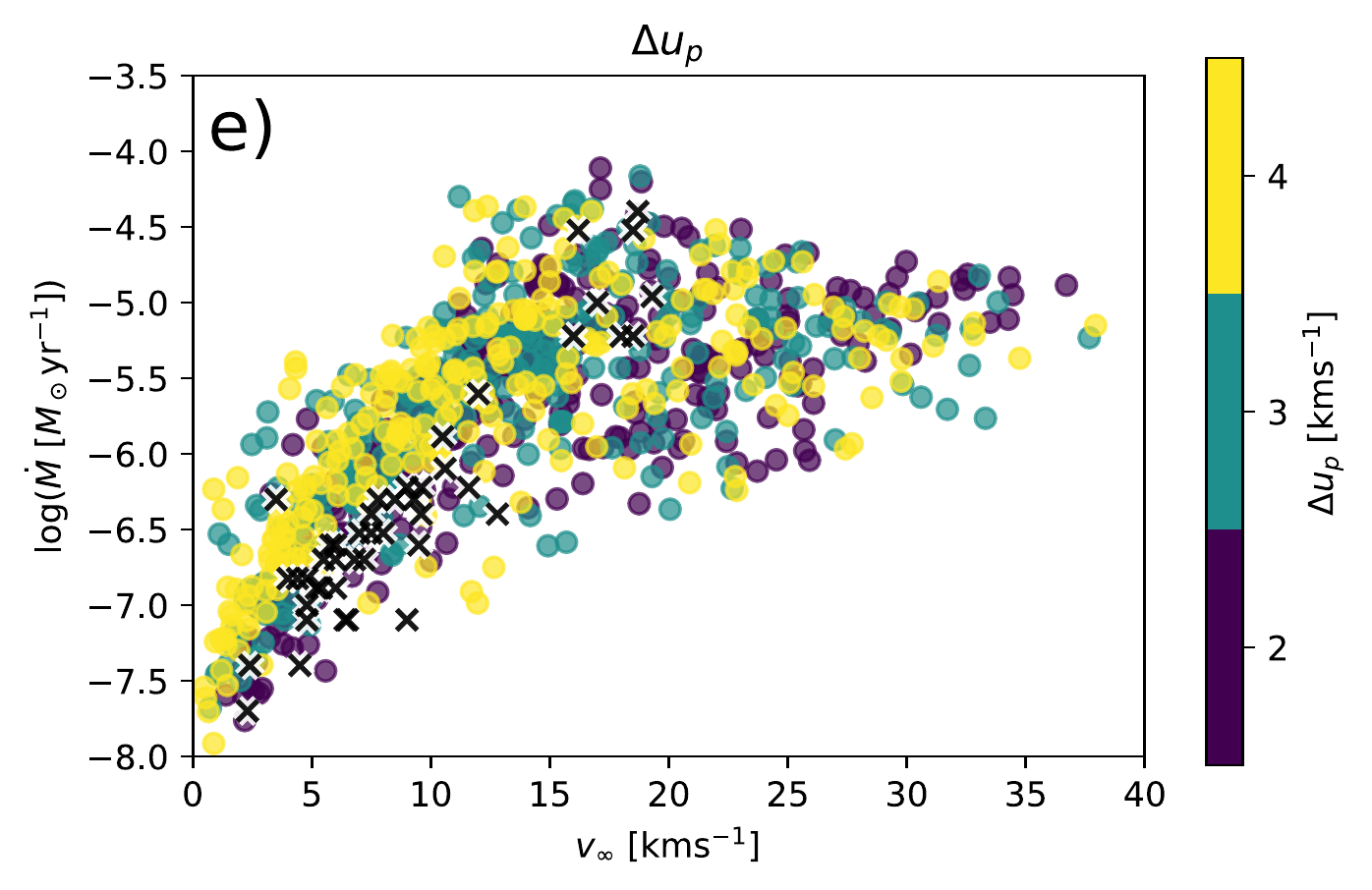}
\includegraphics[width=0.49\textwidth]{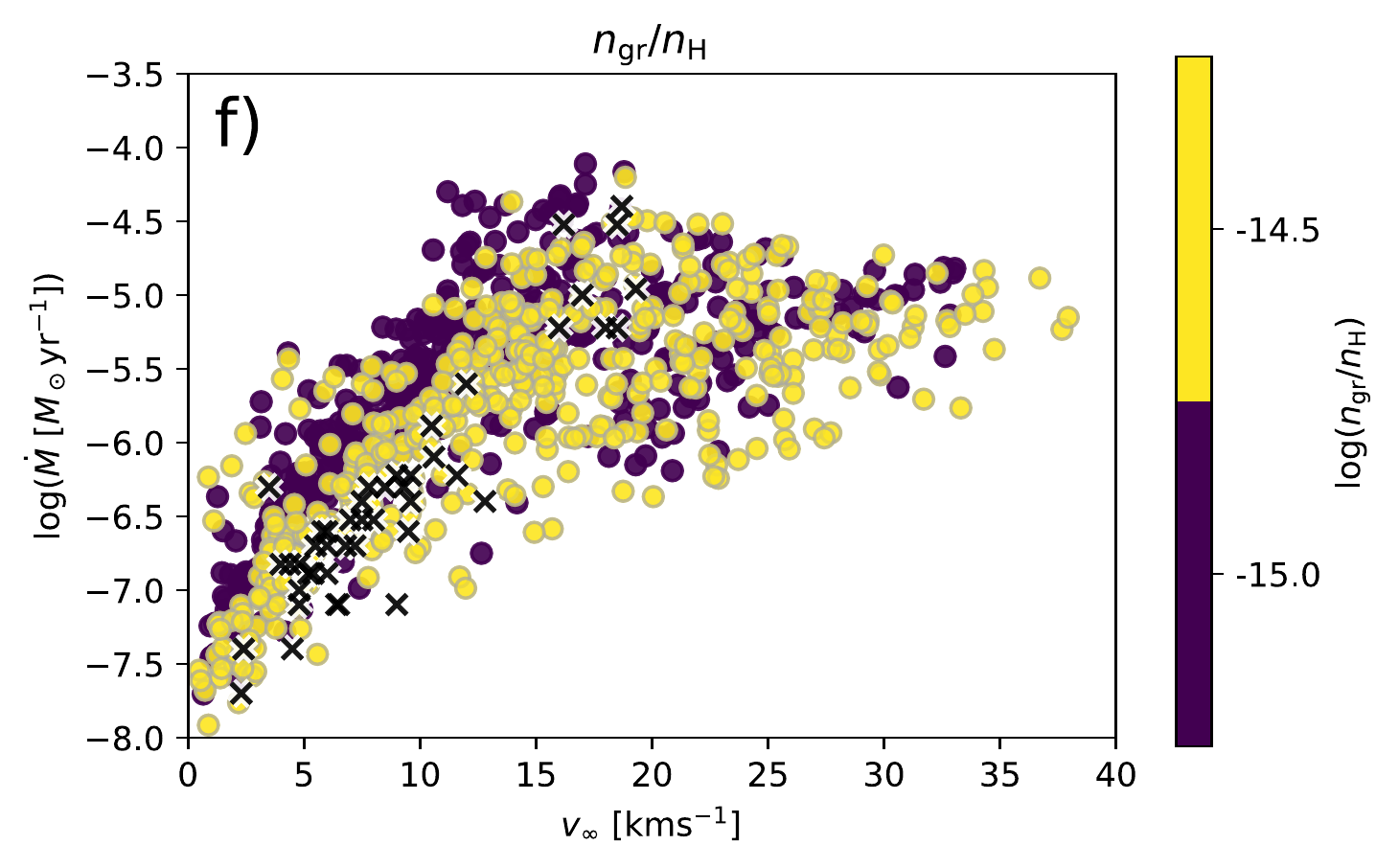}
\includegraphics[width=0.49\textwidth]{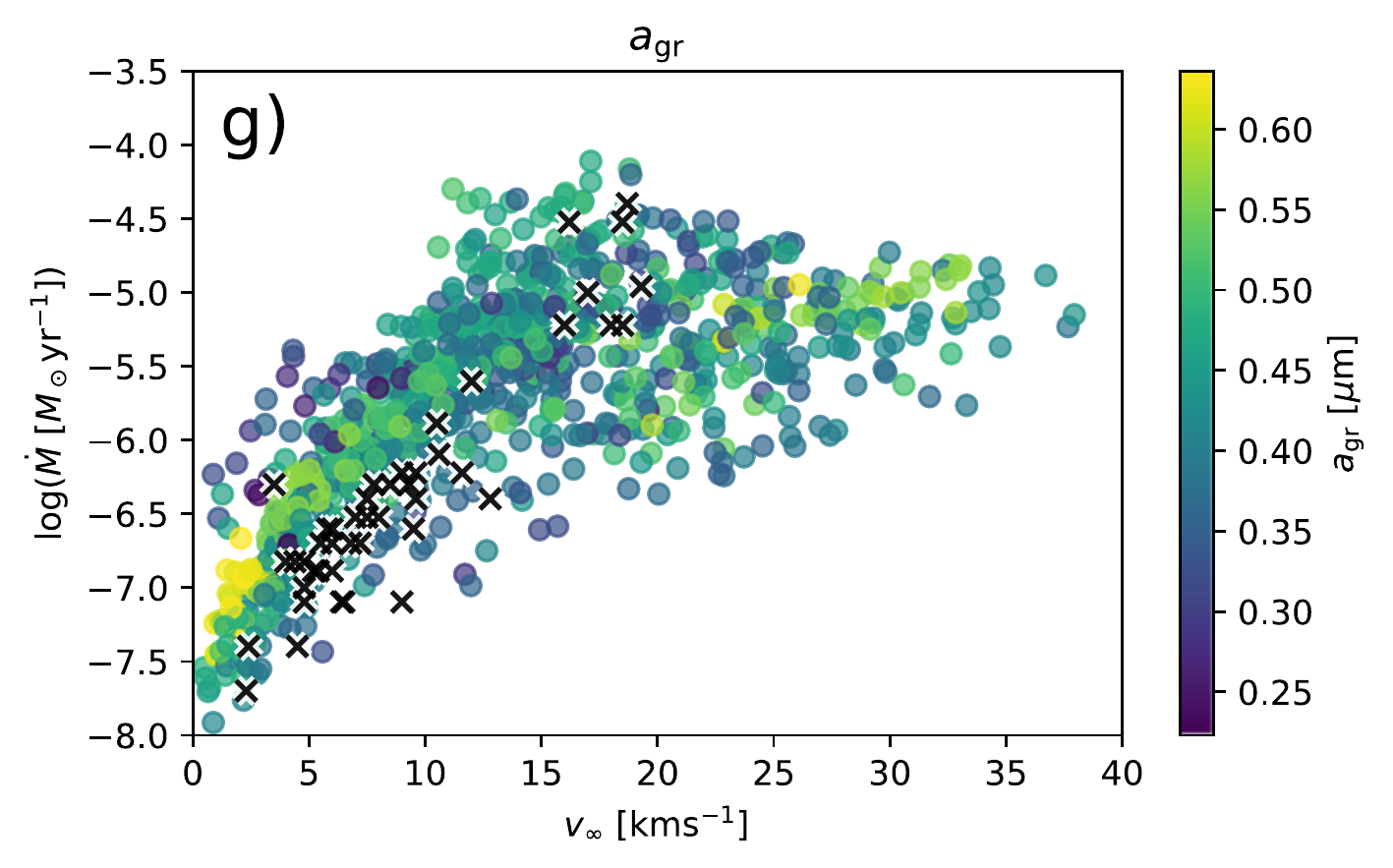}
\includegraphics[width=0.49\textwidth]{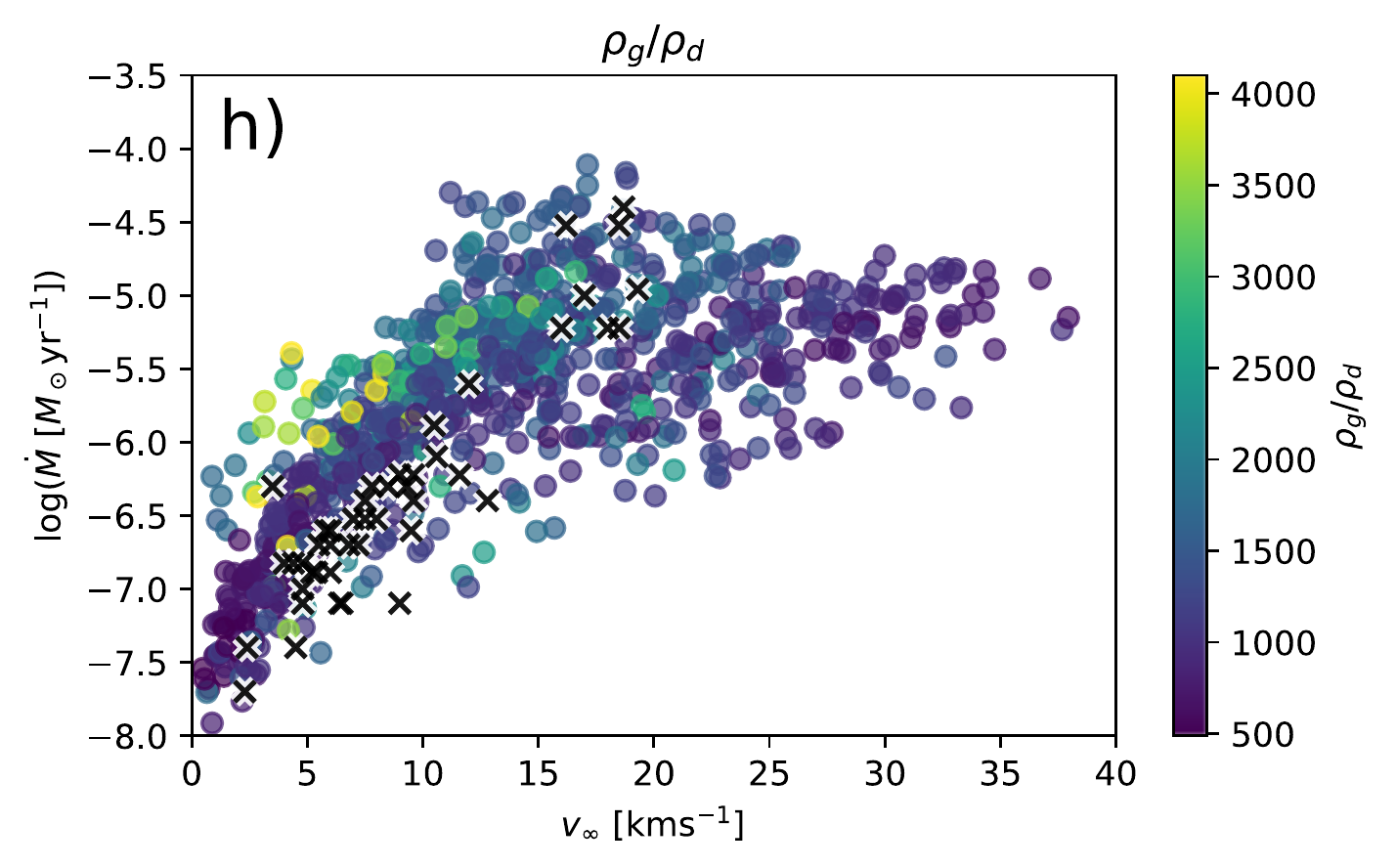}
\caption{Mass-loss rates vs. wind velocities from DARWIN models of M-type AGB stars with a stellar wind (indicated in red in Fig.~\ref{fig:windmaps}) and the corresponding properties from observed M-type AGB stars, derived from CO-lines \citep[][black crosses]{olofsson02,gondel03}. All panels show the same data but with colour-coding according to stellar luminosity, stellar mass, effective temperature, $L_*/M_*$, seed particle abundance, piston velocity amplitude, grain size, and gas-to-dust mass ratio.}
\label{fig:dmdtvel_props}
\end{figure*}

\begin{figure*}
\centering
\includegraphics[width=0.45\textwidth]{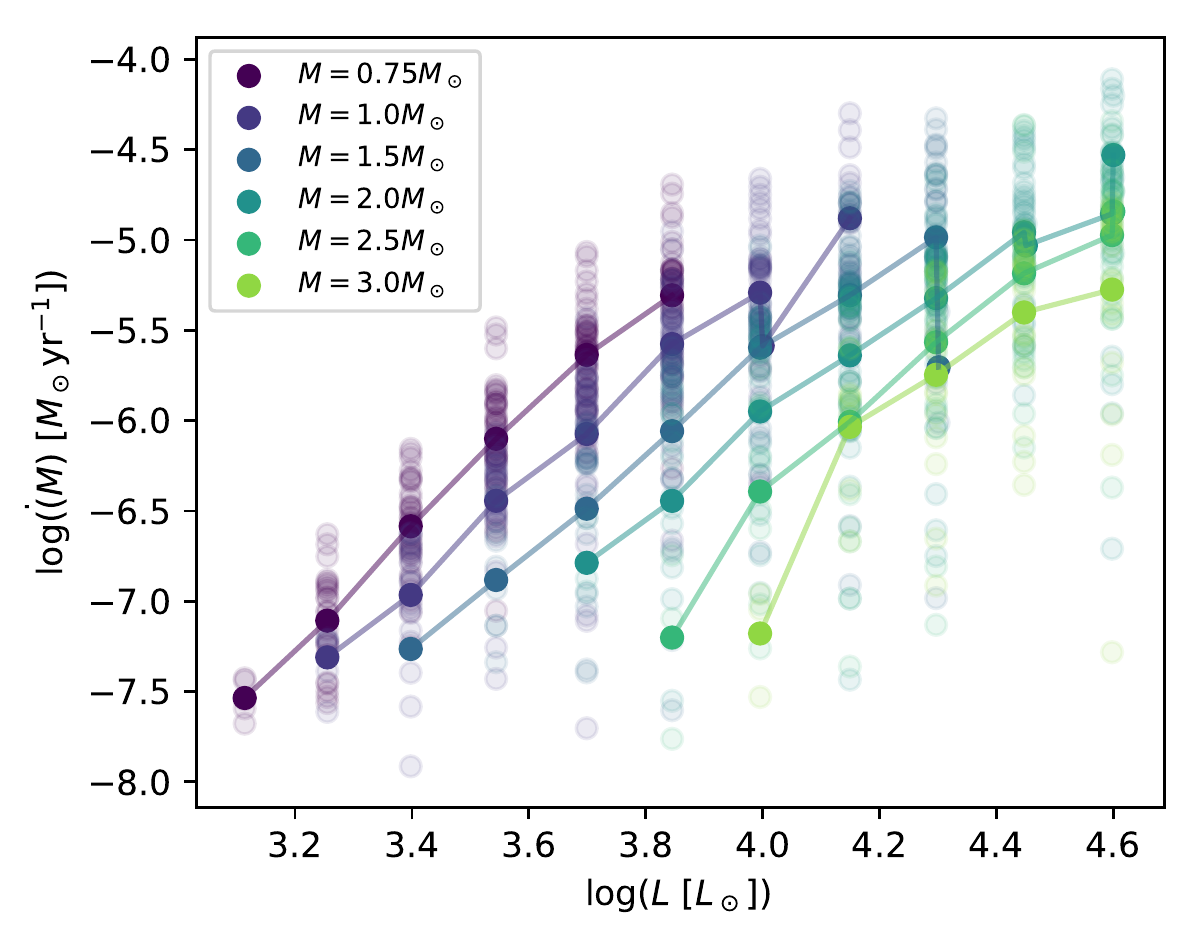} 
\includegraphics[width=0.43\textwidth]{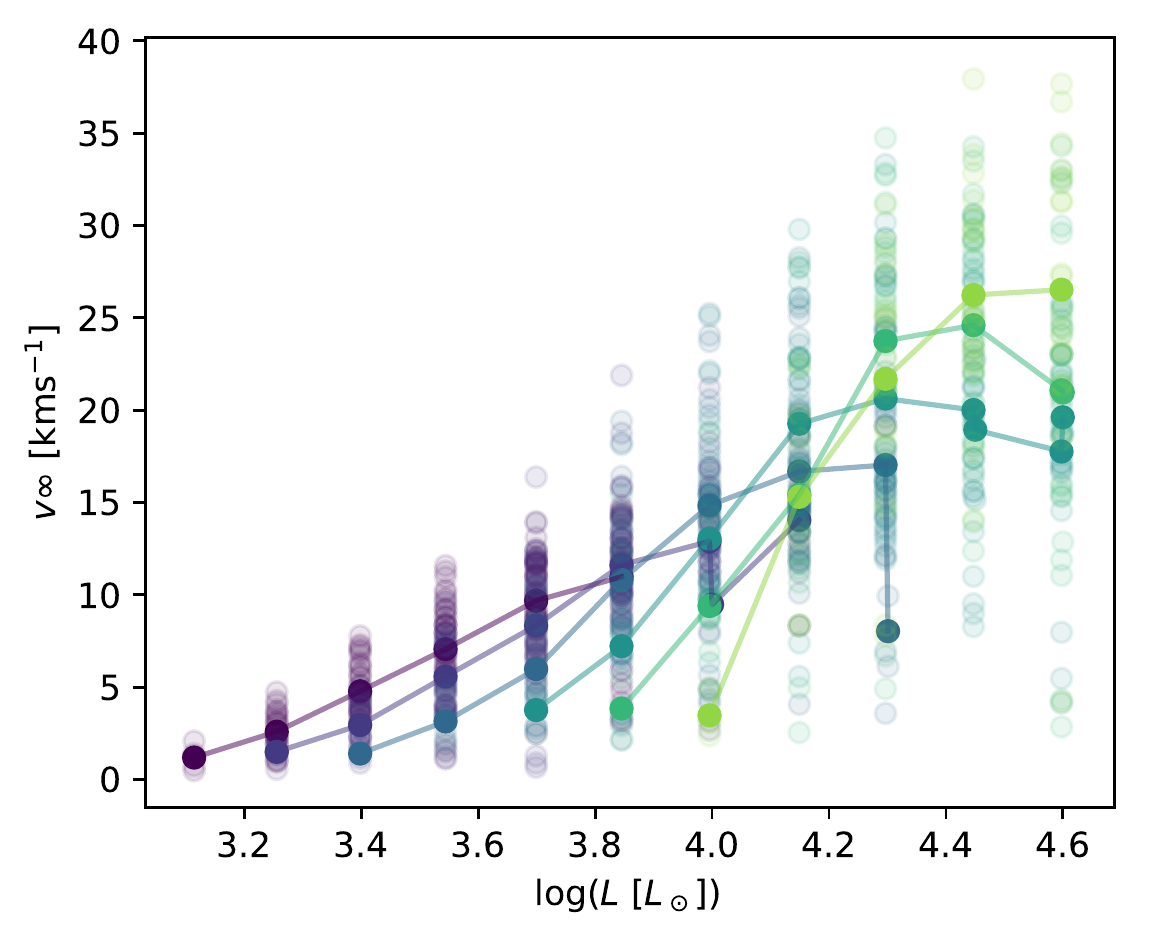}
\includegraphics[width=0.45\textwidth]{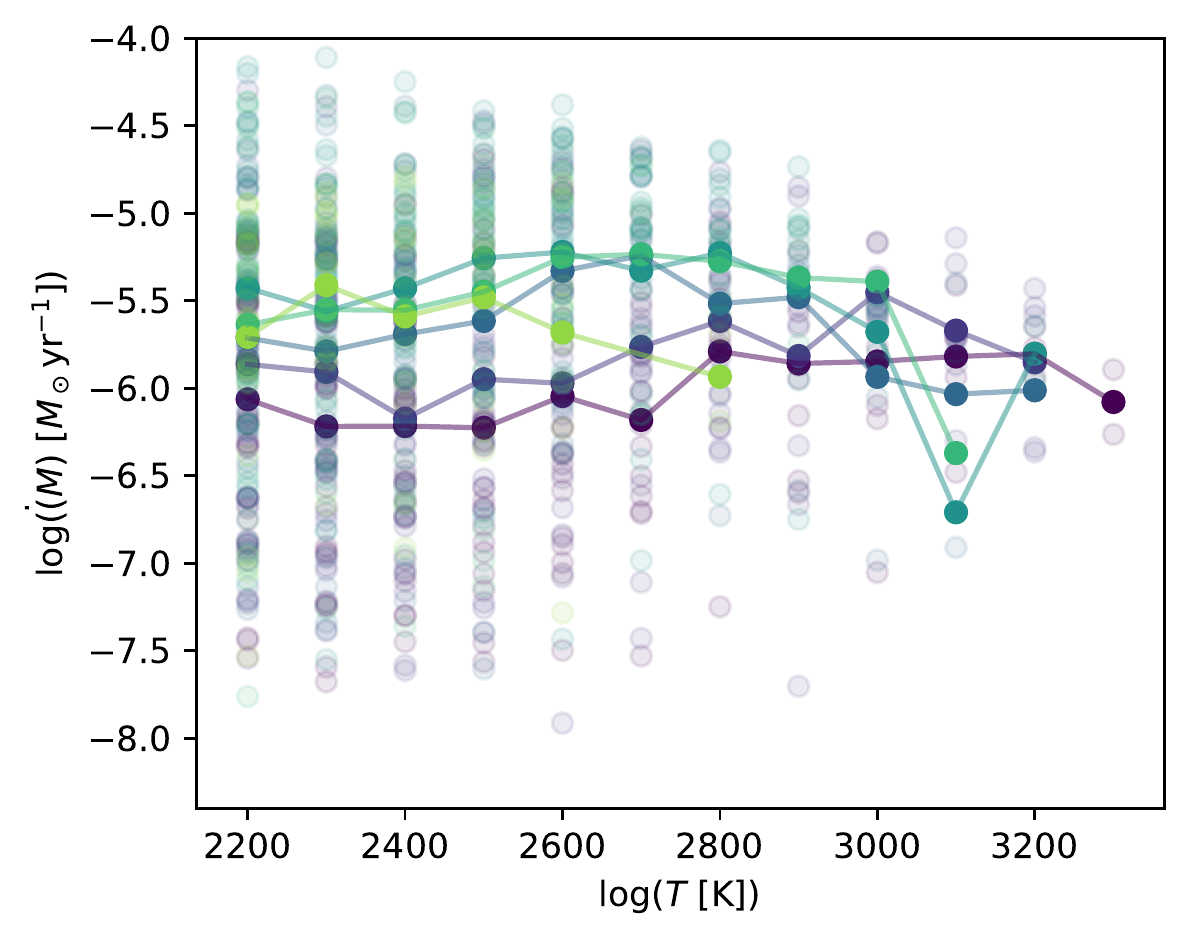}
\includegraphics[width=0.43\textwidth]{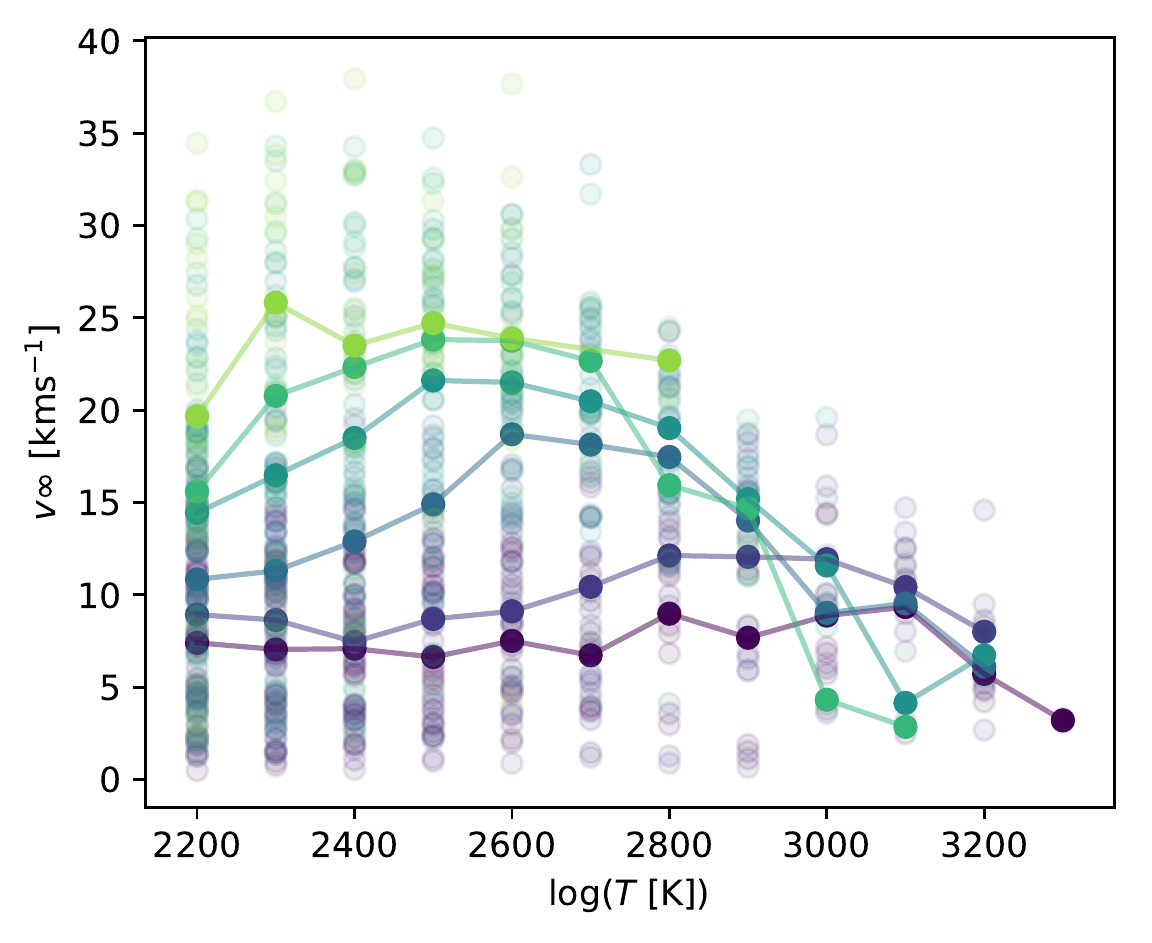}
\caption{Mass-loss rate (left panels) and wind-velocity (right panels) for all models that develop a wind (indicated in red in Fig.~\ref{fig:windmaps}) plotted as a function of luminosity (top panels) and effective temperature (lower panels). The data in all panels are colour-coded according to stellar mass. The linked points show the average mass-loss rate at the indicated luminosities or effective temperature for a specific mass (indicated by the colour-coding).}
\label{fig:per_dmdt1}
\end{figure*}

\subsection{Trends in mass-loss rate vs. stellar parameters}
Figure~\ref{fig:dmdtvel_props} shows the wind velocity and mass-loss rate for all wind-producing models in the grid. The data is colour-coded to illustrate the effects of different input parameters ($L_*$, $M_*$, $T_*$, $L_*/M_*$, $u_{\mathrm{p}}$, $n_{\mathrm{d}}/n_{\mathrm{H}}$, $a_{\mathrm{gr}}$, and $\rho_{\mathrm{p}}/\rho_{\mathrm{d}}$). 

\subsubsection{Luminosity and mass}
The strongest correlation of mass-loss rate with a fundamental parameter is seen in panel a) of Fig.~\ref{fig:dmdtvel_props}, showing the dynamical properties sorted according to luminosity. A higher mass-loss rate correlates with a higher luminosity, especially if the stellar mass is kept fixed. This trend is very clearly illustrated in the top panels of Fig.~\ref{fig:per_dmdt1}, which show the mass-loss rate and wind velocity as a function of luminosity, colour-coded according to stellar mass. The linked points show the average mass-loss rate for the different luminosities at a fixed stellar mass (indicated by the colour-coding). The mass-loss rates from these linked points correlate with luminosity according to $\dot{M}\propto L^{2.3-3.0}$, with the power-law index depending on the stellar mass. The mass-loss formula derived by \cite{wachter02} for models of C-type AGB stars gives a similar trend with luminosity, i.e. $\dot{M}\propto L^{2.47}$ for a fixed effective temperature and mass. However, the mass-loss rates resulting from this formula are about one order of magnitude higher than the mass-loss rates from DARWIN models for similar stellar parameters \citep{eriksson14}. This occurs because the low value chosen for the gas opacity in the grey models by \cite{wachter02} results in much higher gas densities at a given temperature, which in turn translates to higher mass-loss rates.  The top right panel of Fig.~\ref{fig:per_dmdt1} also indicates a weak dependence of the wind velocity on the luminosity, which seems to saturate at higher luminosities.

The dependence on stellar mass has not been explored for DARWIN models of M-type AGB stars. Panel b) of Fig.~\ref{fig:dmdtvel_props} shows the wind properties colour-coded according to mass. There is a strong positive correlation between mass-loss rates and wind velocities for models with a current stellar mass between $0.75$\,M$_{\odot}$ and $1.5$\,M$_{\odot}$, whereas this trend seems to level off and reach a plateau for higher masses. Since the more massive models in the grid ($M_*\geq 1.5$\,M$_{\odot}$) cover luminosities higher than $\log(L_*/L_{\odot})=4.15$, these models will have pulsation periods longer than $\sim$800 days (see Table~\ref{tab:grid}). For the lower mass models ($M_*\leq 1.5$\,M$_{\odot}$) the wind velocity never goes above $\sim$\,15\,km/s. This is partly due to limitations of the grid, e.g. the limitations in luminosity for models with 0.75\,M$_{\odot}$ and 1\,M$_{\odot}$, but is probably also caused by lower densities in the wind acceleration zone for models with lower mass.

\subsubsection{Effective temperature, seed particle abundance, and piston amplitude}
In Fig.~\ref{fig:dmdtvel_props} we also plot the wind properties colour-coded according to effective temperature, piston amplitude, and seed particle abundance. The wind velocity correlates somewhat with seed particle abundance, as can be seen in panel f) of Fig.~\ref{fig:dmdtvel_props}, where diagonal bands shifted in wind velocity are formed when sorted by seed particle abundance, but otherwise there are no distinct trends. The lack of correlation between the mass-loss rate and the effective temperature is also evident in the bottom left panel of Fig.~\ref{fig:per_dmdt1}, where the linked points illustrating the average mass-loss rate of all models with a fixed stellar mass and effective temperature
show very little variation with effective temperature. The linked points form almost horizontal lines, except for the highest effective temperatures where the number statistics is low. This is a specific feature in our models for M-type AGB stars. Models for C-type stars instead show an increase in the mass-loss rate with decreasing effective temperature \citep{wachter02,eriksson14}. This difference is also seen in observations \citep{schoier01,olofsson02,loon05,berche05} and is presumably caused by the strong absorption in the dusty circumstellar envelopes of carbon stars, which may produce significant back-warming effects and thereby affect dust formation. The thermal effects of the near-transparent envelopes of M-type stars on the stellar atmosphere are much smaller. 


\subsubsection{The ratio $L_*/M_*$}

\begin{figure}
\centering
\includegraphics[width=0.44\textwidth]{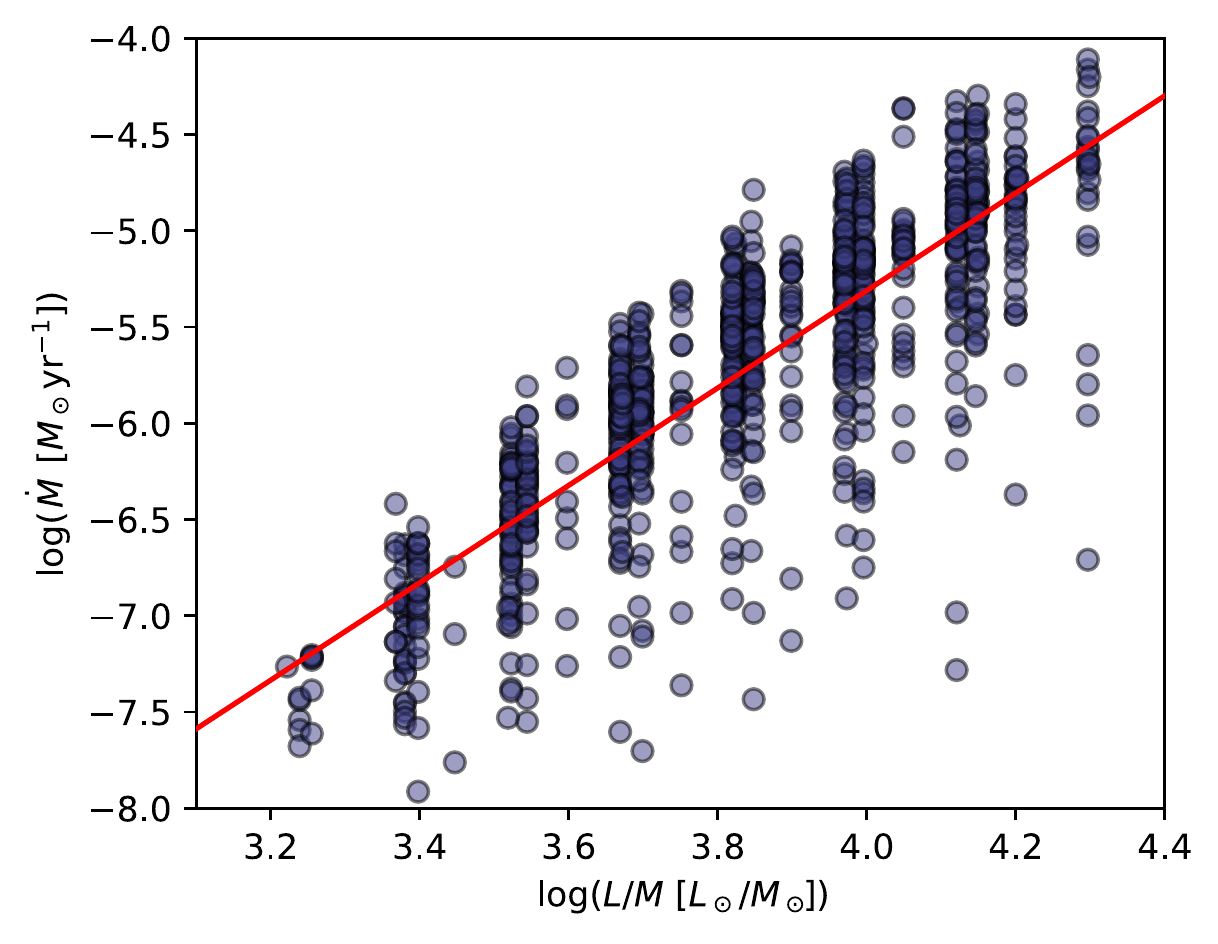}
\caption{Mass-loss rate as a function of $L_*/M_*$. The red line indicates the power-law fit $\dot{M}\propto (L_*/M_*)^{2.5}$.}
\label{fig:per_dmdt3}
\end{figure}

The ratio $L_*/M_*$ is important because it is a measure of the balance between the gravitational acceleration, pulling material towards the star, and the radiative acceleration, driving material away from the star ($a_{\mathrm{rad}}/a_{\mathrm{grav}}\propto L_*/M_*$). Figure~\ref{fig:per_dmdt3} shows the mass-loss rate as a function of $L_*/M_*$ for all models in the grid that produce a stellar wind. The mass-loss rate increases strongly with increasing $L_*/M_*$ (also seen in panel d) of Fig.~\ref{fig:dmdtvel_props}) even though the spread is very large. Increasing the ratio $L_*/M_*$  by an order of magnitude results in increasing the mass-loss rates by about three orders of magnitude, which is substantially steeper than indicated by Reimers’ formula.

It is worth  noting in this context that stellar evolution models tend to favour steeper mass-loss laws than Reimers’ formula \citep{Lattanzio2016}. \cite{renzini81} coined the term superwind to describe the expected high mass-loss rates at the end of the AGB phase, when AGB stars cast off their stellar envelopes and eventually transition to planetary nebulae. From an evolutionary point of view, this stage is characterised by a decreasing current mass as mass layers are blown away, and an increasing luminosity. It is not surprising that the mass-loss rates increase when the gravitational potential decreases at the same time as the radiation pressure increases.


\subsection{Grain properties}
\begin{figure}
\centering
\includegraphics[width=0.44\textwidth]{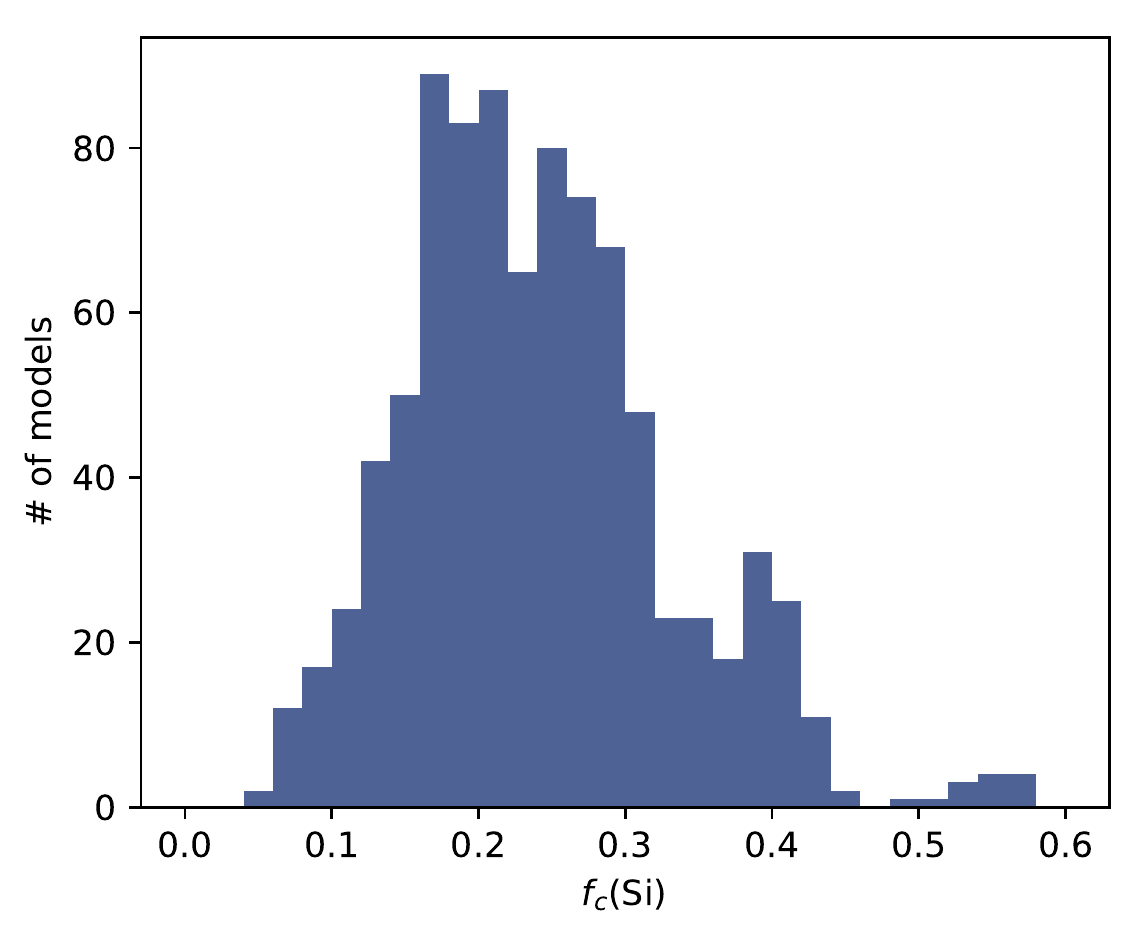}\\
\includegraphics[width=0.44\textwidth]{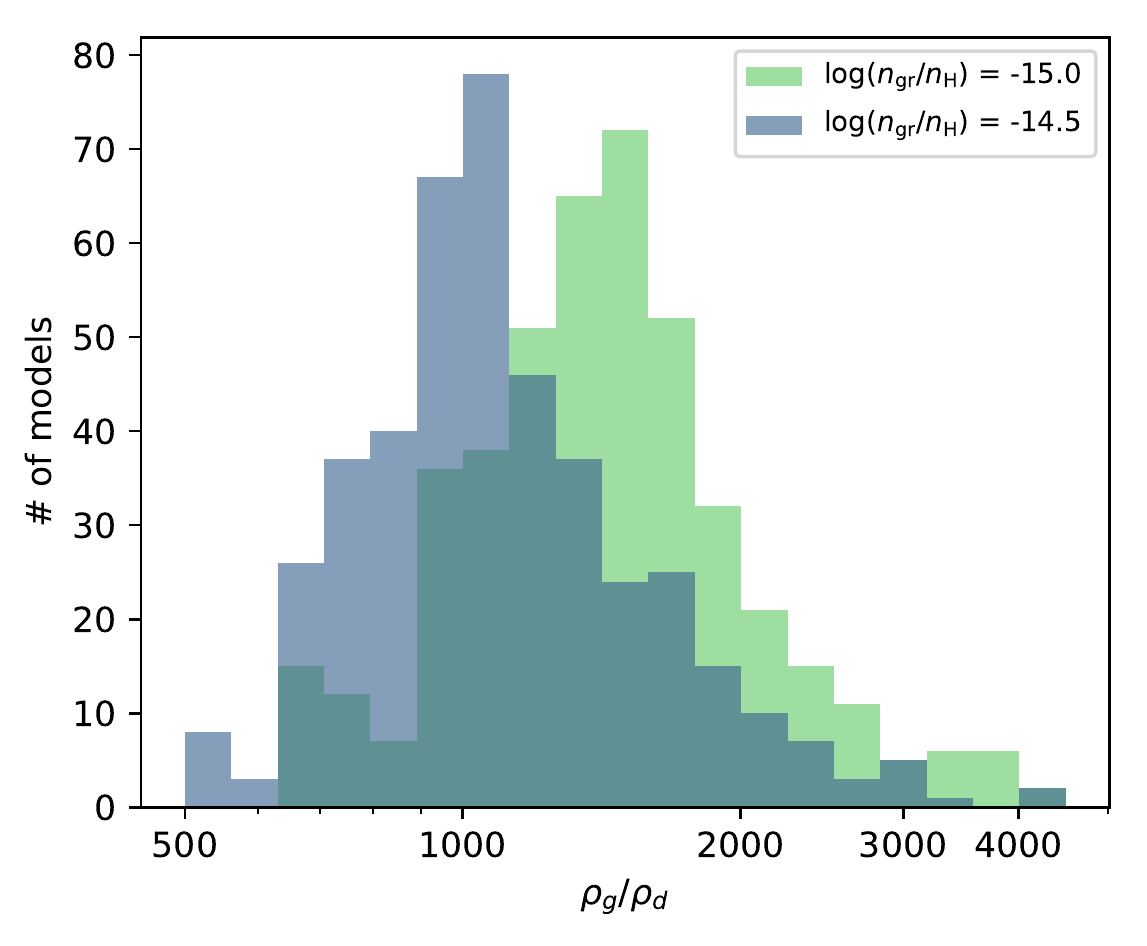}\\  
\includegraphics[width=0.44\textwidth]{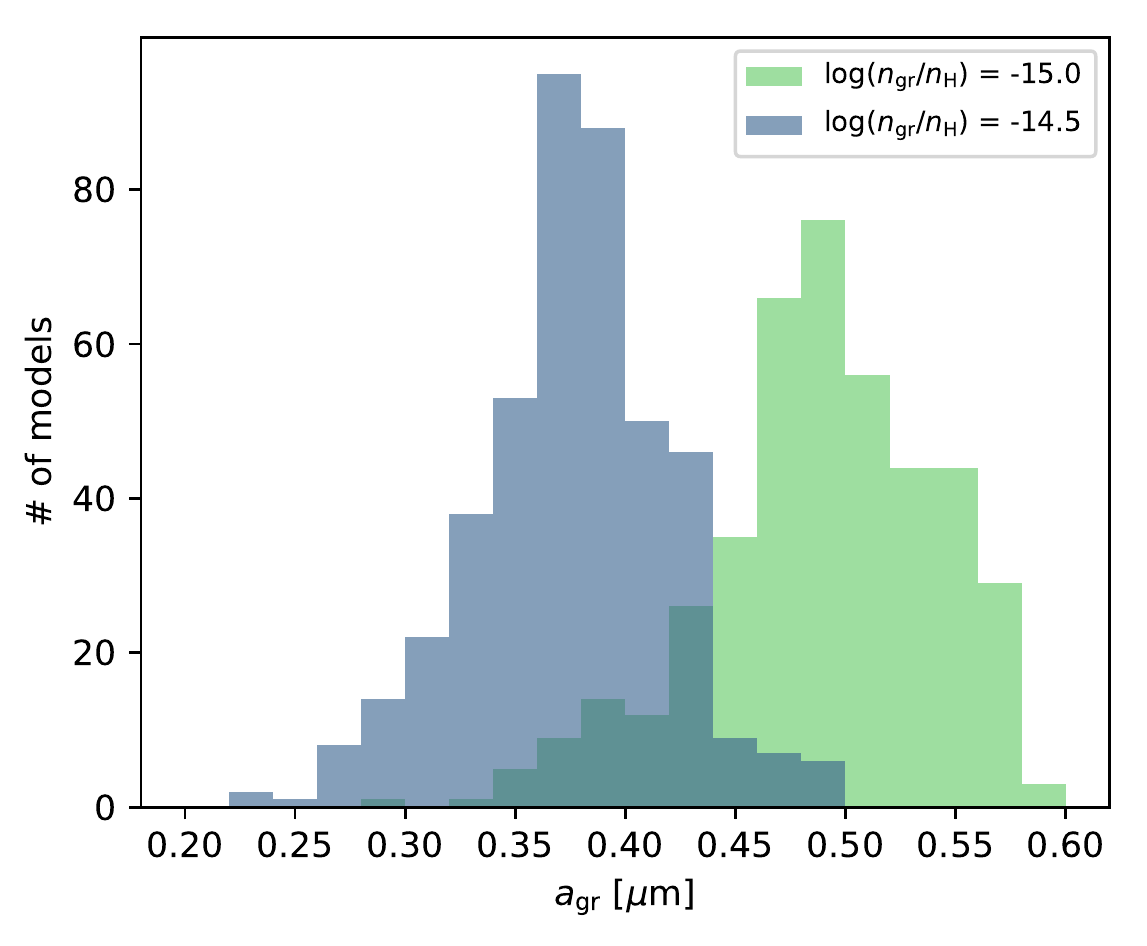}
\caption{Degree of condensed Si (top panel), dust-to-gas ratios (middle panel), and grain sizes (bottom panel) in the outermost atmospheric layers, averaged over time, for all models that develop a wind (indicated in red in Fig.~\ref{fig:windmaps}). The middle and bottom panels are colour-coded according to  seed particle abundance.}
\label{fig:nd_hist}
\end{figure}

\begin{figure}
\centering
\includegraphics[width=0.44\textwidth]{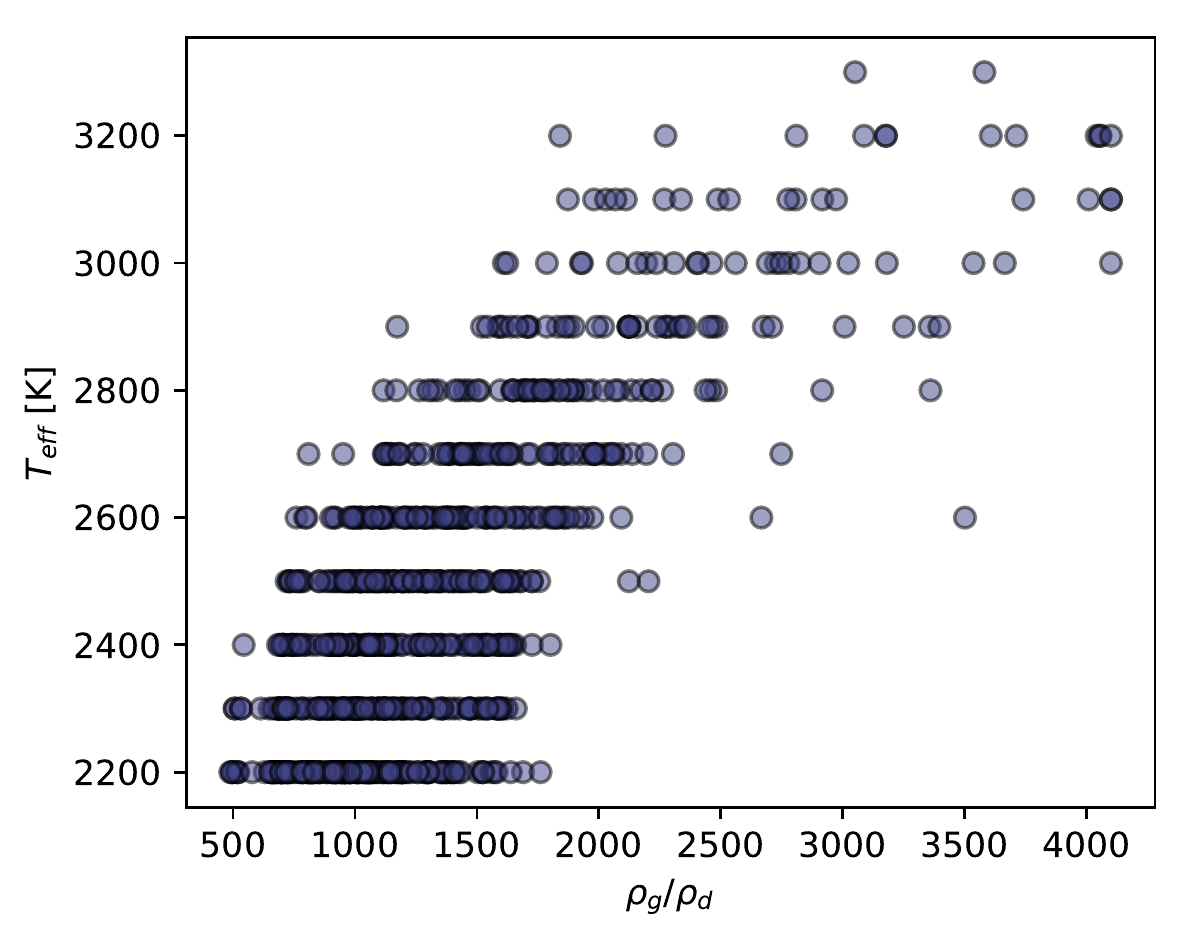}\\
\caption{Gas-to-dust ratio in the outermost atmospheric layers averaged over time vs. the effective temperature of the model.}
\label{fig:nd_model}
\end{figure}

The possible values for the seed particle abundance are limited since photon scattering as a wind-driving mechanism requires grains in a certain size range \citep[see][for a detailed discussion]{hoefner2008}. Fewer seeds make the growth of individual grains more efficient (less competition for material), but may lead to a smaller collective opacity if the grains grow beyond the optimal size range. More seeds lead to stronger competition for condensible material, resulting in smaller grains. This may also lead to a smaller collective opacity if the grains do not grow to the optimal size range for winds predominantly driven by photon scattering. The correlation between grain size and the number of seed particles is seen in the bottom panel of Fig.~\ref{fig:nd_hist}, showing the average grain radius in the outermost layers of models that produce a wind, sorted by seed particle abundance. For a seed particle abundance of $n_{\mathrm{d}}/n_{\mathrm{H}}=1\cdot10^{-15}$ the grain sizes range between $0.3-0.6\,\mu$m, whereas for a seed particle abundance of $n_{\mathrm{d}}/n_{\mathrm{H}}=3\cdot10^{-15}$ the grain sizes range between $0.2-0.5\,\mu$m. 

The degrees of condensed Si in models producing a wind are shown in the top panel, with typical values ranging between 10\% and 40\%. The degrees of condensed magnesium will be about twice as high as for Si since Mg$_2$SiO$_4$ has two magnesium atoms per silicon atom and the elemental abundances of Si and Mg are comparable. This is in good agreement with observations; for example \cite{khouri14} found that about  one-third of Si is condensed into grains for W Hya. The corresponding gas-to-dust mass ratios are shown in the middle panel, with values ranging between 500 and 4000. This result agrees reasonably well with the observational estimates by \cite{Ramstedt2008}, who found dust-to-gas mass ratios around 2000-4000. We note that this is significantly higher than the standard value of 200 used in many studies that derive mass-loss rates from SED fitting.

How grain size and gas-to-dust mass ratios correlate with the wind properties is shown in panels g) and h) of Fig.~\ref{fig:dmdtvel_props}. Models with very low luminosities ($\log (L/L_{\odot})\leq 3.55$) have large grain sizes ($0.4-0.7\,\mu$m) and low gas-to-dust ratios ($500-2000$). The gas-to-dust ratios show almost no dependence on luminosity and mass, but a clear dependence on temperature. This can be seen in Fig.~\ref{fig:nd_model}, showing how the gas-to-dust mass ratios depend on the effective temperature.


\section{Observed and synthetic wind properties}
\label{compobssyn}
\subsection{Trends at pulsation periods below 800 days}
\label{sec:pulswind}

\begin{figure}
\centering
\includegraphics[width=0.47\textwidth]{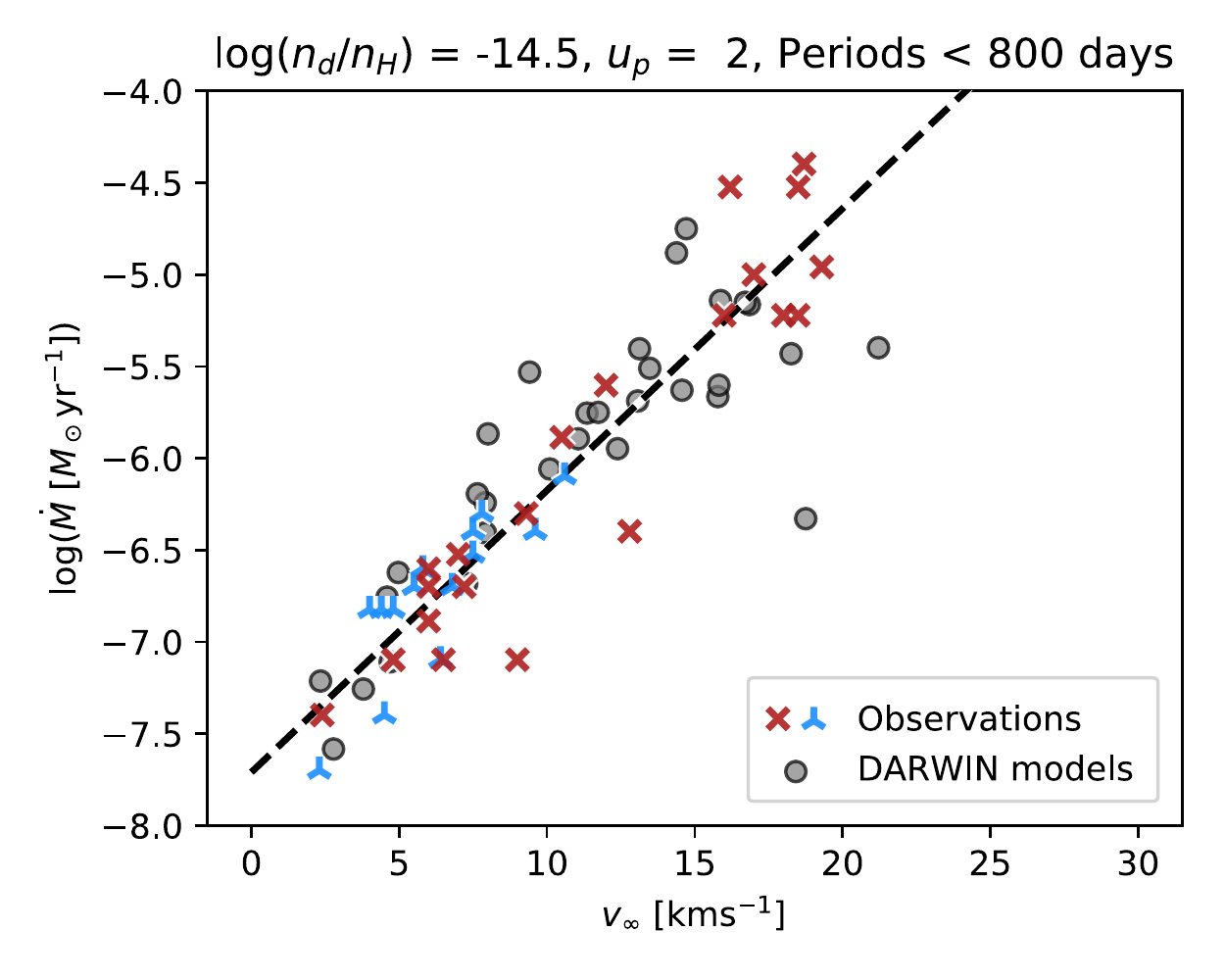}
\includegraphics[width=0.47\textwidth]{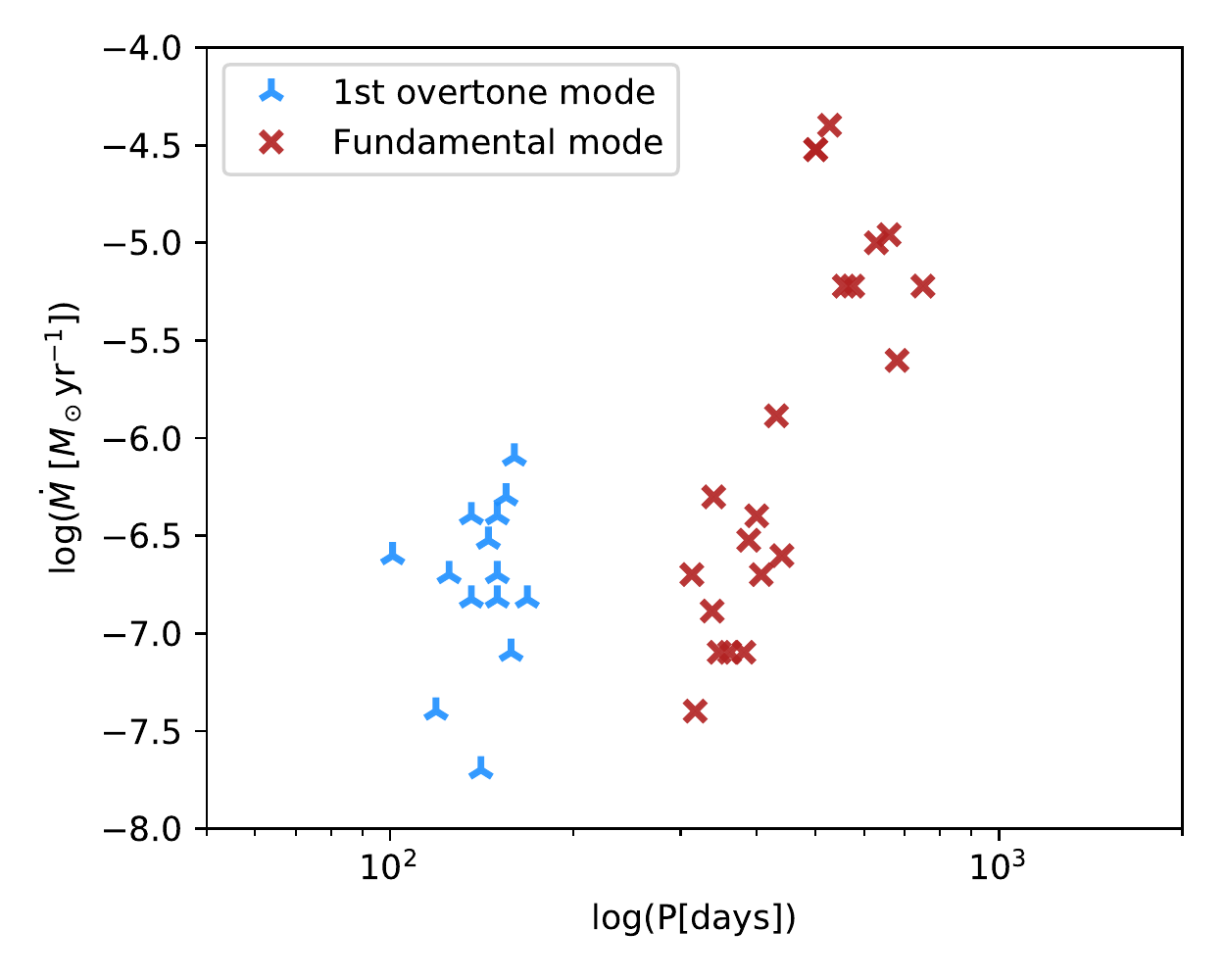}
\includegraphics[width=0.47\textwidth]{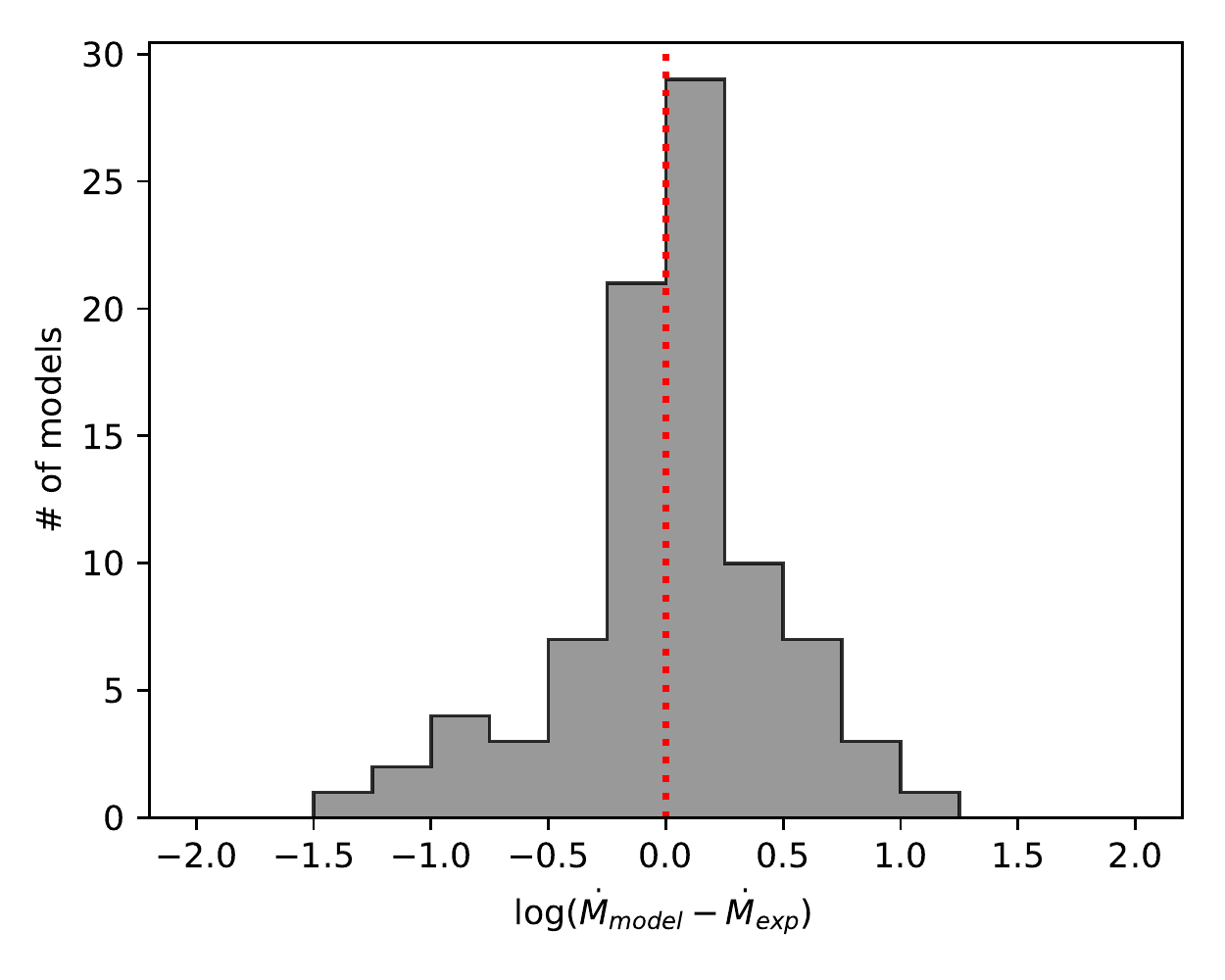}
\caption{\textit{Top panel:} Mass-loss rates and wind velocities of M-type AGB stars derived from observations of several CO lines \citep{olofsson02,gondel03}, and the corresponding properties of models with the combination of log($n_d / n_H) = -14.5$ and $u_p = 2.0$, and with $P<800$ days. The blue and red symbols indicate stars presumably pulsating in first overtone and fundamental mode, respectively. The dashed line is the linear fit to the observed values. \textit{Middle panel:} Mass-loss rate as a function of pulsation period for the same observational sample. \textit{Lower panel:} Difference  between the mass-loss rates of the models and the expected values from the observational linear fit.}
\label{fig:fit}
\end{figure}

The top panel of Fig.~\ref{fig:fit} shows the observed wind properties of nearby AGB stars of M-type, estimated from CO line emission \citep{olofsson02,gondel03}, as well as a linear fit to the observed values. These observed wind properties can be compared with wind properties derived from the model grid in order to evaluate what combinations of seed particle abundance and piston amplitude best reproduce the observational data. Constraining the pulsation period in the model data to less than 800 days to fit the periods of the observed sample, we found that models with log($n_d/n_H) = -14.5$ and $u_p = 2.0$ give the best agreement with observations. The bottom panel of Fig.~\ref{fig:fit} shows the distribution of the differences between model mass-loss rates and the expected mass-loss rates from the linear fit to the observed values.

The middle panel of Fig.~\ref{fig:fit} shows mass-loss rate as a function of pulsation period for the same observational sample as shown in the top panel. One way to asses how sensitive the mass-loss rates are to the pulsation mode is to look at the group of stars indicated by blue symbols in Fig.~\ref{fig:fit}. These stars are semiregular/irregular variables (SRVs), probably pulsating in the first overtone as indicated by their short pulsation periods. Even though these stars presumably pulsate in first overtone, they follow the same correlation between mass-loss rates and wind velocities as the fundamental mode pulsators (red symbols). This is clearly seen in the top panel of Fig.~\ref{fig:fit}, which shows the mass-loss rates and wind velocities for stars pulsating in different modes. Assuming that our reasoning about the stars with shorter periods being first overtone pulsators is correct, it seems that differences in the pulsation modes may not notably affect the mass-loss rates in M-type AGB stars.

\subsection{Trends at the longest pulsation periods}

\begin{figure}
\centering
\includegraphics[width=0.445\textwidth]{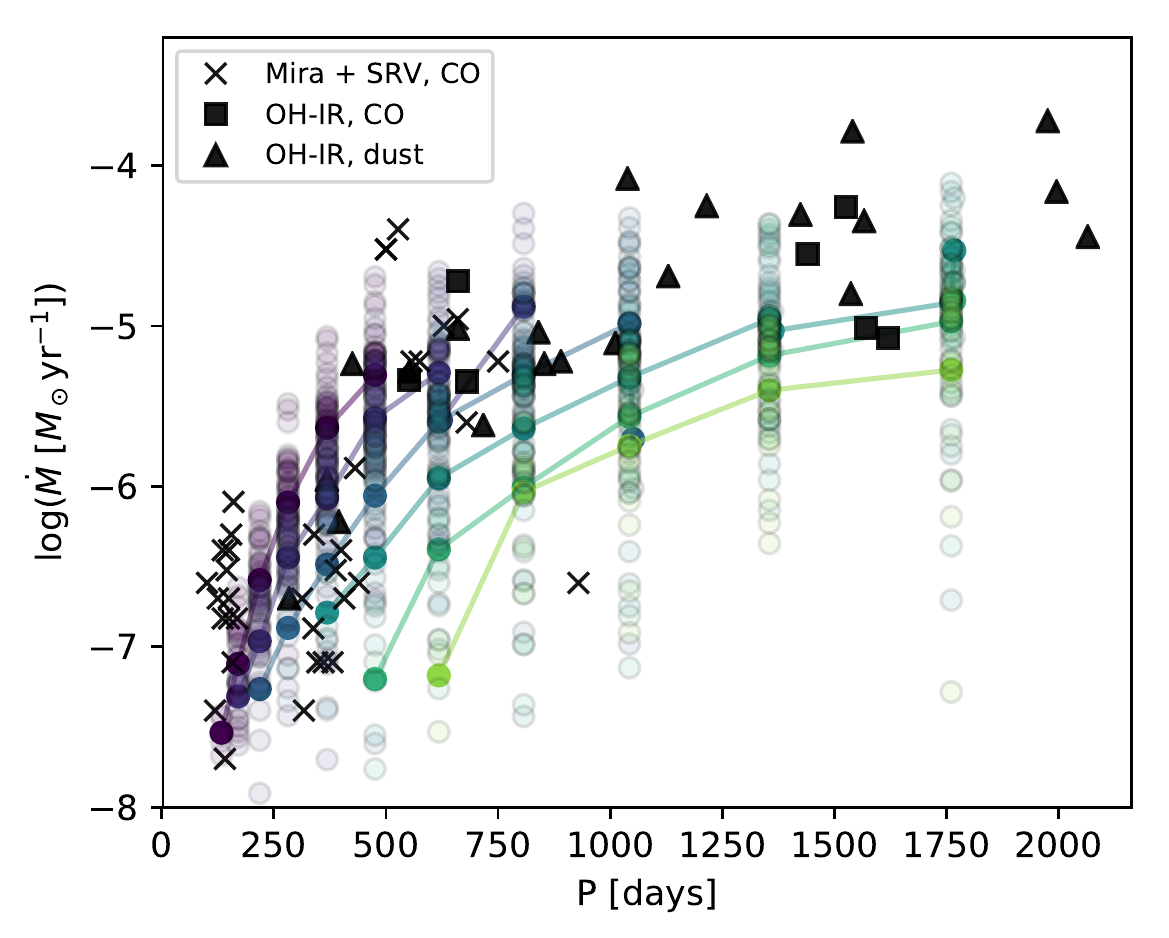} 
\caption{Mass-loss rate as a function of pulsation period for all models that develop a wind (indicated in red in Fig.~\ref{fig:windmaps}), colour-coded according to stellar mass (as in Fig.~\ref{fig:per_dmdt1}). The linked points show the average mass-loss rate at the indicated periods for a specific mass. We also include observations of nearby M-type AGB stars \citep{olofsson02,gondel03} based on CO line emission, as well as a set of OH/IR stars \citep{debeck10,hechen01} with mass-loss estimates based on CO-lines and dust, respectively.}
\label{fig:per_dmdt2}
\end{figure}

 Figure~\ref{fig:per_dmdt2} shows the mass-loss rate as a function of pulsation period for all models in the grid with a stellar wind. Also shown are observed mass-loss rates estimated from CO-line emission against pulsation periods for M-type AGB stars from \cite{olofsson02} and \cite{gondel03}, as well as a sample of OH/IR stars taken from \cite{debeck10} and \cite{hechen01}, with mass-loss rates estimated from CO-line and dust emission, respectively. In  model data and in observations the mass-loss rates level off at longer pulsation periods, even if the models do not reach quite as high as the observed mass-loss rates. However, it is important to remember that the grid is currently biased against models with very high mass-loss rate, where dust formation and wind acceleration is very abrupt, resulting in non-converging dynamical structures (see Sect.~\ref{sec:windmaps}). The exact level where the mass-loss rate flattens in the model values might also depend on the inner boundary conditions of the models, for example the value of $f_{\mathrm{L}}$ or the phase-shift of the luminosity variation with respect to the radial variation of the surface layers \citep{Liljegren16,liljegren17}. It might also be affected by limitations in the parameter space of the grid.

\begin{figure}
\centering
\includegraphics[width=0.44\textwidth]{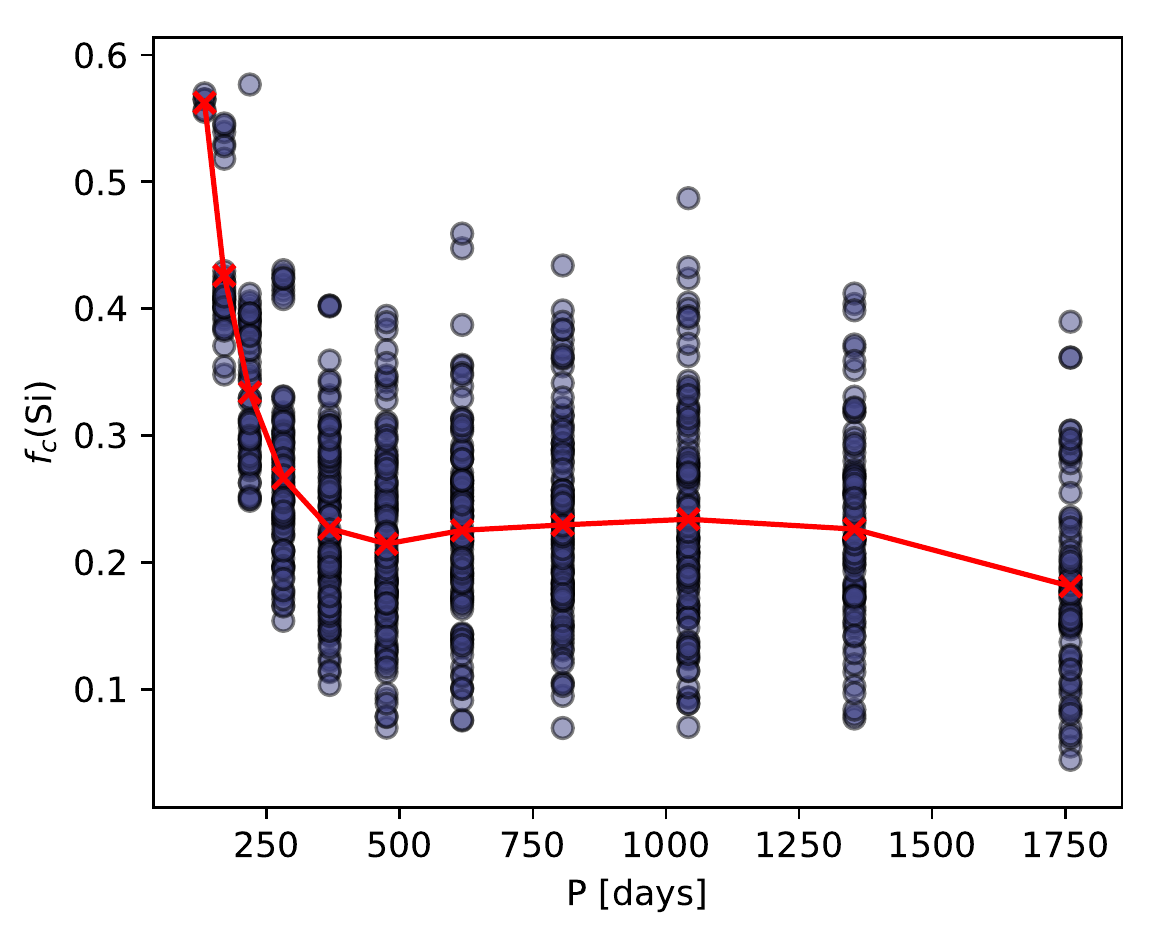}
   \caption{Degree of condensed Si as a function of pulsation period. The linked points show the average degree of condensed Si at the indicated periods.}
      \label{fig:fcvsp}
\end{figure}

To check that the levelling off of the mass-loss rates with increasing period is not simply caused by less efficient dust formation when the pulsation period increases, we also plot the degree of condensed Si as a function of pulsation period in Fig.~\ref{fig:fcvsp}. It appears that the amount of condensed Si does not change significantly at longer pulsation periods. However, it strongly increases at shorter periods (corresponding to lower luminosities), which raises questions about dust-driven winds in AGB stars pulsating in the first overtone. AGB stars pulsating in first overtone do not have as large pulsation amplitudes as Mira variables. Since large pulsation amplitude facilitates dust formation---gas is pushed further away from the star to cooler temperatures---this sets limits for dust-driven winds in such stars. However, first overtone pulsators with high luminosities might still have pulsation amplitudes large enough to produce dust-driven outflows \cite[see e.g. Fig.~1 in][]{Soszy2013}, especially if the efficiency of dust formation increases at shorter pulsation periods.   

\subsection{Summary: comparison of wind properties}

\begin{figure}
\centering
\includegraphics[width=0.49\textwidth]{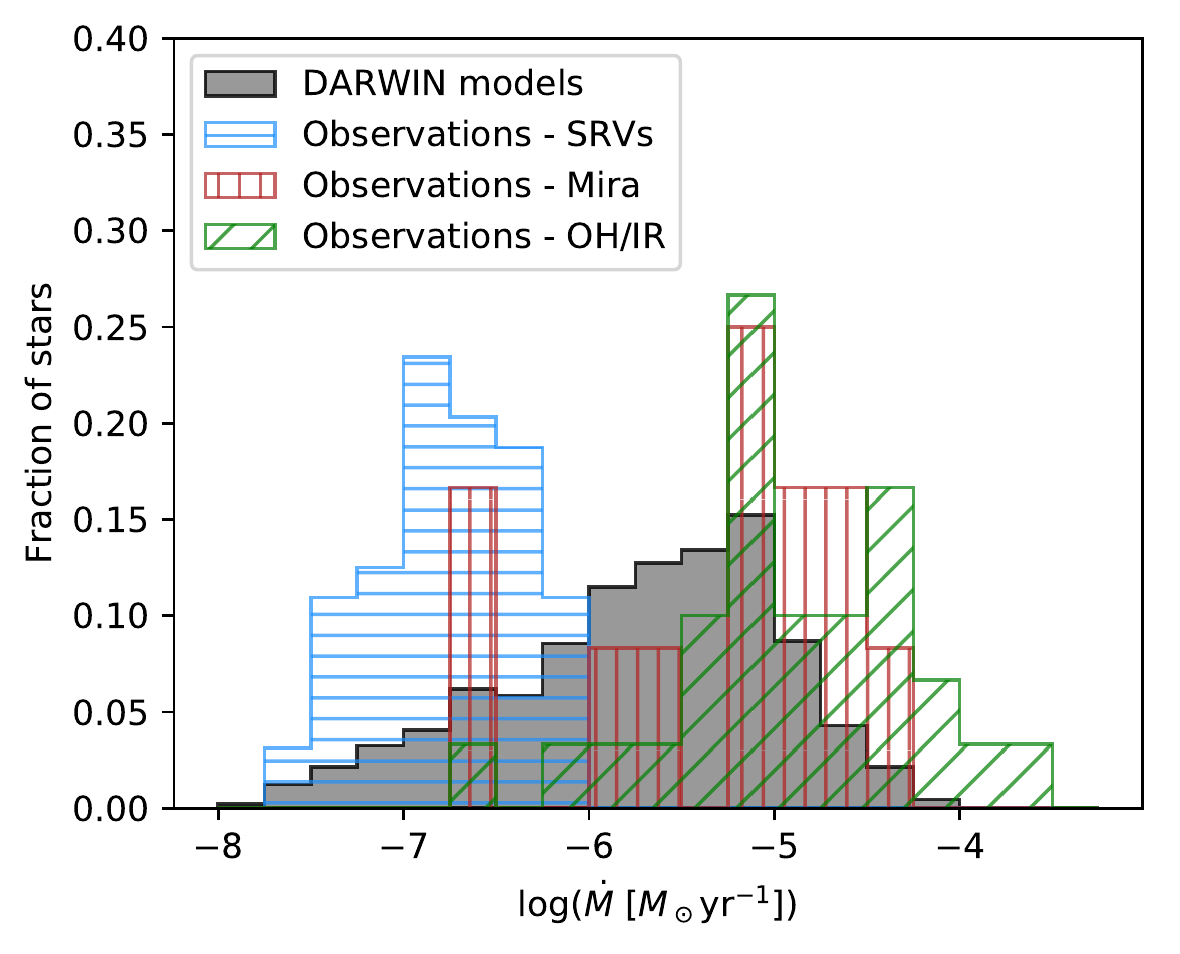}
\includegraphics[width=0.49\textwidth]{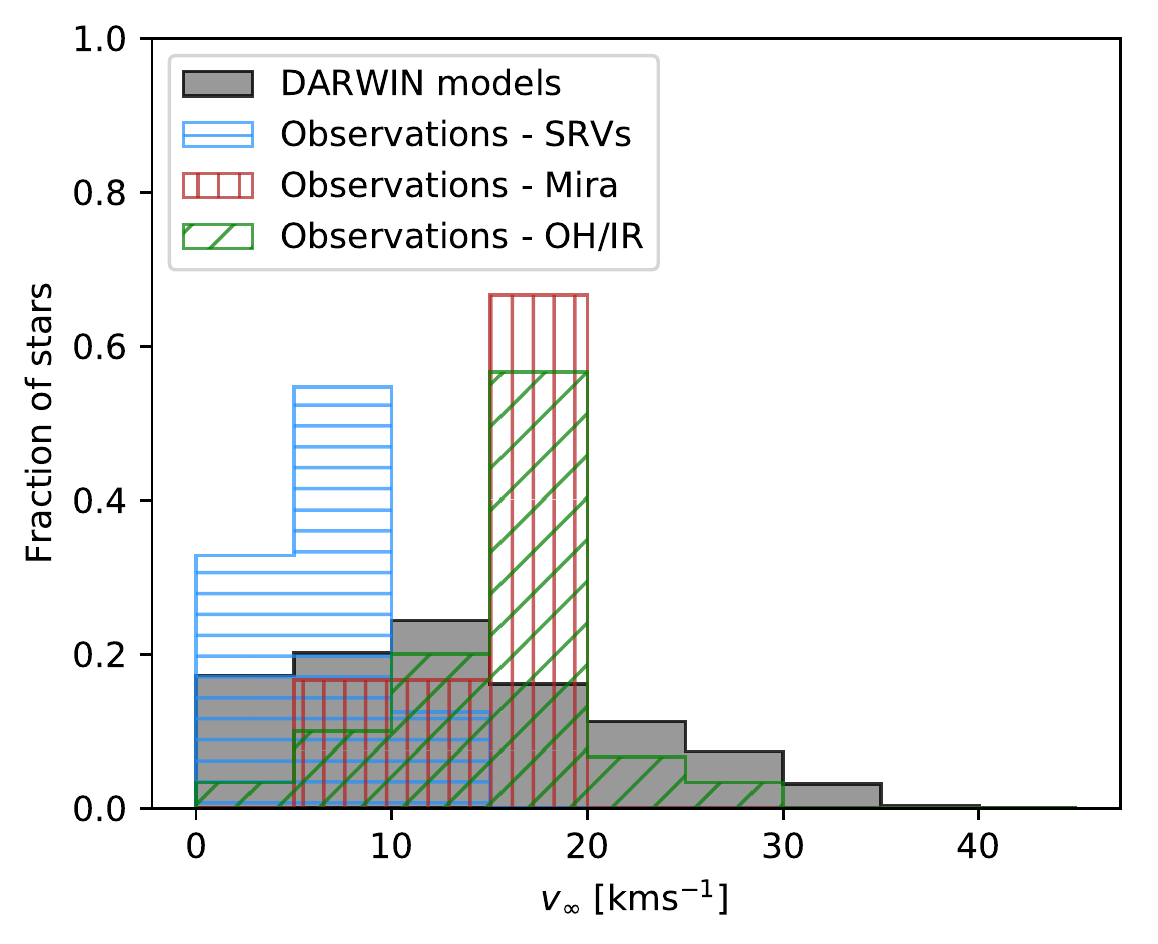}
\caption{Mass-loss rates (upper panel) and wind velocities (lower panel) for all DARWIN models with a stellar wind, and for three observational sets (SRVs, Miras, and OH/IR stars). The wind properties for SRVs and Mira variables are from \cite{olofsson02} and \cite{gondel03} and the wind properties for the OH/IR stars are from \cite{debeck10} and \cite{hechen01}. All the observed wind properties are estimated from CO-line emission, except for the sample from \cite{hechen01}, where the mass-loss rates are estimated from dust emission.}
\label{fig:hist}
\end{figure}

Generally, we manage to reproduce the observed wind properties well, especially of the nearby M-type AGB stars with well-determined wind velocities and mass-loss rates estimated from CO line emission. From Fig.~\ref{fig:per_dmdt2} it is clear that we also reproduce the observed mass-loss rates of OH/IR stars in the sample presented by \cite{debeck10}. However, the models fail to reach the high mass-loss rates observed in the sample by \cite{hechen01}. These results are further confirmed by Fig.~\ref{fig:hist}, which shows a histogram of the mass-loss rates and wind velocities from all DARWIN models with a stellar wind, together with observed wind properties from the nearby field M-type AGB stars and the OH/IR stars.  The observational dataset for the nearby M-type AGB stars is divided into two subsets, one consisting of Mira variables and one of SRVs.  We note that in the model grid every point has equal weight, which does not reflect how probable each combination of stellar parameters is in real AGB stars. Therefore, we  should not compare the frequency of the observational data and model  values in the histograms directly. Even so, it is clear that the models cover the range of observed data for both SRVs and Miras, and for most of the OH/IR stars. The very high mass-loss rates in the sample by \cite{hechen01} are not reproduced. This could be due to uncertainties in the estimated mass-loss rates and the dust-to-gas ratio assumed (which spans a wide range of values; see middle panel of Fig.~\ref{fig:nd_hist}). Moreover, the grid is biased against high mass-loss rates (see Sect.~\ref{gridpar}) which might explain why the models fail to reproduce the highest observed mass-loss rates. The grid also includes models with higher wind velocities than what is observed. The models producing these high wind velocities all have high stellar luminosities ($\log (L_*/L_{\odot})>4.30$) and low effective temperatures ($T_*<2700\,$K), a combination of stellar parameters that is not seen in the evolutionary tracks at solar metallicity calculated with the stellar evolution code COLIBRI (See Fig.~\ref{fig:evo_comp}).
It is interesting to note that DARWIN models for C-type AGB stars \citep{mattsson10,eriksson14} do not reproduce the observed SRVs, as the models for M-type AGB stars seem to do \citep[see Fig.~5 in][]{bladh2019}. The carbon stars not covered by DARWIN models are probably pulsating in first overtone, which suggests that the pulsations periods might affect the mass-loss rate more profoundly in C-type AGB stars.

\section{Mass-loss routine for stellar evolution models}
\label{stellarevo}
We provide a simple routine in FORTRAN for calculating the mass-loss rate as a function of stellar parameters (mass, luminosity, and effective temperature), based on data from the extensive grid of DARWIN models for M-type AGB stars presented in this paper. The resulting mass-loss rates are calculated by interpolation between $M_*, T_*,L_*$. The remaining parameters are selected according to the best fit in Sect.~\ref{compobssyn} (log($n_d / n_H) = -14.5$ and $u_p = 2.0$). The mass-loss routine is available online\footnote{\url{www.astro.uu.se/coolstars/TOOLS/MLR-routines/}} (a modified version of the mass-loss routine for C-type AGB stars presented in \cite{mattsson10} can also be found on this website).

This model grid should be usable for most M-type AGB stars during the AGB phase, at least until the C/O ratio has increased significantly. The surface abundances of evolved stars change as they experience first and second dredge-up, with the most important changes being an increase in N and He and a decrease in the C/O ratio \citep{karakas2014}. These changes affect the gas component, but should not affect the grain growth or the wind properties (see e.g. Figs.~5-6 in \citealt{aringer2016}). Since the abundance changes at first and second dredge-up depend on the stellar mass we assume a solar composition for the model grid for simplicity. The third dredge-up increases the C/O ratio, which eventually will limit the available oxygen and affect the grain growth of Mg$_2$SiO$_4$. This may favour MgSiO$_3$ as a wind-driving dust species in AGB stars of late S type, characterised by C/O close to unity. For AGB stars experiencing a mild hot-bottom burning (4\,\la\, M/M$_{\odot}$\,\la\,5), the effects should be similar to those during first and second dredge-up (increased N and decreased C/O). For AGB stars with efficient hot-bottom burning (typically M\,>\,5\,M$_{\odot}$ at solar metallicity), the activation of the O-N cycle and the Mg-Al chain may lead to a fair production of Al and depletion of O and Mg \citep{marigo13}. This will clearly reduce the grain growth of Mg-rich silicates and may instead favour Al-condensates such as spinel or corundum. For these stars the models should be used with caution.

\section{Outlook}
\label{outlook}
The collective output from the models in the grid (e.g. wind properties and photometry) can be compared to large observational samples of AGB stars, collected via programmes such as the OGLE Catalog of Variable Stars \citep{OLGELMC2009,OGLESMC2011,OGLEBULGE2013}, the Herschel programme MESS \citep{MESS2011}, the Sloan Digital Sky Survey \citep{SDSS2018}, and the forthcoming ALMA project DEATHSTAR\footnote{\url{www.astro.uu.se/deathstar}}. The DARWIN models can also provide mass-loss rates for stellar evolution codes modelling the AGB phase, as well as dust yields and dust-to-gas ratios used to estimate the dust contributions of AGB stars into the interstellar media. A first attempt to include mass-loss rates derived from DARWIN models for C-type AGB stars \citep{mattsson10,eriksson14} when modelling the stellar evolution of TP-AGB stars in the Small Magellanic Clouds using the COLIBRI code has recently been published by \cite{pastorelli2019}.

In addition to investigating the mass-loss mechanism from first principles, DARWIN models can also be used to study individual AGB stars by comparing synthetic spectra and photometry, visibilities, scattered light images, and wind properties to observational data \citep[e.g.][]{sacuto2013,rau17,witt17,aronson17,liljegren17,bladh17}. The availability of a model grid helps when modelling individual stars since the stellar parameters can be quite uncertain; often a range of different stellar parameters needs to be investigated. The new grid also includes more extreme luminosities and higher stellar masses, and can be used to study OH/IR stars. The population of OH/IR stars is not very large, but their high mass-loss rates and dusty envelopes make them important for the overall dust production in the interstellar media. Studies of these stars could also be important for understanding the chemical enrichment from hot-bottom burning or how silicates in M-type AGB stars are enriched with Fe or other impurities.

The episodic wind models in the grid, found in the region between models with stable outflows and models with no wind, are of interest for studying the effects of more complex inner boundary conditions. These boundary conditions can be derived from global 3D radiation-hydrodynamics models of AGB stars in which pulsations emerge \citep{freytag08,liljegren2018}. The 3D models by \cite{freytag17} are inherently non-spherical, with non-uniform shock fronts and luminosity amplitudes that vary from cycle to cycle, in contrast to the spherically symmetric DARWIN models. Boundary conditions derived from such 3D models, accounting for effects of giant convection cells and pulsation, are more irregular than the standard boundary conditions and might trigger intermittent outflows for models which produce no wind with the current set-up. This could shift the boundary between wind-producing models and windless models, which would be of importance for stellar evolution codes modelling the AGB phase.

\section{Conclusions}
\label{conclusions}
We have presented a new extensive grid of DARWIN models for M-type AGB stars, for the first time including models with different current stellar masses, and spanning a wide range in effective temperatures and stellar luminosities. Figure~\ref{fig:grid_comp} gives an overview of how significantly the parameter space has been increased compared to the previous study published in \cite{bladh2015}.
Dynamical properties (mass-loss rates and wind velocities), as well as grain properties (grain sizes, dust-to-gas ratios, and degree of condensed Si) are available online for every converging DARWIN model in the grid that produce a steady stellar wind. The dynamical output from this extensive model grid can be used in stellar evolution codes modelling the AGB phase, and we provide a simple mass-loss routine in FORTRAN for this purpose, and for detailed studies of individual AGB stars.

Schematic graphical representations of the wind properties are given in Fig.~\ref{fig:windmaps}, showing cross sections of the parameter space for a fixed stellar mass, organised as HR diagrams with increasing luminosity in the positive y-direction and decreasing effective temperature in the positive x-direction. 
These `windmaps' illustrate the combinations of stellar parameters (mass, luminosity, and effective temperature) that produce outflows, and where the boundary between models with and without stellar winds is situated. From the windmaps alone it is clear that the DARWIN models for M-type AGB stars produce outflows for a wide range of stellar parameters, although it is generally more difficult to drive winds for high stellar masses, high effective temperatures and low luminosities. There is also a distinct trend regarding the occurrence of stellar winds at higher effective temperatures and lower luminosities for smaller current mass.

The wind properties show trends similar to those of previous studies \citep{hoefner2008,bladh2015}. The mass-loss rate correlates strongly with luminosity, especially for a fixed stellar mass, as can be seen in panel a) of Fig.~\ref{fig:dmdtvel_props} and Fig.~\ref{fig:per_dmdt1}. The mass-loss rate also correlates strongly with the ratio $L_*/M_*$: increasing $L_*/M_*$ by an order of magnitude results in increasing the mass-loss rates by about three orders of magnitude, even though the spread is quite large. This trend is substantially steeper than indicated by Reimers’ formula, which may naturally lead to a superwind regime in evolution models. Lack of correlation between mass-loss rate and effective temperature is clearly seen in the bottom left panel of Fig.~\ref{fig:per_dmdt1}. This is a specific feature of our models for M-type AGB stars, whereas models for C-type stars show an increase in the mass-loss rate with decreasing effective temperature \citep{wachter02,eriksson14}. This difference is also seen in observations \citep[see discussion in][and references therein]{hoefner18}. It is presumably caused by the strong absorption by dust in the circumstellar environment of C-rich stars, giving rise to back-warming effects that affect dust formation. The thermal effects of the almost transparent visual and near-IR circumstellar environment of M-type models are much smaller. The dust grain radii in the grid range from 0.25\,\micron to 0.6\,\micron. The amount of condensed Si is typically between 10\% and 40\%, with gas-to-dust ratios between 500 and 4000.

The model grid reveals a levelling out of the mass-loss rate at higher luminosities, and consequently at longer pulsation periods ($P>800$ days; see Fig.~\ref{fig:per_dmdt2}). We investigated whether this phenomenon is simply caused by less efficient dust formation when the pulsation period increases (see Fig.~\ref{fig:fcvsp}), but the amount of condensed Si does not decrease significantly at longer pulsation periods. However, it strongly increases at the shortest periods, which raises questions about limits for dust-driven winds in AGB stars pulsating in the first overtone: first overtone pulsators with high luminosity might still have pulsation amplitudes large enough to produce dust-driven outflows if the efficiency of dust formation increases at shorter pulsation periods. Finally, the grid of DARWIN models generally reproduces the observed wind characteristics well, both for SRVs and Miras of M-type, and for most OH/IR stars. Only the very high mass-loss rates ($\dot{M}>10^{-4}\,\mathrm{M}_{\odot}$/yr) inferred for some OH/IR stars are not reached by the current grid.

\begin{acknowledgements}
This work has been supported by the Swedish Research Council (Vetenskapsrådet), by the Schönberg donation, and by the ERC Consolidator Grant funding scheme (project project STARKEY, G.A. No. 615604). The computations were performed on resources provided by the Swedish National Infrastructure for Computing (SNIC) at UPPMAX.
\end{acknowledgements}


\bibliographystyle{aa}
\bibliography{biblio}

\begin{appendix}
\section{A}
\label{appendixA} 

\begin{table*}
\caption{Fundamental model parameters (stellar, pulsation, and dust parameters) and the resulting wind and dust characteristics for DARWIN models producing a stellar wind (marked in red in Fig.~\ref{fig:windmaps}).}     
\label{tab:online}      
\centering          
\begin{tabular}{c c c  c  c  c c c  c c  c c c}   
\hline\hline     
$M_*$ & $\log (L_*/L_{\odot})$ & $T_*$ & $P$ & $\Delta u_{\mathrm{p}}$ & $f_{\mathrm{L}}$ & $\log (n_{\mathrm{d}}/n_{\mathrm{H}})$ & $\dot{M}$ & $u_{\mathrm{inf}}$ & $a_{\mathrm{gr}}$ & $f_{\mathrm{Si}}$ & $\rho_{\mathrm{g}}/\rho_{\mathrm{d}}$ \\ 
~[M$_{\odot}$]  &                        & [K] & [d] & [km/s] & &  & [M$_{\odot}$/yr] & [km/s] & [$\mu$m] & & \\
\hline

\hline
0.75 &3.55& 2500 & 283 &3.0 & 2.0 & -14.5 &6.80E-07& 8.3&0.37&0.25&1137.8

\end{tabular}
\tablefoot{This table is available in its entirety in the online journal. A portion is shown here for guidance regarding its form and content.}
\end{table*}

\end{appendix} 

\end{document}